\documentclass[a4paper,12pt]{article}

\usepackage{amsmath,amsthm,amsbsy,amssymb,amsfonts,graphicx,mathrsfs,bm,accents,cite}
\usepackage[all]{xy}

\makeatletter
\catcode`\@=11
\@addtoreset{equation}{section}

\makeatother

\renewcommand{\thefootnote}{\arabic{footnote}}

\newcommand{\Exp}[1]{\operatorname{e}^{#1}}
\newcommand{\abs}[1]{\lvert{#1} \rvert}
\newcommand{\rmd}{{\mathrm{d}}}
\newcommand{\nn}{\nonumber}
\newcommand{\Lie}{\pounds}
\newcommand{\gLie}{\hat{\pounds}}

\newcommand{\cA}{\mathcal A}

\newcommand{\cE}{\mathcal E}
\newcommand{\cF}{\mathcal F}
\newcommand{\cH}{\mathcal H}
\newcommand{\cL}{\mathcal L}
\newcommand{\cM}{\mathcal M}
\newcommand{\cP}{\mathcal P}
\newcommand{\cQ}{\mathcal Q}
\newcommand{\cZ}{\mathcal Z}

\newcommand{\sfh}{\mathsf{h}}
\newcommand{\sfm}{\mathsf{m}}
\newcommand{\sfn}{\mathsf{n}}

\newcommand{\sfx}{\mathsf{x}}
\newcommand{\sfy}{\mathsf{y}}
\newcommand{\sfB}{\mathsf{B}}
\newcommand{\sfC}{\mathsf{C}}
\newcommand{\sfM}{\mathsf{M}}
\newcommand{\sfN}{\mathsf{N}}

\newcommand{\tr}{\mathrm{tr}}
\newcommand{\rmT}{\mathrm{T}}
\newcommand{\rmM}{\text{\tiny M}}

\newcommand{\SL}{\mathrm{SL}}
\newcommand{\GL}{\mathrm{GL}}
\newcommand{\OO}{\mathrm{O}}

\newcommand{\BEG}{\mathsf{G}}

\newcommand{\Wa}{a}
\newcommand{\Wb}{b}
\newcommand{\Wc}{c}
\newcommand{\Wd}{d}
\newcommand{\We}{e}
\newcommand{\Wf}{f}
\newcommand{\Wg}{g}
\newcommand{\Wh}{h}

\newcommand{\Aphi}{\phi}
\newcommand{\Bphi}{\varphi}
\newcommand{\AG}{\mathfrak{g}}
\newcommand{\AB}{\mathfrak{B}}
\newcommand{\AC}{\mathfrak{C}}

\newcommand{\sMA}{{\mathtt{I}}}
\newcommand{\sMB}{{\mathtt{J}}}

\newcommand{\sBA}{{\mathtt{M}}}

\newcommand{\BA}{C}
\newcommand{\GA}{A}
\newcommand{\EE}{\cE}

\newcommand{\By}{{\sfy}}
\newcommand{\bbeta}{{\boldsymbol\eta}}
\newcommand{\bdelta}{{\boldsymbol\delta}}

\newcommand{\EPSpos}{}
\newcommand{\EPSneg}{-}
\newcommand{\EPSplus}{+}
\newcommand{\EPSminus}{-}

\allowdisplaybreaks[3]

\begin{document}

\begin{titlepage}
\renewcommand{\thefootnote}{\fnsymbol{footnote}}

\vspace*{1.0cm}

\begin{center}
\Large\textbf{Exceptional M-brane sigma models and $\eta$-symbols}
\end{center}

\vspace{1.0cm}

\centerline{
{Yuho Sakatani}%
\footnote{E-mail address: \texttt{yuho@koto.kpu-m.ac.jp}}
\ \ and \ \ 
{Shozo Uehara}%
\footnote{E-mail address: \texttt{uehara@koto.kpu-m.ac.jp}}
}

\vspace{0.2cm}

\begin{center}
{\it Department of Physics, Kyoto Prefectural University of Medicine,}\\
{\it Kyoto 606-0823, Japan}
\end{center}

\vspace*{1cm}

\begin{abstract}
We develop the M-brane actions proposed in arXiv:1607.04265 by using $\eta$-symbols determined in arXiv:1708.06342. Introducing $\eta$-forms that are defined with the $\eta$-symbols, we present $U$-duality-covariant M-brane actions which describe the known brane worldvolume theories for M$p$-branes with $p=0,2,5$. We show that the self-duality relation known in the double sigma model is naturally generalized to M-branes. In particular, for an M5-brane, the self-duality relation is nontrivially realized, where the Hodge star operator is defined with the familiar M5-brane metric while the $\eta$-form contains the self-dual 3-form field strength. The action for a Kaluza-Klein monopole is also partially reproduced. Moreover, we explain how to treat type IIB branes in our general formalism. As a demonstration, we reproduce the known action for a $(p,q)$-string. 
\end{abstract}
\thispagestyle{empty}
\end{titlepage}

\tableofcontents

\newpage

\setcounter{footnote}{0}

\section{Introduction}

String theory compactified on a $d$-torus has the $\OO(d,d)$ $T$-duality symmetry, but the duality is not manifest in the conventional formulation. 
A $T$-duality manifest formulation for strings, called the double sigma model (DSM), was originally developed in \cite{Duff:1989tf,Tseytlin:1990nb,Tseytlin:1990va,Siegel:1993xq,Hull:2004in,Hull:2006va}, where the dimensions of the target spacetime are doubled by introducing the dual winding coordinates. 
Utilizing the idea of the doubled spacetime, a manifestly $T$-duality-covariant formulation of low-energy superstrings was developed in \cite{Siegel:1993xq,Siegel:1993th,Siegel:1993bj,Hull:2009mi,Hohm:2010pp}, which is nowadays known as the double field theory (DFT). 
More recent studies on the DSM include \cite{Copland:2011wx,Lee:2013hma,Driezen:2016tnz,Park:2016sbw}. 
Other than the fundamental string, higher-dimensional objects also transform covariantly under $T$-duality. 
A $T$-duality-covariant action for D-branes was constructed in \cite{Asakawa:2012px} (see also \cite{Hull:2004in}) and a covariant action for a family of type II 5-branes [i.e.~NS5-brane, Kaluza--Klein Monopole (KKM), and the exotic $5^2_2$-brane] was constructed in \cite{Blair:2017hhy}. 

In fact, string theory compactified on a $(d-1)$-torus or M-theory on a $d$-torus has a larger duality symmetry generated by the $E_{d(d)}$ $U$-duality group. 
As a natural generalization of the $T$-duality-covariant string theory, $U$-duality-covariant membrane theory was first investigated in \cite{Duff:1990hn}. 
Moreover, by generalizing the idea of DFT, a manifestly $U$-duality-covariant formulation of supergravity, called the exceptional field theory (EFT), was developed in \cite{West:2000ga,West:2001as,Hillmann:2009pp,Berman:2010is,Berman:2011jh,Berman:2012vc,West:2012qz,Hohm:2013pua,Hohm:2013vpa,Hohm:2013uia,Hohm:2014fxa,Hohm:2015xna,Abzalov:2015ega,Musaev:2015ces,Berman:2015rcc}. 
Utilizing DFT/EFT, unified treatments of brane solutions were studied in \cite{Berkeley:2014nza,Berman:2014jsa,Berman:2014hna,Bakhmatov:2016kfn,Lee:2016qwn}. 
Further attempts at $U$-duality-manifest M-brane theories were made in \cite{Bengtsson:2004nj,West:2004iz,Linch:2015lwa,Linch:2015fya,Linch:2015qva,Linch:2015fca,Linch:2016ipx,Linch:2017eru}, but some obstacles to the manifestation of the whole $U$-duality symmetry are reported in \cite{Percacci:1994aa,Duff:2015jka,Hu:2016vym} (see Sect.~\ref{sec:duality} for more details on this point). 
Thus, it remains to be investigated whether we can formulate brane actions in a $U$-duality-covariant manner. 

In this paper, we develop the worldvolume theories for M-branes proposed in \cite{Sakatani:2016sko}. 
The proposed theory is based on the geometry of the exceptional spacetime (introduced in EFT) and can reproduce the conventional worldvolume theories for the M2-brane and M5-brane in a uniform manner. 
The action for an M$p$-brane takes the form
\begin{align}
 S = \EPSneg \frac{1}{p+1} \int_{\Sigma_{p+1}} \Bigl[\,\frac{1}{2}\,\cM_{IJ}(X)\, \cP^I\wedge *_\gamma \cP^J \EPSplus \Omega_{p+1}\,\Bigr] \,.
\end{align}
However, the $U$-duality covariance has not been manifest in $\Omega_{p+1}$. 
In this paper, by using the $\eta$-symbols recently determined in \cite{Sakatani:2017xcn}, we introduce a covariant object $\bbeta_{IJ}$, to be called the $\eta$-form, and propose a duality-covariant action that reproduces the above action. 
As we shall argue later, the $\eta$-form can be regarded as a natural generalization of the $\OO(d,d)$-invariant metric $\eta_{IJ}$ in DFT or DSM. 
Indeed, we show that the self-duality relation in DSM,
\begin{align}
 \eta_{IJ}\, \cP^J = \EPSneg \cH_{IJ} \, *_\gamma \cP^J \,,
\end{align}
can be naturally generalized to
\begin{align}
 \bbeta_{IJ}\wedge \cP^J = \EPSpos \cM_{IJ}\,*_\gamma \cP^J
\end{align}
for an M$p$-brane. 
Moreover, we argue that the action for a KKM can also be naturally reproduced in our formalism, although the whole action is not reproduced due to limitations of our analysis. 
We also demonstrate that our formalism can reproduce brane actions for type IIB branes. 

The present paper is organized as follows. 
In Sect.~\ref{sec:DSM-review}, we briefly review the DSM constructed in \cite{Lee:2013hma} and explain a slight difference from our approach. 
In Sect.~\ref{sec:M-branes}, we apply our approach to M-branes; M0, M2, M5-branes and KKM. 
In Sect.~\ref{sec:IIB-branes}, we explain how to apply our formalism to type IIB branes and reproduce the action for a $(p,q)$-string. 
A possible application to exotic branes is discussed in Sect.~\ref{sec:exotic}. 
Section \ref{sec:Conclusion} is devoted to conclusions and discussion.

\section{Double sigma model}
\label{sec:DSM-review}

In this section, we review the standard construction of the DSM and explain a slight difference from our approach. 
The difference is not significant in the DSM, but it becomes important when we consider higher dimensional objects in the following sections. 

\subsection{A brief review of double sigma model}

Let us begin with a brief review of Lee and Park's DSM \cite{Lee:2013hma} (known as the string sigma model on the doubled-yet-gauged spacetime). 
The action takes the form
\begin{align}
 S = \EPSneg \frac{1}{2} \int_{\Sigma_2} \Bigl[\,\frac{1}{2}\,\cH_{IJ}(X)\, D X^I\wedge *_\gamma D X^J \EPSplus \eta_{IJ}\,D X^I \wedge \cA^J \,\Bigr] \,,
\label{eq:L-P-action}
\end{align}
where $\eta_{IJ}$ is the $\OO(d,d)$-invariant metric, $\gamma_{\Wa\Wb}(\sigma)$ is the intrinsic metric on the worldsheet, $X^I(\sigma)$ is the embedding function of the string into the doubled spacetime, and $\cH_{IJ}(X)$ is the generalized metric satisfying the section condition $\partial^K\partial_K \cH_{IJ} = 0$\,. 
According to the section condition (or equivalently the coordinate gauge symmetry \cite{Park:2013mpa}), there are $d$ generalized Killing vectors, which take the form $\tilde{\partial}^i$ ($i=1,\dotsc,d$) when $\cH_{IJ}$ depends only on the $x^i$ coordinates. 
Associated to the isometries, we introduce 1-form gauge fields $\cA^I(\sigma)$ satisfying
\begin{align}
 \cA^I(\sigma)\,\partial_I T(x) = 0 
\label{eq:A-partial-0}
\end{align}
for an arbitrary supergravity field $T(x)$\,, and define the covariant derivative $D X^I(\sigma)\equiv \rmd X^I(\sigma) -\cA^I(\sigma)$\,. 

In order to see the equivalence to the conventional string sigma model, let us consider a duality frame where $\tilde{\partial}^k\cH_{IJ} = 0$ is realized. 
In such frame, \eqref{eq:A-partial-0} requires $\cA^I$ and $D X^I$ to have the following form:
\begin{align}
 \cA^I(\sigma) = \begin{pmatrix} 0 \\ \cA_i(\sigma) \end{pmatrix} , \qquad 
 D X^I = \rmd X^I - \cA^I = \begin{pmatrix} \rmd X^i \\ \rmd \tilde{X}_i - \cA_i \end{pmatrix} .
\label{eq:A-DX-canonical-section}
\end{align}
By further using the parameterization of the generalized metric
\begin{align}
 (\cH_{IJ}) = \begin{pmatrix} (G-B\,G^{-1}\,B)_{ij} & (B\,G^{-1})_i{}^j \\ -(G^{-1}\,B)^i{}_j & G^{ij} \end{pmatrix} ,
\end{align}
the action becomes
\begin{align}
 S &= \EPSneg \frac{1}{2} \int_{\Sigma_2} \Bigl[G_{ij}\,\rmd X^i\wedge *_\gamma \rmd X^j \EPSminus B_{ij}\,\rmd X^i\wedge \rmd X^j \EPSminus \rmd \tilde{X}_i\wedge \rmd X^i
\\
 &\quad + \frac{1}{2}\, G^{ij}\,\bigl(\cA_i -\rmd \tilde{X}_i + B_{ik}\,\rmd X^k \EPSminus G_{ik} *_\gamma \rmd X^k\bigr)\wedge *_\gamma\bigl(\cA_j-\rmd \tilde{X}_j + B_{jl}\,\rmd X^l \EPSminus G_{jl} *_\gamma \rmd X^l\bigr) \Bigr] \,. 
\nn
\end{align}
Eliminating the gauge fields $\cA_i$, we obtain the action
\begin{align}
 S = \EPSneg \frac{1}{2} \int_{\Sigma_2} \Bigl[\,G_{ij}(X) \,\rmd X^i\wedge *_\gamma \rmd X^j \EPSminus B_{ij}(X) \,\rmd X^i\wedge \rmd X^j \EPSminus \rmd \tilde{X}_i\wedge \rmd X^i\,\Bigr] \,,
\label{eq:DSMtoStandard-tilde}
\end{align}
which is the familiar sigma model action for the bosonic string up to a total-derivative term. 
The DSM is thus classically equivalent to the conventional string sigma model. 
The action \eqref{eq:L-P-action} is manifestly invariant under global $\OO(d,d)$ rotations and worldsheet diffeomorphisms. 
It is also invariant under generalized diffeomorphisms in the target doubled spacetime \cite{Lee:2013hma}. 

\subsection{Our approach}
\label{sec:DSM-ours}

In this paper, we basically follow the approach of Lee and Park, but there are slight differences. 
Following \cite{Sakatani:2017xcn}, we introduce a set of null generalized vectors $\lambda^a$ ($a=1,\dotsc,d$) satisfying
\begin{align}
 \lambda^a_I\,\eta^{IJ}\,\lambda^b_J = 0 \,. 
\end{align}
These $\lambda^a$ specify a solution of the section condition, and an arbitrary supergravity field $T(x)$ must satisfy the linear section equation \cite{Sakatani:2017xcn}
\begin{align}
 \lambda^a_I\,\eta^{IJ}\,\partial_J T(x) = 0 \,. 
\label{eq:double-linear-section}
\end{align}
For example, a choice
\begin{align}
 \lambda^a \equiv (\lambda^a_I) \equiv \begin{pmatrix} \lambda^a_i \\ \lambda^{i;\,a} \end{pmatrix} = \begin{pmatrix} \delta^a_i \\ 0 \end{pmatrix}
\label{eq:DSM-canonical-section}
\end{align}
corresponds to the section where supergravity fields satisfy $\tilde{\partial}^iT(x)=0$\,. 
For a given set of null generalized vectors $\lambda^a$ that specifies a section, we express the condition \eqref{eq:A-partial-0} for $\cA^I$ as
\begin{align}
 \cA^I(\sigma)\,\lambda^a_I = 0 \,.
\label{eq:A-lambda-0}
\end{align}
This is a minor difference (though it becomes important when we consider brane actions). 

\begin{figure}
 \centering
\includegraphics[width=0.6\linewidth]{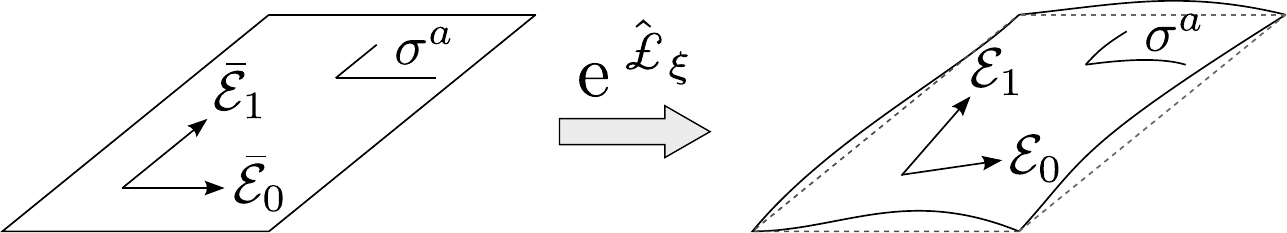}
\caption{Fluctuations of a string as active generalized diffeomorphisms.}
\label{fig:fluct}
\end{figure}

A major difference is in the parameterization of fluctuations. 
In the DSMs known in the literature, fluctuations of a string are described by the embedding function $X^I(\sigma)$, but we take a different approach, which is important in the generalization to branes. 
We first choose the section \eqref{eq:DSM-canonical-section}, where all fields and gauge parameters depend only on the physical coordinates $x^i$\,. 
We then prepare a static string worldsheet, where the tangent vectors to the worldsheet take the form
\begin{align}
 \bar{\EE}_a \equiv (\bar{\EE}_a^I) = \begin{pmatrix} \delta_a^i \\ 0 \end{pmatrix} \qquad (a=0,1)\,,
\end{align}
where the bar represents that the string is static. 
If we introduce a 1-form
\begin{align}
 \bar{\EE}^I \equiv \bar{\EE}_a^I\,\rmd \sigma^a = {\small \begin{pmatrix} \rmd \sigma^a \\ 0 \\ \vdots \\ 0 \end{pmatrix}} ,
\label{eq:DSM-Ebar}
\end{align}
it corresponds to $\rmd X^I(\sigma)$ of a string in the static gauge, $X^0(\sigma)=\sigma^0$ and $X^1(\sigma)=\sigma^1$. 
In order to describe a fluctuation of the string, we perform a finite active diffeomorphism along a gauge parameter $\xi^I(x)=(\xi^i,\,\tilde{\xi}_i)$ satisfying $\tilde{\partial}^i\xi^I=0$ (see Figure \ref{fig:fluct}). 
Under the section \eqref{eq:DSM-canonical-section}, a generalized diffeomorphism $\Exp{\gLie_\xi}$ can be decomposed into a $B$-field gauge transformation and a usual diffeomorphism
\begin{align}
 \Exp{\gLie_\xi} = \begin{pmatrix} \delta^i_k & 0 \\ F_{ik}(x') & \delta_i^k \end{pmatrix} \begin{pmatrix} \frac{\partial x'^k}{\partial x^j} & 0 \\ 0 & \frac{\partial x^j}{\partial x'^k} \end{pmatrix} ,
\label{eq:finite-transf}
\end{align}
where $x'^i=\Exp{\xi^j\partial_j}x^i$, $F_{ij}\equiv \partial_i A_j -\partial_j A_i$, and $A_i$ are complicated functions of $\xi^I$ (which coincide with $\tilde{\xi}_i$ when $\xi^i=0$). 
The usual diffeomorphism maps $\bar{\EE}^I$ as
\begin{align}
 \bar{\EE}^I(\sigma) \, \to \, \begin{pmatrix} \rmd X'^i(\sigma) \\ 0 \end{pmatrix} ,
\end{align}
and the $B$-field gauge transformation further maps it as
\begin{align}
 \bar{\EE}^I(\sigma) \, \overset{\text{diffeo.}}{\to} \, \begin{pmatrix} \rmd X'^i(\sigma) \\ 0 \end{pmatrix} \, \overset{\text{$B$-field gauge trsf.}}{\to}\, \begin{pmatrix} \rmd X'^i(\sigma) \\ F_{ij}(X'(\sigma))\, \rmd X'^j(\sigma) \end{pmatrix} . 
\end{align}
We thus introduce a generalized vector $\EE^I(\sigma)$, which describes fluctuations of a string, as
\begin{align}
 \EE^I(\sigma) \equiv \begin{pmatrix} \rmd X^i(\sigma) \\ F_{ij}(X(\sigma))\, \rmd X^j(\sigma) \end{pmatrix} , 
\label{eq:E-parameterization}
\end{align}
where the prime has been removed for simplicity. 
The scalar fields $X^i$ describe fluctuations of a string inside the $d$-dimensional physical subspace of the doubled target space (with coordinates $x^i$), while the 1-form $A_i$ describes the fluctuation along the dual directions in the doubled spacetime. 
In general, since the integrability condition is violated, i.e.~$(\partial_\tau\EE_\sigma-\partial_\sigma\EE_\tau)^I\neq 0$, we cannot find the embedding functions $X^I(\sigma)$ that realize $\EE^I(\sigma)=\rmd X^I(\sigma)$\,. 
However, inside the physical subspace, the integrability condition, $(\partial_\tau\EE_\sigma-\partial_\sigma\EE_\tau)^i=0$, is satisfied and the worldsheet is a manifold described by $X^i(\sigma)$ as usual. 
Thus, the violation in the dual components may be related to the gerbe structure discussed in \cite{Berman:2014jba}. 
In this paper, instead of assuming the existence of the embedding functions $X^I(\sigma)$, we parameterize fluctuations of a string by using the diffeomorphism parameters $\xi^I$, or equivalently $\{X^i(\sigma),\,A_i\bigl(X(\sigma)\bigr)\}$. 

Since $\EE^I$ is obtained by acting a generalized diffeomorphism on a generalized vector $\bar{\EE}^I$, $\EE^I$ also transforms as a generalized vector. 
Such behavior of $\EE^I$ is ensured as long as $F_{ij}$ transforms (like the $B$-field) as
\begin{align}
 \delta_V F_{ij} = \Lie_v F_{ij} + \partial_i \tilde{v}_j - \partial_j \tilde{v}_i \qquad
 \bigl(\delta_V A_i = v^k F_{ki} + \tilde{v}_i\bigr)\,, 
\end{align}
under an infinitesimal diffeomorphism. 
It should also be noted that $\EE^I$ is a null generalized vector; $\eta_{IJ}\,\EE^I\,\EE^J=0$. 
Assuming the null property, our parameterization \eqref{eq:E-parameterization} is the most general parameterization up to duality rotations. 

Now, our action is given by
\begin{align}
 S = \EPSneg \frac{1}{2} \int_{\Sigma_2} \Bigl[\,\frac{1}{2}\,\cH_{IJ}(X)\, \cP^I\wedge *_\gamma \cP^J \EPSplus \eta_{IJ}\,\cP^I \wedge \cA^J \,\Bigr] \,,
\label{eq:DSM-ours1}
\end{align}
which is simply obtained from \eqref{eq:L-P-action} with the replacements
\begin{align}
 \rmd X^I \to\, \EE^I\,,\qquad DX^I \to\, \cP^I \equiv \EE^I -\cA^I \,. 
\end{align}
In the duality frame \eqref{eq:DSM-canonical-section}, the condition \eqref{eq:A-lambda-0} leads to
\begin{align}
 \cA^I = \begin{pmatrix} 0 \\ \cA_i \end{pmatrix}  , \qquad 
 \cP^I = \EE^I - \cA^I = \begin{pmatrix} \rmd X^i \\ F_{ij}\bigl(X(\sigma)\bigr)\,\rmd X^j - \cA_i \end{pmatrix} \equiv \begin{pmatrix} \rmd X^i \\ \cP_i \end{pmatrix} ,
\end{align}
and, in the following, we consider $\cP_i$ as the fundamental variable rather than $\cA_i$\,. 
If we rewrite the action as
\begin{align}
 S = \EPSneg \frac{1}{2} \int_{\Sigma_2} \Bigl[\,\frac{1}{2}\,\cH_{IJ}(X)\, \cP^I\wedge *_\gamma \cP^J \EPSplus \eta_{IJ}\,\cP^I \wedge \EE^J \,\Bigr] \,,
\label{eq:DSM-ours1_5}
\end{align}
we observe that $F_{ij}$ appears only in the second term. 
In the second term, since the only quantity with an upper index is $\rmd X^i$ (other than the Kronecker delta), we see that $F_{ij}$ appears only through the pullback,
\begin{align}
 F_2(\sigma)\equiv \rmd A_1(\sigma)\,,\qquad A_1(\sigma)\equiv A_i\bigl(X(\sigma)\bigr)\,\rmd X^i \,.
\end{align}
Indeed, we can explicitly expand the second term as
\begin{align}
 S = \EPSneg \frac{1}{2} \int_{\Sigma_2} \Bigl[\,\frac{1}{2}\,\cH_{IJ}(X)\, \cP^I\wedge *_\gamma \cP^J \EPSplus \cP_i \wedge \rmd X^i - 2\,F_2 \,\Bigr] \,. 
\label{eq:DSM-ours2}
\end{align}
Therefore, the fundamental fields in our action are
\begin{align}
 \{X^i(\sigma),\,A_1(\sigma),\,\cP_i(\sigma),\,\gamma_{\Wa\Wb}(\sigma)\} \,. 
\end{align}
Namely, not all components of $A_i\bigl(X(\sigma)\bigr)$ appear in the action---only the pullback $A_1(\sigma)$ does. 
Eliminating the auxiliary fields $\cP_i(\sigma)$ by using their equations of motion, we obtain
\begin{align}
 S = \EPSneg \frac{1}{2} \int_{\Sigma_2} \Bigl[\,G_{ij}(X)\,\rmd X^i\wedge *_\gamma \rmd X^j \EPSminus B_{ij}(X)\,\rmd X^i\wedge \rmd X^j\,\Bigr] - \int_{\Sigma_2} F_2 \,. 
\end{align}
The main difference from Lee and Park's action is in the last term. 
The last term in \eqref{eq:DSMtoStandard-tilde} reproduces our $F_2$ if we regard $\rmd \tilde{X}_i$ as $F_{ij} \,\rmd X^j$.

Let us comment on the symmetry of the action \eqref{eq:DSM-ours1}. 
The invariance under the worldsheet diffeomorphism is manifest. 
Under an infinitesimal generalized diffeomorphism, $\EE^I$ transforms as a generalized vector and $\cA^I$ is also supposed to transform as a generalized vector. 
Then, since $\cH_{IJ}$ and $\eta_{IJ}$ are generalized tensors, \eqref{eq:DSM-ours1} is manifestly invariant under generalized diffeomorphism. 
The action is also formally covariant under global $\OO(d,d)$ rotations. 
In the $\OO(d,d)$ rotated frame, $\lambda^a_I$ no longer takes the form \eqref{eq:DSM-canonical-section}, and supergravity fields depend on another set of $d$ coordinates, which may contain the dual coordinates. 
The parameterization of the generalized vector $\EE^I$ is also changed since the physical subspace and the generalized diffeomorphism are changed (see Sect.~\ref{sec:duality} for more details). 

For later convenience, let us also comment on the self-dual relation \cite{Duff:1989tf,Hull:2004in,Hull:2006va,Lee:2013hma}. 
The equations of motion for the auxiliary fields can be written as
\begin{align}
 \cP_i = \EPSneg G_{ij}\,*_\gamma \rmd X^j + B_{ij}\, \rmd X^j \,. 
\end{align}
A duality-covariant rewriting of this equation is known as the self-dual relation, and takes the form
\begin{align}
 \eta_{IJ}\, \cP^J = \EPSneg \cH_{IJ}(X)\, *_\gamma \cP^J \,. 
\end{align}
In this paper, we find a similar self-dual relation for M-branes that determines all of the auxiliary fields in terms of the conventional fields. 

\section{M-branes in exceptional spacetime}
\label{sec:M-branes}

In this section, we consider worldvolume actions for M-branes. 
We decompose the eleven-dimensional spacetime into an $(11-d)$-dimensional ``external space'' and a $d$-dimensional ``internal space,'' and enlarge the internal space into an exceptional space with dimension $D=\dim R_1$, where $R_1$ is a fundamental representation of the $E_{d(d)}$ group (see Appendix \ref{app:Edd}). 
For simplicity, we disregard the external space and consider dynamics of branes in the internal space only. 
This assumption becomes less restrictive as $d$ becomes larger. 
In order to describe the time evolution, we include the time direction in the internal space. 

In Sect.~\ref{sec:Mp-branes}, we construct the brane actions for M$p$-branes $(p=0,2,5)$. 
The detailed properties and the equivalence to the conventional theories are studied in Sects.~\ref{sec:M0}--\ref{sec:M5}. 
The action for a KKM is discussed in Sect.~\ref{sec:mKKM}. 
In Sect.~\ref{sec:duality}, we discuss the $U$-duality covariance of our actions. 

\subsection{Action for an M$p$-brane}
\label{sec:Mp-branes}

In order to describe M-branes, we parameterize the generalized coordinates in the $E_{d(d)}$ exceptional spacetime ($d\leq 7$) as
\begin{align}
 (x^I) = \Bigl(x^i,\,\frac{y_{i_1i_2}}{\sqrt{2!}},\,\frac{y_{i_1\cdots i_5}}{\sqrt{5!}},\,\frac{y_{i_1\cdots i_7,\,i}}{\sqrt{7!}}\Bigr) \,,
\end{align}
where $i=1,\dotsc,d$ and $I=1,\dotsc,\dim R_1$. 
We also parameterize the generalized metric $\cM_{IJ}$ as follows by using the fields in the eleven-dimensional supergravity \cite{Berman:2011jh,Lee:2016qwn,Sakatani:2017nfr}:
\begin{align}
 &\cM_{IJ} = (L^\rmT\,\hat{\cM}\,L)_{IJ} \equiv \abs{G}^{\frac{1}{9-d}}\,\bar{\cM}_{IJ}\,,\quad 
 L\equiv \Exp{\frac{1}{3!}\,\BA_{i_1i_2i_3}\,R^{i_1i_2i_3}} \Exp{\frac{1}{6!}\,\BA_{i_1\cdots i_6}\,R^{i_1\cdots i_6}}\,, 
\\
 &\hat{\cM} \equiv \abs{G}^{\frac{1}{9-d}}\,
 \begin{pmatrix}
  G_{ij} & 0 & 0 & 0 \\
  0 & G^{i_1i_2,\,j_1j_2} & 0 & 0 \\
  0 & 0 & G^{i_1\cdots i_5,\,j_1\cdots j_5} & 0 \\
  0 & 0 & 0 & G^{i_1\cdots i_7,\,j_1\cdots j_7}\,G^{ij}
 \end{pmatrix} , 
\\
 &\Exp{\frac{1}{3!}\,\BA_{i_1i_2i_3}\,R^{i_1i_2i_3}} 
\nn\\
 &\equiv {\footnotesize
 \begin{pmatrix}
 \delta^i_j & 0 & 0 & 0 
\\
 -\frac{\BA_{i_1i_2 j}}{\sqrt{2!}} & \delta_{i_1i_2}^{j_1j_2} & 0 & 0 
\\
 -\frac{5!\delta_{i_1\cdots i_5}^{k_1\cdots k_5}\BA_{k_1k_2k_3} \BA_{k_4k_5 j}}{2!\,3!\,2!\sqrt{5!}} 
 & \frac{5!\delta^{j_1j_2k_1k_2k_3}_{i_1\cdots i_5} \BA_{k_1k_2k_3}}{3!\sqrt{2!\,5!}} 
 & \delta_{i_1\cdots i_5}^{j_1\cdots j_5} & 0 \\
 -\frac{7!\delta_{i_1\cdots i_7}^{k_1k_1p_1p_2p_3q_1q_2} \BA_{i k_1k_2}\BA_{p_1p_2p_3}\BA_{q_1q_2 j}}{3!\,2!\,3!\,2!\sqrt{7!}} 
 & \frac{7!\delta_{i_1\cdots i_7}^{j_1j_2 k_1k_2l_1l_2l_3} \BA_{i k_1k_2}\BA_{l_1l_2l_3}}{2!\,2!\,3!\sqrt{2!\,7!}} 
 & \frac{7!\delta^{j_1\cdots j_5k_1k_2}_{i_1\cdots i_7}\BA_{i k_1k_2}}{2!\sqrt{5!\,7!}} 
 & \!\!\delta_{i_1\cdots i_7}^{j_1\cdots j_7}\delta^j_i
 \end{pmatrix}},
\\
 &\Exp{\frac{1}{6!}\,\BA_{i_1\cdots i_6}\,R^{i_1\cdots i_6}} \equiv {\footnotesize
 \begin{pmatrix}
 \delta^i_j & 0 & 0 & 0 
\\
 0 & \delta_{i_1i_2}^{j_1j_2} & 0 & 0 
\\
 \frac{\BA_{i_1\cdots i_5 j}}{\sqrt{5!}} & 0 & \delta_{i_1\cdots i_5}^{j_1\cdots j_5} & 0
\\
 0 & -\frac{7!\,\delta^{j_1j_2k_1\cdots k_5}_{i_1\cdots i_7}\,\BA_{i k_1\cdots k_5}}{5!\sqrt{2!\,7!}} & 0 & \delta_{i_1\cdots i_7}^{j_1\cdots j_7}\,\delta^j_i
 \end{pmatrix}} ,
\\
 &L = \! {\footnotesize
 \begin{pmatrix}
 \delta^i_j & 0 & 0 & 0 
\\
 -\frac{\BA_{i_1i_2 j}}{\sqrt{2!}} & \delta_{i_1i_2}^{j_1j_2} & 0 & 0 
\\
 \frac{\BA_{i_1\cdots i_5 j}-5 \BA_{[i_1i_2i_3} \BA_{i_4i_5] j}}{\sqrt{5!}} 
 & \frac{20 \delta^{j_1j_2}_{[i_1i_2} \BA_{i_3i_4i_5]}}{\sqrt{2!\,5!}} 
 & \delta_{i_1\cdots i_5}^{j_1\cdots j_5} & 0 \\
 \frac{21\BA_{i[i_1i_2} (\BA_{i_3\cdots i_7] j}-\frac{5}{3}\BA_{i_3i_4i_5}\BA_{i_6i_7] j})}{\sqrt{7!}} 
 &\! -\frac{42\delta_{[i_1i_2}^{j_1j_2} (\BA_{|i| i_3\cdots i_7]}-5 \BA_{|i| i_3i_4}\BA_{i_5i_6i_7]})}{\sqrt{2!\,7!}} \!
 &\!\! \frac{7!\delta_{[i_1\cdots i_5}^{j_1\cdots j_5}\BA_{i_6i_7] i}}{2!\sqrt{5!\,7!}} \!\!
 &\!\! \delta_{i_1\cdots i_7}^{j_1\cdots j_7}\delta^j_i
 \end{pmatrix}} . 
\end{align}
Here, $(G_{ij},\,\BA_{i_1i_2i_3},\,\BA_{i_1\cdots i_6})$ are the conventional fields in the eleven-dimensional supergravity, $\abs{G}\equiv\det (G_{ij})$, and $(R^{i_1i_2i_3})^I{}_J$ and $(R^{i_1\cdots i_6})^I{}_J$ are $E_{d(d)}$ generators in the $R_1$-representation (see Appendix \ref{app:Edd} for the details). 
We also defined $\delta^{i_1\cdots i_p}_{j_1\cdots j_p}\equiv \delta^{[i_1}_{[j_1} \cdots \delta^{i_p]}_{j_p]}$ and $G^{i_1\cdots i_p,\,j_1\cdots j_p} \equiv G^{i_1k_1}\cdots G^{i_pk_p}\,\delta_{k_1\cdots k_p}^{j_1\cdots j_p}$\,. 

Similar to the case of the DSM, we specify the section by introducing a set of null generalized vectors $\lambda^a_I$ ($a=1,\dotsc,d$) satisfying \cite{Sakatani:2017xcn}
\begin{align}
 \lambda^a_I\,\eta^{IJ;\,\sMA}\,\lambda^b_J = 0 \,, \qquad 
 \lambda^a_I\,\Omega^{IJ}\,\lambda^b_J = 0 \,, 
\end{align}
where the explicit forms of $\eta^{IJ;\,\sMA}$ and $\Omega^{IJ}$ are given in Appendix \ref{app:eta-symbol}. 
For a given $\lambda^a_I$, the linear section equations for arbitrary supergravity fields and gauge parameters $T(x)$ become
\begin{align}
 \lambda^a_I\,\eta^{IJ;\,\sMA}\,\partial_J T(x) = 0 \,, \qquad 
 \lambda^a_I\,\Omega^{IJ}\,\partial_J T(x) = 0 \,. 
\label{eq:EFT-linear-section}
\end{align}
Using the same $\lambda^a$, we can express a condition for $\cA^I$ as
\begin{align}
 \cA^I(\sigma) \, \lambda^a_I = 0 \,, 
\end{align}
which corresponds to \eqref{eq:A-lambda-0} in the DSM. 
For a natural choice of $\lambda^a$,
\begin{align}
 (\lambda^a_I) = \begin{pmatrix} \lambda^a_i\\ \frac{\lambda^{i_1i_2;\,a}}{\sqrt{2!}}\\ \frac{\lambda^{i_1\cdots i_5;\,a}}{\sqrt{5!}}\\ \frac{\lambda^{i_1\cdots i_7,\,i;\,a}}{\sqrt{7!}} \end{pmatrix}
 = \begin{pmatrix} \delta^a_i\\ 0\\ 0\\ 0 \end{pmatrix} ,
\label{eq:M-theory-section}
\end{align}
supergravity fields depend only on the physical coordinates $x^i$ and $\cA^I$ takes the form
\begin{align}
 (\cA^I) = \begin{pmatrix} 0 \\ \frac{\cA_{i_1i_2}}{\sqrt{2!}} \\ \frac{\cA_{i_1\cdots i_5}}{\sqrt{5!}} \\ \frac{\cA_{i_1\cdots i_7,\,i}}{\sqrt{7!}} \end{pmatrix} . 
\end{align}

Similar to the DSM, we describe fluctuations of a $p$-brane by using the 1-form-valued null generalized vector $\EE^I(\sigma)$\,. 
In the case of the exceptional sigma model, we parameterize the null generalized vector as
\begin{align}
 \EE^I(\sigma) = (\cL^I{}_J)\, {\footnotesize \begin{pmatrix}
 \rmd X^j \\ 0 \\ 0 \\ 0 \end{pmatrix}}
 = {\footnotesize \begin{pmatrix}
 \rmd X^i 
\\
 \frac{F_{i_1i_2 j}\,\rmd X^j}{\sqrt{2!}} 
\\
 \frac{-(F_{i_1\cdots i_5 j}+5\, F_{[i_1i_2i_3}\, F_{i_4i_5] j})\,\rmd X^j}{\sqrt{5!}} \\
 \frac{21\,F_{i[i_1i_2}\, (F_{i_3\cdots i_7] j}+\frac{5}{3}\,F_{i_3i_4i_5}\,F_{i_6i_7] j})\,\rmd X^j}{\sqrt{7!}} 
 \end{pmatrix}} ,
\label{eq:E-M-brane}
\end{align}
where we defined
\begin{align}
\begin{split}
 &F_{i_1i_2i_3}(x) \equiv 3\,\partial_{[i_1} \GA_{i_2i_3]}(x)\,,\qquad
 F_{i_1\cdots i_6}(x) \equiv 6\,\partial_{[i_1} \GA_{i_2\cdots i_6]}(x)\,,
\\
 &\cL \equiv (\cL^I{}_J) \equiv \Exp{-\frac{1}{3!}\,F_{i_1i_2i_3}\,R^{i_1i_2i_3}} \Exp{-\frac{1}{6!}\,F_{i_1\cdots i_6}\,R^{i_1\cdots i_6}}\,. 
\end{split}
\end{align}
As in the string case, $X^i$, $\GA_{i_1i_2}$, and $\GA_{i_1\cdots i_5}$ are understood as functions of the diffeomorphism parameters $\xi^I$ that fluctuate a static brane. 
In order for $\EE^I$ to transform as a generalized vector, $F_{i_1i_2i_3}$ and $F_{i_1\cdots i_6}$ should transform as
\begin{align}
\begin{split}
 &\delta_V F_{i_1i_2i_3} = \Lie_v F_{i_1i_2i_3} - 3\,\partial_{[i_1} v_{i_2i_3]} \qquad 
 \bigl(\delta_V A_{i_1i_2} = v^k F_{ki_1i_2} - v_{i_1i_2}\bigr)\,,
\\
 &\delta_V F_{i_1\cdots i_6} = \Lie_v F_{i_1\cdots i_6} -30\,\partial_{[i_1} v_{i_2i_3}\,F_{i_4i_5i_6]}- 6\,\partial_{[i_1} v_{i_2\cdots i_5]}
\\
 &\bigl(\delta_V A_{i_1\cdots i_5} = v^k F_{ki_1\cdots i_5} -5\, v_{[i_1i_2}\,F_{i_3i_4i_5]}- v_{i_1\cdots i_5}\bigr)\,,
\end{split}
\label{eq:F3-F6-diffeo}
\end{align}
under an infinitesimal generalized diffeomorphism along $(V^I)=\bigl(v^i,\,\frac{v_{i_1i_2}}{\sqrt{2!}},\,\frac{v_{i_1\cdots i_5}}{\sqrt{5!}},\,\frac{v_{i_1\cdots i_7,\,k}}{\sqrt{7!}}\bigr)$. 
Now, we define the generalized vector $\cP^I(\sigma)$ as
\begin{align}
 \cP^I \equiv \EE^I - \cA^I 
 = {\footnotesize
 \begin{pmatrix}
 \rmd X^i 
\\
 \frac{F_{i_1i_2 j}\,\rmd X^j -\cA_{i_1i_2}}{\sqrt{2!}} 
\\
 \frac{-(F_{i_1\cdots i_5 j}+5\, F_{[i_1i_2i_3}\, F_{i_4i_5] j})\,\rmd X^j - \cA_{i_1\cdots i_5}}{\sqrt{5!}} \\
 \frac{21\,F_{i[i_1i_2}\, (F_{i_3\cdots i_7] j}+\frac{5}{3}\,F_{i_3i_4i_5}\,F_{i_6i_7] j})\,\rmd X^j -\cA_{i_1\cdots i_7,\,i}}{\sqrt{7!}} 
 \end{pmatrix}}
 \equiv \begin{pmatrix} \rmd X^i \\ \frac{\cP_{i_1i_2}}{\sqrt{2!}} \\ \frac{\cP_{i_1\cdots i_5}}{\sqrt{5!}} \\ \frac{\cP_{i_1\cdots i_7,\,i}}{\sqrt{7!}} 
 \end{pmatrix} ,
\end{align}
and regard $\{\cP_{i_1i_2},\, \cP_{i_1\cdots i_5},\, \cP_{i_1\cdots i_7,\,i}\}$ as the fundamental fields instead of $\{\cA_{i_1i_2},\, \cA_{i_1\cdots i_5},\, \cA_{i_1\cdots i_7,\,i}\}$\,. 

Unlike the doubled case, the $\eta$-symbols $\eta_{IJ;\,\sMA}$ in the $E_{d(d)}$ exceptional spacetime contain an additional index $\sMA$ \cite{Sakatani:2017xcn}. 
Then, in order to describe a $p$-brane, we introduce a $(p-1)$-form $\cQ^{\sMA}$ that transforms in the $R_2$-representation, and define a $(p-1)$-form-valued $\eta$-symbol
\begin{align}
 \bbeta_{IJ} \equiv \eta_{IJ;\,\sMA}\, \cQ^{\sMA}\,, 
\end{align}
which we call the $\eta$-form. 
In particular, when we consider an M$p$-brane ($p=0,2,5$), we choose $\cQ^{\sMA}$ as follows:
\begin{align}
 \cQ_{(\text{\tiny M0})}^{\sMA} \equiv {\footnotesize\begin{pmatrix} 0 \\ 0 \\ 0 \\ 0 \\ 0 \end{pmatrix}}, \qquad 
 \cQ_{(\text{\tiny M2})}^{\sMA} \equiv \frac{\mu_2}{2}{\footnotesize\begin{pmatrix} \rmd X^i \\ 0 \\ 0 \\ 0 \\ 0 \end{pmatrix}}, \qquad 
 \cQ_{(\text{\tiny M5})}^{\sMA} \equiv \frac{\mu_5}{5}{\footnotesize\begin{pmatrix} F_3\wedge\rmd X^i \\ \frac{\rmd X^{i_1\cdots i_4}}{\sqrt{4!}} \\ 0 \\ 0 \\ 0\end{pmatrix}}, 
\label{eq:M-Q-charge}
\end{align}
where $F_3\equiv \frac{1}{3!}\,F_{i_1i_2i_3}\,\rmd X^{i_1i_2i_3}$, $\mu_p$ are constants representing the brane charge, and we have introduced an abbreviated notation
\begin{align}
 \rmd X^{i_1\cdots i_p}\equiv \rmd X^{i_1}\wedge\cdots\wedge\rmd X^{i_p} \,.
\end{align}
Then, we propose the following actions:
\begin{align}
\begin{split}
 S_0 &= \EPSneg \frac{1}{2}\int_{\Sigma_1} \cM_{IJ}(X)\, \cP^I\wedge *_\gamma \cP^J \,,
\\
 S_2 &= \EPSneg \frac{1}{3}\int_{\Sigma_3} \Bigl[\,\frac{1}{2}\, \cM_{IJ}(X)\, \cP^I\wedge *_\gamma \cP^J \EPSminus \cP^I \wedge \bbeta^{(\text{\tiny M2})}_{IJ}\, \wedge \EE^J\,\Bigr] \,,
\\
 S_5 &= \EPSneg \frac{1}{6}\int_{\Sigma_6} \Bigl[\, \frac{1}{2}\, \cM_{IJ}(X)\, \cP^I\wedge *_\gamma \cP^J \EPSminus \cP^I \wedge \bbeta^{(\text{\tiny M5})}_{IJ}\, \wedge \EE^J \,\Bigr] \,, 
\end{split}
\label{eq:Mp-action}
\end{align}
where $\bbeta^{(\text{\tiny M$p$})}_{IJ}\equiv \eta_{IJ;\,\sMA}\, \cQ_{(\text{\tiny M$p$})}^{\sMA}$\,.

Note that the M5-brane charge \eqref{eq:M-Q-charge} has been obtained from the static ``purely M5-brane charge'' $\bar{\cQ}_{(\text{\tiny M5})}^{\sMA}$ through the active generalized diffeomorphism \eqref{eq:E-M-brane},
\begin{align}
 \bar{\cQ}_{(\text{\tiny M5})}^{\sMA} \equiv \frac{\mu_5}{5}{\footnotesize\begin{pmatrix} 0 \\ \frac{\rmd X^{i_1\cdots i_4}}{\sqrt{4!}} \\ 0 \\ 0 \\ 0\end{pmatrix}} \quad \to \quad 
 \cQ_{(\text{\tiny M5})}^{\sMA} = \cL^\sMA{}_\sMB \,\bar{\cQ}_{(\text{\tiny M5})}^{\sMB} 
 = \cQ_{(\text{\tiny M5})}^{\sMA} \equiv \frac{\mu_5}{5}{\footnotesize\begin{pmatrix} F_3\wedge\rmd X^i \\ \frac{\rmd X^{i_1\cdots i_4}}{\sqrt{4!}} \\ 0 \\ 0 \\ 0\end{pmatrix}} ,
\end{align}
where the transformation matrix $\cL^\sMA{}_\sMB$ for the $R_2$-representation is given by (see Appendix \ref{app:Edd})
\begin{align}
 \cL^\sMA{}_\sMB \equiv (\Exp{-\frac{1}{3!}\,F_{i_1i_2i_3}\,R^{i_1i_2i_3}} \Exp{-\frac{1}{6!}\,F_{i_1\cdots i_6}\,R^{i_1\cdots i_6}})^\sMA{}_\sMB\,.
\label{eq:cL-R2}
\end{align}
The M2-brane charge is invariant under the active diffeomorphism, $\cL^\sMA{}_\sMB\,\cQ_{(\text{\tiny M2})}^{\sMB}=\cQ_{(\text{\tiny M2})}^{\sMA}$\,. 
As long as $F_{i_1i_2i_3}$ and $F_{i_1\cdots i_6}$ behave as in \eqref{eq:F3-F6-diffeo}, $\cQ_{(\text{\tiny M$p$})}^{\sMA}$ transforms as a generalized vector in the $R_2$-representation and hence the $\eta$-form $\bbeta_{IJ}$ transforms as a generalized tensor.

In our actions, the generalized metric $\cM_{IJ}(X)$ includes an overall factor $\abs{G(X)}^{\frac{1}{9-d}}$, which is important for the duality covariance in EFT. 
However, it does not play an important role in the worldvolume theory because it can be absorbed into the intrinsic metric $\gamma_{\Wa\Wb}$. 
For convenience, we introduce an independent scalar field $\Exp{\bar{\omega}(\sigma)}$ inside $\cM_{IJ}(X)$ and regard the combination, $\Exp{\omega(\sigma)} \equiv \Exp{\bar{\omega}(\sigma)} \,\abs{G(X)}^{\frac{1}{9-d}}$, as a new fundamental field. 
Namely, in the following, when we denote $\cM_{IJ}(X)$ in the worldvolume action, it means
\begin{align}
 \cM_{IJ}(X) = \Exp{\omega(\sigma)} \bar{\cM}_{IJ}(X)\,,
\label{eq:GM-scalar-M}
\end{align}
and $\Exp{\omega(\sigma)}$ is an independent field. 
For a $p$-brane ($p\neq 1$), the action has a local symmetry,
\begin{align}
 \Exp{\omega(\sigma)} \to \Omega^{1-p}(\sigma)\,\Exp{\omega(\sigma)}\,,\qquad 
 \gamma_{\Wa\Wb}(\sigma) \to \Omega^{2}(\sigma)\,\gamma_{\Wa\Wb}(\sigma) \,, 
\end{align}
and $\Exp{\omega(\sigma)}$ is not a dynamical field. 
Indeed, as we see later, $\Exp{\omega}$ disappears from the action after eliminating $\gamma_{\Wa\Wb}$ by using the equation of motion. 
Only for the case of a string ($p=1$) in type IIB theory does the new scalar field $\Exp{\omega}$ play an important role (see Sect.~\ref{sec:pq-string}). 

Let us summarize the fundamental fields in our M-brane actions. 
There are always scalar fields $X^i(\sigma)$, auxiliary fields $\{\cP_{i_1i_2}(\sigma),\,\cP_{i_1\cdots i_5}(\sigma),\,\cP_{i_1\cdots i_7,\,i}(\sigma)\}$, and the intrinsic metric $\gamma_{\Wa\Wb}(\sigma)$. 
In addition, the generalized vector $\EE^I$ contains quantities like $F_{i_1i_2 j}\,\rmd X^j$ and $F_{i_1\cdots i_5 j}\,\rmd X^j$\,. 
As we explained in the doubled case, since all of the indices of $F_{i_1\cdots i_{p+1}}$ are contracted with $\rmd X^i$ in the action, only their pullbacks
\begin{align}
\begin{split}
 F_3(\sigma)&\equiv \rmd \GA_2(\sigma)\,,\qquad \GA_2(\sigma)\equiv \frac{1}{2!}\,\GA_{i_1i_2}\bigl(X(\sigma)\bigr)\,\rmd X^{i_1i_2}\,,
\\
 F_6(\sigma)&\equiv \rmd \GA_5(\sigma)\,,\qquad \GA_5(\sigma)\equiv \frac{1}{5!}\,\GA_{i_1\cdots i_5}\bigl(X(\sigma)\bigr)\,\rmd X^{i_1\cdots i_5}\,,
\end{split}
\end{align}
can appear in the action. 
Then, from the dimensionality, for example, $F_6$ cannot appear in the M2-brane action, and the fundamental fields can be summarized as follows:
\begin{align}
\begin{split}
 \text{M0-brane}:&\quad \{X^i(\sigma),\,\cP_{i_1i_2}(\sigma),\,\cP_{i_1\cdots i_5}(\sigma),\,\cP_{i_1\cdots i_7,\,i}(\sigma),\,\gamma_{\Wa\Wb}(\sigma),\, \omega(\sigma) \} \,,
\\
 \text{M2-brane}:&\quad \{X^i(\sigma),\,\cP_{i_1i_2}(\sigma),\,\cP_{i_1\cdots i_5}(\sigma),\,\cP_{i_1\cdots i_7,\,i}(\sigma),\,\gamma_{\Wa\Wb}(\sigma),\, \omega(\sigma) ,\,\GA_2(\sigma)\} \,,
\\
 \text{M5-brane}:&\quad \{X^i(\sigma),\,\cP_{i_1i_2}(\sigma),\,\cP_{i_1\cdots i_5}(\sigma),\,\cP_{i_1\cdots i_7,\,i}(\sigma),\,\gamma_{\Wa\Wb}(\sigma),\, \omega(\sigma) ,\,\GA_2(\sigma),\,\GA_5(\sigma)\} \,. 
\end{split}
\end{align}

Our action for an M$p$-brane ($p=0,2,5$) can be summarized as
\begin{align}
\begin{split}
 &S_p = \EPSneg \frac{1}{p+1}\int_{\Sigma_{p+1}} \Bigl[\,\frac{1}{2}\,\cM_{IJ}(X)\, \cP^I\wedge *_\gamma \cP^J \EPSminus \cP^I \wedge \bbeta^{(\text{\tiny M$p$})}_{IJ}\, \wedge \EE^J\, \Bigr] \,,\qquad 
 \bbeta^{(\text{\tiny M0})}_{IJ} = 0\,,
\\
 &\bbeta^{(\text{\tiny M2})}_{IJ} = \frac{\mu_2}{2}\, \eta_{IJ;\,k} \,\rmd X^k\,,\qquad
 \bbeta^{(\text{\tiny M5})}_{IJ} = \frac{\mu_5}{5}\, \Bigl(\frac{1}{4!}\,\eta_{IJ;\,k_1\cdots k_4}\,\rmd X^{k_1\cdots k_4}+ F_3\wedge\eta_{IJ;\,k}\,\rmd X^{k}\Bigr) \,.
\end{split}
\end{align}
This action is manifestly invariant under a generalized diffeomorphism along $V^I$,
\begin{align}
\begin{split}
 &\delta_V \cM_{IJ} = \gLie_V \cM_{IJ}\,,\qquad
 \delta_V X^i = v^i\,,
\\
 &\delta_V A_2 = \iota_v F_3 - v_2\,, \qquad
 \delta_V A_5 = \iota_v F_6 - \frac{1}{2}\,v_2\wedge F_3 - v_5 \,, 
\end{split}
\end{align}
where $v^i$ is restricted to be tangent to the worldvolume and we have defined $v_2\equiv \frac{1}{2!}\,v_{i_1i_2}\,\rmd X^{i_1i_2}$ and $v_5\equiv \frac{1}{5!}\,v_{i_1\cdots i_5}\,\rmd X^{i_1\cdots i_5}$\,. 
The covariance of our action under global $U$-duality rotations is discussed in Sect.~\ref{sec:duality}.

In order to expand the action explicitly, it is convenient to define the untwisted vector
\begin{align}
 &(\hat{\cP}^I) \equiv \begin{pmatrix}
 \rmd X^i \\ \frac{\hat{\cP}_{i_1i_2}}{\sqrt{2!}}\\ \frac{\hat{\cP}_{i_1\cdots i_5}}{\sqrt{5!}}\\ \frac{\hat{\cP}_{i_1\cdots i_7,i}}{\sqrt{7!}} 
 \end{pmatrix} \equiv L^I{}_J\,\cP^J \,,
\end{align}
where
\begin{align}
\begin{split}
 \hat{\cP}_{i_1i_2} &= \cP_{i_1i_2}-\BA_{i_1i_2 j}\,\rmd X^j \,,
\\
 \hat{\cP}_{i_1\cdots i_5}&= \cP_{i_1\cdots i_5} + 10\, \cP_{[i_1i_2}\, \BA_{i_3i_4i_5]} + \bigl(\BA_{i_1\cdots i_5 j}-5\, \BA_{[i_1i_2i_3}\, \BA_{i_4i_5] j}\bigr)\,\rmd X^j \,,
\\
 \hat{\cP}_{i_1\cdots i_7,i}&=\cP_{i_1\cdots i_7,\,i} + 21\,\cP_{[i_1\cdots i_5}\,\BA_{i_6i_7] i} - 21\,\cP_{[i_1i_2}\,\bigl(\BA_{|i| i_3\cdots i_7]}-5\, \BA_{|i| i_3i_4}\,\BA_{i_5i_6i_7]}\bigr) 
\\
 &\quad + 21\,\BA_{i[i_1i_2}\,\bigl(\BA_{i_3\cdots i_7] j}-\tfrac{5}{3}\,\BA_{i_3i_4i_5}\,\BA_{i_6i_7] j}\bigr)\,\rmd X^j \,. 
\end{split}
\end{align}
Then, we can expand the first term of the action as
\begin{align}
 \cM_{IJ}\,\cP^I\wedge *_\gamma\cP^J
 &= \Exp{\omega}\Bigl[G_{ij}\,\rmd X^i\wedge *_\gamma \rmd X^j
 +\frac{1}{2!}\,G^{i_1i_2,\,j_1j_2}\,\hat{\cP}_{i_1i_2}\wedge *_\gamma\hat{\cP}_{j_1j_2}
\nn\\
 &\qquad 
 +\frac{1}{5!}\,G^{i_1\cdots i_5,\,j_1\cdots j_5}\,\hat{\cP}_{i_1\cdots i_5}\wedge *_\gamma\hat{\cP}_{j_1\cdots j_5}
\nn\\
 &\qquad 
 +\frac{1}{7!}\,G^{i_1\cdots i_7,\,j_1\cdots j_7}\,G^{ij}\,\hat{\cP}_{i_1\cdots i_7,\,i}\wedge *_\gamma\hat{\cP}_{j_1\cdots j_7,\,j} \Bigr]\,.
\end{align}
We can also calculate the second term of the action as
\begin{align}
 &\cP^I \wedge \bbeta^{(\text{\tiny M2})}_{IJ}\, \wedge \EE^J =\frac{1}{2!}\,\cP_{i_1i_2}\wedge \rmd X^{i_1i_2} - 3 \,F_3 
\nn\\
 &\quad = \frac{1}{2!}\,\hat{\cP}_{i_1i_2}\wedge \rmd X^{i_1i_2} + 3 \,(\BA_3 - F_3) \,,
\\
 &\cP^I \wedge \bbeta^{(\text{\tiny M5})}_{IJ}\, \wedge \EE^J = \frac{1}{5!}\,\cP_{i_1\cdots i_5}\wedge \rmd X^{i_1\cdots i_5} + \frac{1}{2!}\, \cP_{i_1i_2}\wedge \rmd X^{i_1i_2}\wedge F_3 - 6 \,F_6 
\nn\\
 &\quad = \frac{1}{5!}\,\hat{\cP}_{i_1\cdots i_5}\wedge \rmd X^{i_1\cdots i_5} + \frac{1}{2!}\, \hat{\cP}_{i_1i_2}\wedge \rmd X^{i_1i_2}\wedge H_3 + 6 \,(\BA_6-F_6) + 3 \,\BA_3\wedge F_3 \,,
\end{align}
where
\begin{align}
 H_3\equiv F_3-\BA_3 \,. 
\end{align}
Note that $\cP^I \wedge \bbeta^{(\text{\tiny M2})}_{IJ}\, \wedge \EE^J$ and $\cP^I \wedge \bbeta^{(\text{\tiny M5})}_{IJ}\, \wedge \EE^J$ expressed in the above forms are the same as $\Omega_2$ and $\Omega_5$ introduced in \cite{Sakatani:2016sko} (up to conventions), and the actions presented above can be understood as a rewriting of the actions in \cite{Sakatani:2016sko} making the duality covariance manifest. 

For later convenience, we also define
\begin{align}
 \cZ_I \equiv \cM_{IJ}\,\cP^J\,,\qquad \hat{\cZ}_I \equiv (L^{-\rmT})_I{}^J\,\cZ_J = \cM_{IJ}\,\hat{\cP}^J \,. 
\end{align}

\subsection{M0-brane}
\label{sec:M0}

Let us consider the simplest example, the action for a particle in M-theory, sometimes called the M0-brane. 
The action is simply given by (see also \cite{Ko:2016dxa,Blair:2017gwn} for particle actions in extended spacetimes)
\begin{align}
 S_0 = \EPSneg \frac{1}{2}\int_{\Sigma_1} \cM_{IJ}(X)\, \cP^I\wedge *_\gamma \cP^J \,. 
\end{align}
The equations of motion for the auxiliary fields $\cP_{i_1i_2}$, $\cP_{i_1\cdots i_5}$, and $\cP_{i_1\cdots i_7,\,i}$ give
\begin{align}
 \hat{\cP}_{i_1i_2}=0\,,\qquad \hat{\cP}_{i_1\cdots i_5}=0\,,\qquad \hat{\cP}_{i_1\cdots i_7,\,i}=0\,,
\end{align}
and by eliminating the auxiliary fields, we obtain
\begin{align}
 S_0 = - \int \rmd \tau \,\frac{1}{2v}\, G_{ij}(X)\, \partial_\tau X^i\, \partial_\tau X^j \,,
\label{eq:S0-conventional}
\end{align}
where $v\equiv \Exp{-\omega}\sqrt{\abs{\gamma_{\tau\tau}}}\, \gamma^{\tau\tau}$\,. 
By considering $v$ as the fundamental variable (instead of the redundantly introduced fields $\omega$ and $\gamma_{\tau\tau}$), this is the bosonic part of the superparticle action discussed in \cite{Bergshoeff:1996tu}.

\subsubsection*{Type IIA branes: D0-brane}

For completeness, we review how to reproduce the D0-brane action from the above particle action \cite{Bergshoeff:1996tu}. 
By considering the reduction ansatz
\begin{align}
 (G_{ij})&\equiv \begin{pmatrix}
 G_{rs} & G_{r \rmM} \\ G_{\rmM s} & G_{\rmM\rmM}
 \end{pmatrix}
 =\begin{pmatrix}
 \Exp{-\frac{2}{3}\,\Aphi}\,\AG_{rs}+\Exp{\frac{4}{3}\,\Aphi}\,\AC_r\, \AC_s & \Exp{\frac{4}{3}\,\Aphi}\, \AC_r \\ \Exp{\frac{4}{3}\,\Aphi}\, \AC_s & \Exp{\frac{4}{3}\,\Aphi} \end{pmatrix}
\nn\\
 &=
 \begin{pmatrix}
 \delta_r^t & \AC_r \\ 0 & 1 \end{pmatrix}
 \begin{pmatrix}
 \Exp{-\frac{2}{3}\,\Aphi}\,\AG_{tu} & 0 \\ 0 & \Exp{\frac{4}{3}\,\Aphi} \end{pmatrix}
\begin{pmatrix}
 \delta^u_s & 0 \\ \AC_s & 1 \end{pmatrix} ,
\end{align}
where $r,s=1,\dotsc,d-1$ and $x^\rmM$ represents the M-theory direction, the action \eqref{eq:S0-conventional} becomes
\begin{align}
 S_0 = - \int \rmd \tau \,\frac{1}{2v}\, \Bigl[\Exp{-\frac{2}{3}\,\Aphi}\,\AG_{rs}(X)\, \partial_\tau X^r\, \partial_\tau X^s + \Exp{\frac{4}{3}\,\Aphi}\,\bigl(\partial_\tau X^\rmM + \AC_r\,\partial_\tau X^r\bigr)^2\Bigr] \,.
\end{align}
From the equations of motion for $X^\rmM$, we obtain
\begin{align}
 \Exp{\frac{4}{3}\,\Aphi}\, \bigl(\partial_\tau X^\rmM + \AC_r\,\partial_\tau X^r\bigr) = \mu\,v \,,
\end{align}
where $\mu$ is the integration constant, and using this, the action becomes
\begin{align}
 S_0 = - \int \rmd \tau \,\frac{1}{2}\, \Bigl[\frac{\Exp{-\frac{2}{3}\,\Aphi}}{v}\,\AG_{rs}(X)\, \partial_\tau X^r\, \partial_\tau X^s - v\,\mu^2\,\Exp{-\frac{4}{3}\,\Aphi} \Bigr] - \mu \int \rmd \tau \,\AC_r\,\partial_\tau X^r \,.
\end{align}
Here, we have added a total-derivative term $\mu\,\partial_\tau X^\rmM$. 
Using the equation of motion for $v$,
\begin{align}
 v^2\,\mu^2 = - \Exp{\frac{2}{3}\,\Aphi} \,\AG_{rs}(X)\, \partial_\tau X^r\, \partial_\tau X^s \,,
\end{align}
we obtain the standard D0-brane action
\begin{align}
 S_0 = - \abs{\mu}\int \rmd \tau \Exp{-\Aphi} \sqrt{-\AG_{rs}(X)\, \partial_\tau X^r\, \partial_\tau X^s} - \mu \int \AC_1 \,,
\end{align}
where $\AC_1\equiv \AC_r \,\partial_\tau X^r\,\rmd \tau$\,. 

\subsection{M2-brane}
\label{sec:M2}

Let us next consider the action for an M2-brane
\begin{align}
 S_2 &= \EPSneg \frac{1}{3} \int_{\Sigma_3} \Bigl[\,\frac{1}{2}\,\cM_{IJ}\,\cP^I\wedge *_\gamma\cP^J \EPSminus \cP^I \wedge \bbeta^{(\text{\tiny M2})}_{IJ}\, \wedge \EE^J\,\Bigr] 
\nn\\
 &= \EPSneg \frac{1}{3} \int_{\Sigma_3} \Bigl[\,\frac{1}{2}\,\cM_{IJ}\,\cP^I\wedge *_\gamma\cP^J \EPSminus \frac{\mu_2}{2!}\,\hat{\cP}_{i_1i_2}\wedge \rmd X^{i_1i_2} \,\Bigr] + \mu_2 \int_{\Sigma_3} (\BA_3 - F_3) \,. 
\end{align}
We derive the conventional action for the usual fields $X^i$ by using the equations of motion for auxiliary fields $\cP_{i_1\cdots i_3}$, $\cP_{i_1\cdots i_5}$, $\cP_{i_1\cdots i_7,\,i}$, and $\gamma_{\Wa\Wb}$. 
The equations of motion for $\cP_{i_1\cdots i_5}$ and $\cP_{i_1\cdots i_7,\,i}$ can be written as
\begin{align}
 \hat{\cP}_{i_1\cdots i_5}=0\,,\qquad \hat{\cP}_{i_1\cdots i_7,\,i} =0\,. 
\end{align}
Using these, the equation of motion for $\cP_{i_1i_2}$ becomes
\begin{align}
 \Exp{\omega}G^{i_1i_2,\,j_1j_2}\, *_\gamma\hat{\cP}_{j_1j_2} \EPSminus \mu_2\,\rmd X^{i_1i_2} = 0\,.
\end{align}
These equations of motion completely determine $\hat{\cP}^I$ and $\hat{\cZ}_I$ in terms of $X^i$ and $\gamma_{\Wa\Wb}$,
\begin{align}
 (\hat{\cP}^I) = \begin{pmatrix}
 \rmd X^i \\
 \frac{\EPSneg \mu_2\Exp{-\omega}G_{i_1i_2,\,j_1j_2}*_\gamma\rmd X^{j_1j_2}}{\sqrt{2!}} \\ 0 \\ 0
 \end{pmatrix} ,\qquad 
 (\hat{\cZ}_I) = \begin{pmatrix}
 \Exp{\omega}G_{ij}\,\rmd X^j \\
 \EPSneg \frac{\mu_2\,*_\gamma\rmd X^{i_1i_2}}{\sqrt{2!}} \\ 0 \\ 0
 \end{pmatrix} .
\end{align}
The intrinsic metric $\gamma_{\Wa\Wb}$ can also be determined by using its equation of motion,
\begin{align}
 \cM_{IJ}\,\cP_\Wa^I\,\cP_\Wb^J =0\,. 
\end{align}
Indeed, from this and the above solutions for $\hat{\cP}^I$, we obtain
\begin{align}
 h_{\Wa\Wb} &\equiv G_{ij}\,\partial_\Wa X^i\,\partial_\Wb X^j 
 = - \frac{1}{2!}\,G^{i_1i_2,\,j_1j_2}\,\hat{\cP}_{\Wa;\, i_1i_2}\,\hat{\cP}_{\Wb;\, j_1j_2} 
\nn\\
 &= - \frac{(\mu_2\Exp{-\omega})^2}{2!}\,G_{i_1i_2,\,j_1j_2}\,\varepsilon_\Wa{}^{\Wc_1\Wc_2}\,\varepsilon_\Wb{}^{\Wd_1\Wd_2}\,\partial_{\Wc_1}X^{i_1}\,\partial_{\Wc_2}X^{i_2} \,\partial_{\Wd_1}X^{j_1}\,\partial_{\Wd_2}X^{j_2} 
\nn\\
 &= (\mu_2\Exp{-\omega})^2 \,\frac{\det h}{\det\gamma}\,\bigl(\gamma\,h^{-1}\,\gamma\bigr)_{\Wa\Wb} \,.
\end{align}
This leads to
\begin{align}
 \frac{\det\gamma}{\det h} = (\mu_2\Exp{-\omega})^6 \,,\qquad 
 (\mu_2\Exp{-\omega})^4 \,\bigl(h\gamma^{-1}\,h\,\gamma^{-1}\bigr)_{\Wa}{}^{\Wb} = \delta_\Wa^\Wb \,. 
\end{align}
Note that if we define a matrix $R_\Wa{}^\Wb\equiv (\mu_2\Exp{-\omega})^2\,(h\gamma^{-1})_\Wa{}^\Wb$, it cannot vary (i.e.~$\delta R_\Wa{}^\Wb=0$) because of $(R^2)_\Wa{}^\Wb=\delta_\Wa^\Wb$\,. 
Therefore, if $R_\Wa{}^\Wb=\delta_\Wa^\Wb$ is satisfied at an initial time, it must be always satisfied, namely
\begin{align}
 \gamma_{\Wa\Wb} = (\mu_2\Exp{-\omega})^2 \,h_{\Wa\Wb} \,. 
\end{align}

Using the above equations of motion, the action for $X^i$ becomes
\begin{align}
 S_2 &= \frac{\mu_2}{3} \int_{\Sigma_3} \frac{1}{2!}\,\hat{\cP}_{i_1i_2}\wedge \rmd X^{i_1i_2} + \mu_2\int_{\Sigma_3} (\BA_3 - F_3)
\nn\\
 &= \EPSpos \frac{1}{3} \int_{\Sigma_3} \frac{\Exp{\omega}}{2}\,G^{i_1i_2,\,j_1j_2} \,\hat{\cP}_{i_1i_2}\wedge *_\gamma\hat{\cP}_{j_1j_2} + \mu_2\int_{\Sigma_3} (\BA_3 - F_3)
\nn\\
 &= \EPSneg \frac{1}{3} \int_{\Sigma_3} \Exp{\omega}G_{ij}\,\rmd X^i\,\wedge *_\gamma\rmd X^j + \mu_2\int_{\Sigma_3} (\BA_3 - F_3) 
\nn\\
 &= - \abs{\mu_2} \int_{\Sigma_3} \rmd^3\sigma\, \sqrt{-\det h} + \mu_2\int_{\Sigma_3} (\BA_3 - F_3) \,.
\label{eq:M2-conventional}
\end{align}
This is the bosonic part of the well-known membrane action \cite{Bergshoeff:1987cm}. 

Now, let us show the self-duality relation. 
Using the equations of motion, we can show
\begin{align}
 \bbeta^{(\text{\tiny M2})}_{IJ} \wedge \hat{\cP}^J 
 &= \begin{pmatrix}
 \EPSneg \frac{\mu_2^2\Exp{-2\omega}}{2}\,G_{ki,\,k_1k_2}\,\rmd X^k\wedge *_\gamma\rmd X^{k_1k_2} \\
 \frac{\mu_2\,\rmd X^{i_1i_2}}{\sqrt{2!}} \\ 0 \\ 0
 \end{pmatrix}
= \begin{pmatrix}
 \EPSpos \Exp{\omega}G_{ij}\,*_\gamma\rmd X^j \\
 \frac{\mu_2\,\rmd X^{i_1i_2}}{\sqrt{2!}} \\ 0 \\ 0
 \end{pmatrix} 
 = \EPSpos *_\gamma \hat{\cZ}_I \,. 
\label{eq:M2-Zhat}
\end{align}
This relation straightforwardly leads to the self-duality relation
\begin{align}
 \fbox{\quad$\displaystyle \bbeta^{(\text{\tiny M2})}_{IJ}\wedge \cP^J = \EPSpos \cM_{IJ} *_\gamma \cP^J \,.$\quad} 
\end{align}

\subsubsection*{Type IIA branes: D2-brane and F-string}

For completeness, let us review the derivation of the actions for a D2-brane and a fundamental string from the M2-brane action. 
In order to obtain the D2-brane action, we follow the procedure of \cite{Bergshoeff:1996tu}. 
We first rewrite the action \eqref{eq:M2-conventional} as
\begin{align}
 S_2 = \frac{\abs{\mu_2}}{2} \int_{\Sigma_3} \rmd^3\sigma\, \Bigl(\frac{\det h}{v} -v \Bigr) + \mu_2\int_{\Sigma_3} (\BA_3 - F_3) 
\end{align}
by introducing an auxiliary field $v$\,. 
Under the reduction ansatz
\begin{align}
\begin{split}
 (G_{ij})&= \begin{pmatrix}
 G_{rs} & G_{r \rmM} \\ G_{\rmM s} & G_{\rmM\rmM}
 \end{pmatrix}
 =\begin{pmatrix}
 \Exp{-\frac{2}{3}\,\Aphi}\,\AG_{rs}+\Exp{\frac{4}{3}\,\Aphi}\,\AC_r\, \AC_s & \Exp{\frac{4}{3}\,\Aphi}\, \AC_r \\ \Exp{\frac{4}{3}\,\Aphi}\, \AC_s & \Exp{\frac{4}{3}\,\Aphi} \end{pmatrix} , 
\\
 \BA_3 &= \AC_3 - \AB_2\wedge \AC_1 + \AB_2 \wedge (\rmd x^\rmM + \AC_1) \,, 
\end{split}
\end{align}
the action becomes
\begin{align}
 S_2 &= \frac{\abs{\mu_2}}{2} \int_{\Sigma_3} \rmd^3\sigma\, \biggl[\frac{\Exp{-2\Aphi}(\det \sfh)}{v}\,\bigl(1+\Exp{2\Aphi}\,\sfh^{\Wa\Wb}\,Y_\Wa\,Y_\Wb\bigr) -v \biggr] 
\nn\\
 &\quad + \mu_2\int_{\Sigma_3} \bigl[\AC_3 + \AB_2 \wedge (Y_1- \AC_1) - F_3\bigr] \,,
\end{align}
where $Y_1\equiv \rmd X^\rmM + \AC_1$ and we used the identity
\begin{align}
\begin{split}
 \det h &= \Exp{-2\Aphi}\det (\sfh_{\Wa\Wb}+\Exp{2\Aphi}\,Y_\Wa\,Y_\Wb) 
\\
 &= \Exp{-2\Aphi}(\det \sfh)\bigl(1+\Exp{2\Aphi}\,\sfh^{\Wa\Wb}\,Y_\Wa\,Y_\Wb\bigr) \qquad 
 \bigl(\sfh_{\Wa\Wb}\equiv \AG_{rs}\,\partial_\Wa X^r\,\partial_\Wb X^s\bigr) \,. 
\end{split}
\end{align}
By introducing a Lagrange multiplier $A_1$ that imposes the constraint $\rmd Y_1\equiv \rmd\AC_1$, we can rewrite the action as
\begin{align}
 S_2 &= \frac{\abs{\mu_2}}{2} \int_{\Sigma_3} \rmd^3\sigma\, \biggl[\frac{\Exp{-2\Aphi}(\det \sfh)}{v}\,\bigl(1+\Exp{2\Aphi}\,\sfh^{\Wa\Wb}\,Y_\Wa\,Y_\Wb\bigr) -v \biggr] 
\nn\\
 &\quad + \mu_2\int_{\Sigma_3} \bigl[\AC_3 + (\AB_2-F_2) \wedge (Y_1- \AC_1) - F_3\bigr] 
\nn\\
&= \frac{\abs{\mu_2}}{2} \int_{\Sigma_3} \rmd^3\sigma\, \biggl[\frac{\Exp{-2\Aphi} \det \sfh}{v} -v\,\Bigl(1+\frac{1}{2} \sfh^{\Wc\Wd}\,\sfh^{\We\Wf}\,\cF_{\Wc\Wd} \, \cF_{\We\Wf}\Bigr) \biggr] 
  + \mu_2\int_{\Sigma_3} \bigl(\AC_3 +\cF_2 \wedge \AC_1 - F_3\bigr)
\nn\\
 &\quad + \frac{\abs{\mu_2}}{2} \int_{\Sigma_3} \rmd^3\sigma\, \frac{(\det \sfh)}{v}\, \sfh^{\Wa\Wb}\,\biggl[Y_\Wa \EPSplus \frac{\frac{\mu_2}{2\abs{\mu_2}}\,v}{(\det \sfh)}\, \sfh_{\Wa\Wc}\,\epsilon^{\Wc\Wd\We}\,\cF_{\Wd\We}\biggr]\,\biggl[Y_\Wb \EPSplus \frac{\frac{\mu_2}{2\abs{\mu_2}}\,v}{(\det \sfh)}\, \sfh_{\Wb\Wf}\,\epsilon^{\Wf\Wg\Wh}\,\cF_{\Wg\Wh}\biggr] \,,
\end{align}
where we defined $F_2\equiv \rmd A_1$ and $\cF_2\equiv \rmd A_1 - \AB_2$, and $Y_1$ is regarded as a fundamental field. 
By eliminating $Y_\Wa$ and using
\begin{align}
 1+\frac{1}{2} \sfh^{\Wc\Wd}\,\sfh^{\We\Wf}\,\cF_{\Wc\Wd} \, \cF_{\We\Wf} = \frac{\det(\sfh+\cF)}{\det\sfh} \,,
\end{align}
we obtain
\begin{align}
 S_2 = \frac{\abs{\mu_2}}{2} \int_{\Sigma_3} \rmd^3\sigma\, \biggl[\frac{\Exp{-2\Aphi} \det \sfh}{v} -v\,\frac{\det(\sfh+\cF)}{\det\sfh} \biggr] 
  + \mu_2\int_{\Sigma_3} \bigl(\AC_3 +\cF_2 \wedge \AC_1 - F_3\bigr) \,.
\end{align}
Finally, using the equation of motion for $v$, we obtain the well-known D2-brane action
\begin{align}
 S_2 = -\abs{\mu_2} \int_{\Sigma_3} \rmd^3\sigma\, \Exp{-\Aphi} \sqrt{-\det(\sfh+\cF)} + \mu_2\int_{\Sigma_3} \bigl(\AC_3 +\cF_2 \wedge \AC_1 - F_3\bigr) \,.
\end{align}

On the other hand, when we derive the string action, we first make an ansatz,
\begin{align}
 X^r(\sigma^a) = X^r(\sigma^0,\,\sigma^1)\,,\qquad
 X^\rmM(\sigma^a) = \sigma^2\,,\qquad 
 \iota_{\frac{\partial}{\partial\sigma^2}} A_2(\sigma^a) = - A_1(\sigma^0,\,\sigma^1) \,.
\end{align}
Then, we can easily reproduce the Nambu--Goto-type action for a fundamental string
\begin{align}
 S_1 = - \abs{\mu_1} \int_{\Sigma_2} \rmd^2\sigma\, \sqrt{-\det \tilde{\sfh}} + \mu_1\int_{\Sigma_2} (\AB_2-F_2) \,,
\end{align}
where $\mu_1\equiv \mu_2\,(2\pi R_\rmM)$, $F_2\equiv \rmd A_1$, and $\det\tilde{\sfh}\equiv \det (\sfh_{\tilde{\Wa}\tilde{\Wb}})$ ($\tilde{\Wa},\,\tilde{\Wb}=0,1$).

\subsection{M5-brane}
\label{sec:M5}

Let us next consider an M5-brane action
\begin{align}
 S_5 &= \EPSneg\frac{1}{6} \int_{\Sigma_6} \Bigl[\,\frac{1}{2}\,\cM_{IJ}\,\cP^I\wedge *_\gamma\cP^J \EPSminus \cP^I \wedge \bbeta^{(\text{\tiny M5})}_{IJ}\, \wedge \EE^J\,\Bigr] 
\nn\\
 &= \EPSneg\frac{1}{6} \int_{\Sigma_6} \Bigl[\,\frac{1}{2}\,\cM_{IJ}\,\cP^I\wedge *_\gamma\cP^J \EPSminus \frac{\mu_5}{5!}\,\hat{\cP}_{i_1\cdots i_5}\wedge \rmd X^{i_1\cdots i_5} \EPSminus \frac{\mu_5}{2!}\, \hat{\cP}_{i_1i_2}\wedge \rmd X^{i_1i_2}\wedge H_3 \,\Bigr] 
\nn\\
 &\quad + \mu_5 \int_{\Sigma_6} \Bigl(\BA_6-F_6+ \frac{1}{2}\,\BA_3\wedge F_3\Bigr) \,,
\end{align}
where $H_3= F_3-\BA_3$\,. 

The equations of motion for $\cP_{i_1\cdots i_7,\,i}$ and $\cP_{i_1\cdots i_5}$ give
\begin{align}
 \hat{\cP}_{i_1\cdots i_7,\,i} =0\,, \qquad 
 \Exp{\omega}G^{i_1\cdots i_5,\,j_1\cdots j_5} *_\gamma\hat{\cP}_{j_1\cdots j_5} \EPSminus \mu_5 \,\rmd X^{i_1\cdots i_5} = 0\,.
\end{align}
From these, the equation of motion for $\cP_{i_1i_2}$ takes the following form:
\begin{align}
 0&= \sqrt{2!} \, \cM^{i_1i_2}{}_{J}\, *_\gamma\cP^J \EPSminus \mu_5 \,\rmd X^{i_1i_2}\wedge F_3 
\nn\\
 &= \Exp{\omega}G^{i_1i_2,\,j_1j_2} \, *_\gamma\hat{\cP}_{j_1j_2}
 + \frac{\Exp{\omega}}{3!}\,\delta^{i_1i_2}_{[j_1j_2}\,A_{j_3j_4j_5]}\,G^{j_1\cdots j_5,\,k_1\cdots k_5} \, *_\gamma\hat{\cP}_{k_1\cdots k_5}
  \EPSminus \mu_5 \, \rmd X^{i_1i_2}\wedge F_3 
\nn\\
 &= \Exp{\omega}G^{i_1i_2,\,j_1j_2} \, *_\gamma\hat{\cP}_{j_1j_2} \EPSminus \mu_5 \,\rmd X^{i_1i_2}\wedge H_3 \,.
\end{align}
Then, the equations of motion for auxiliary fields can be summarized as
\begin{align}
(\hat{\cP}^I) = \begin{pmatrix}
 \rmd X^i 
\\
 \EPSpos\frac{\mu_5\Exp{-\omega}G_{i_1i_2,\,j_1j_2}*_\gamma (\rmd X^{j_1j_2}\wedge H_3)}{\sqrt{2!}} 
\\
 \EPSpos\frac{\mu_5\Exp{-\omega}G_{i_1\cdots i_5,\,j_1\cdots j_5}\,*_\gamma\rmd X^{j_1\cdots j_5}}{\sqrt{5!}}
\\
 0
 \end{pmatrix} ,\qquad 
 (\hat{\cZ}_I) = \begin{pmatrix}
 \Exp{\omega}G_{ij}\,\rmd X^j 
\\
 \EPSpos\frac{\mu_5\,*_\gamma (\rmd X^{i_1i_2}\wedge H_3)}{\sqrt{2!}} 
\\
 \EPSpos\frac{\mu_5\,*_\gamma\rmd X^{i_1\cdots i_5}}{\sqrt{5!}}
\\
 0
 \end{pmatrix} .
\end{align}
It should be noted that if we compute $\cZ_I = (L^{\rmT})_I{}^J\,\hat{\cZ}_J$ for $d\leq 6$ as
\begin{align}
 (\cZ_I) = \begin{pmatrix}
 \Exp{\omega}G_{ij}\,\rmd X^j 
  \EPSminus \mu_5\,*_\gamma\bigl[\iota_i \BA_3 \wedge H_3 + \iota_i \BA_6 + \tfrac{1}{2}\,\iota_i \BA_3\wedge \BA_3 \bigr] 
\\
 \EPSpos\frac{\mu_5\,*_\gamma (\rmd X^{i_1i_2}\wedge F_3)}{\sqrt{2!}} 
\\
 \EPSpos\frac{\mu_5\,*_\gamma\rmd X^{i_1\cdots i_5}}{\sqrt{5!}}
 \end{pmatrix} ,
\end{align}
its time component appears to be reproducing the generalized momenta, Eq.~(2.9) in \cite{Hatsuda:2013dya}, obtained in the Hamiltonian analysis. 
The equation of motion for $\gamma_{\Wa\Wb}$ and the above solution for $\hat{\cP}^I$ give
\begin{align}
 h_{\Wa\Wb} &\equiv G_{ij}\,\partial_\Wa X^i\,\partial_\Wb X^j 
 = - \frac{1}{2!}\,G^{i_1i_2,\,j_1j_2}\,\hat{\cP}_{\Wa i_1i_2}\,\hat{\cP}_{\Wb j_1j_2}
  - \frac{1}{5!}\,G^{i_1\cdots i_5,\,j_1\cdots j_5}\,\hat{\cP}_{\Wa i_1\cdots i_5}\,\hat{\cP}_{\Wb j_1\cdots j_5}
\nn\\
 &= - \frac{1}{2!}\,G_{i_1i_2,\,j_1j_2}\,\frac{\mu_5\Exp{-\omega}}{3!}\,\varepsilon^{\Wc_1\cdots\Wc_5}{}_\Wa \,\partial_{\Wc_1}X^{i_1}\,\partial_{\Wc_2}X^{i_2}\,H_{\Wc_3\Wc_4\Wc_5} \,
 \frac{\mu_5\Exp{-\omega}}{3!}\,\varepsilon^{\Wd_1\cdots\Wd_5}{}_\Wb\,\partial_{\Wd_1}X^{j_1}\,\partial_{\Wd_2}X^{j_2}\,H_{\Wd_3\Wd_4\Wd_5}
\nn\\
 &\quad - \frac{(\mu_5\Exp{-\omega})^2}{5!}\,G_{i_1\cdots i_5,\,j_1\cdots j_5}\,\varepsilon^{\Wc_1\cdots\Wc_5}{}_\Wa \,\partial_{\Wc_1}X^{i_1}\cdots\partial_{\Wc_5}X^{i_5}\,\varepsilon^{\Wd_1\cdots\Wd_5}{}_\Wb \,\partial_{\Wd_1}X^{j_1}\cdots\partial_{\Wd_5}X^{j_5}
\nn\\
 &= (\mu_5\Exp{-\omega})^2 \,\frac{\det h}{\det\gamma}\, \Bigl[\frac{2}{3}\,h^{\Wc_3\Wc_4\Wc_5\We,\,\Wd_3\Wd_4\Wd_5\Wf}\,H_{\Wc_3\Wc_4\Wc_5}\,H_{\Wd_3\Wd_4\Wd_5}\,\gamma_{\Wa\We}\,\gamma_{\Wb\Wf}
  + (\gamma\,h^{-1}\,\gamma)_{\Wa\Wb}\Bigr] 
\nn\\
 &= (\mu_5\Exp{-\omega})^2 \,\frac{\det h}{\det\gamma}\, \gamma_{\Wa\Wc}\,\theta^{\Wc}{}_\Wd\,(h^{-1}\gamma)^\Wd{}_\Wb \,,
\end{align}
where we have defined $h^{\Wa_1\cdots\Wa_n,\,\Wb_1\cdots\Wb_n}\equiv h^{\Wa_1\Wc_1}\cdots h^{\Wa_n\Wc_n}\,\delta_{\Wc_1\cdots\Wc_n}^{\Wb_1\cdots\Wb_n}$ and
\begin{align}
 \theta^{\Wa}{}_\Wb \equiv \Bigl(1+\frac{\tr(H^2)}{6}\Bigr)\delta^\Wa_\Wb -\frac{1}{2}\,(H^2)^\Wa{}_\Wb \,,\qquad 
 (H^2)^\Wa{}_\Wb \equiv h^{\Wa\We}\,h^{\Wc_1\Wd_1}\,h^{\Wc_2\Wd_2}\,H_{\We\Wc_1\Wc_2}\,H_{\Wb\Wd_1\Wd_2} \,.
\end{align}
By rewriting this as
\begin{align}
 \sqrt{-\gamma}^2\,(\gamma^{-1}\,h\,\gamma^{-1}\,h)^\Wa{}_\Wb = (\mu_5\Exp{-\omega})^2\,\sqrt{-h}^2\,\theta^{\Wa}{}_\Wb \,,
\end{align}
and taking the square root, we obtain
\begin{align}
 \sqrt{-\gamma} \,(\gamma^{-1}\,h)^\Wa{}_\Wb = \abs{\mu_5}\Exp{-\omega} \sqrt{-h} \,(\theta^{\frac{1}{2}})^{\Wa}{}_\Wb\,,
\label{eq:M5-gamma-inv-h}
\end{align}
or
\begin{align}
 \gamma_{\Wa\Wb} = \bigl(\abs{\mu_5} \Exp{-\omega}\bigr)^{\frac{1}{2}}\,(\det\theta^{\Wa}{}_\Wb)^{\frac{1}{8}}\,\bigl(\theta^{-\frac{1}{2}}\bigr)_{\Wa\Wb} \,,\qquad 
 \biggl(\frac{\sqrt{-\gamma}}{\sqrt{-h}}\biggr)^2 =\bigl(\abs{\mu_5}\Exp{-\omega}\bigr)^3\,(\det\theta^{\Wa}{}_\Wb)^{\frac{1}{4}}\,.
\end{align}
Here and hereafter, we raise or lower the worldvolume indices $\Wa,\,\Wb$ by using the induced metric $h_{\Wa\Wb}$\,. 
The trace of \eqref{eq:M5-gamma-inv-h} gives
\begin{align}
 G_{ij}\,\rmd X^i\wedge *_\gamma \rmd X^j 
 = \EPSpos \sqrt{-\gamma}\,\rmd^6\sigma\,(\gamma^{-1}\,h)^{\Wa}{}_{\Wa} 
 = \EPSpos \abs{\mu_5}\Exp{-\omega} \sqrt{-h}\,\rmd^6\sigma\, \tr(\theta^{\frac{1}{2}}) \,,
\end{align}
and the action becomes
\begin{align}
 S_5 &=\frac{\mu_5}{6} \int_{\Sigma_6} \Bigl( \frac{1}{5!}\,\hat{\cP}_{i_1\cdots i_5}\wedge \rmd X^{i_1\cdots i_5} + \frac{1}{2!}\, \hat{\cP}_{i_1i_2}\wedge \rmd X^{i_1i_2}\wedge H_3\Bigr)
\nn\\
 &\quad + \mu_5 \int_{\Sigma_6}\Bigl(\BA_6 + \frac{1}{2}\,\BA_3\wedge F_3 -F_6\Bigr) 
\nn\\
 &=\EPSpos \frac{1}{6} \int_{\Sigma_6}\Bigl(\frac{\Exp{\omega}}{5!}\,G^{i_1\cdots i_5,\,j_1\cdots j_5}\,\hat{\cP}_{i_1\cdots i_5}\wedge *_\gamma \hat{\cP}_{j_1\cdots j_5} + \frac{\Exp{\omega}}{2!}\,G^{i_1 i_2,\,j_1 j_2}\, \hat{\cP}_{i_1i_2}\wedge *_\gamma \hat{\cP}_{j_1j_2}\Bigr) 
\nn\\
 &\quad + \mu_5 \int_{\Sigma_6}\Bigl(\BA_6 + \frac{1}{2}\,\BA_3\wedge F_3 -F_6\Bigr) 
\nn\\
 &=\EPSneg \frac{1}{6} \int_{\Sigma_6} \Exp{\omega}G_{ij}\,\rmd X^i\wedge *_\gamma \rmd X^j + \mu_5 \int_{\Sigma_6} \Bigl(\BA_6 + \frac{1}{2}\,\BA_3\wedge F_3 -F_6\Bigr) 
\nn\\
 &= -\abs{\mu_5} \int_{\Sigma_6}\rmd^6\sigma\sqrt{-h}\,\frac{\tr(\theta^{\frac{1}{2}})}{6} 
 + \mu_5 \int_{\Sigma_6}\Bigl(\BA_6 + \frac{1}{2}\,\BA_3\wedge F_3 -F_6\Bigr) \,.
\label{eq:M5-action-nonstandard}
\end{align}
This is the action obtained in \cite{Sakatani:2016sko}, and, as was shown there, at least in the weak field limit $\abs{H_3}\ll 1$, this theory is equivalent to the conventional M5-brane theory. 
In the following, we will check the equivalence at the non-linear level. 

\subsubsection{Results from the superembedding approach}
\label{eq:superembedding}

A variation of our action \eqref{eq:M5-action-nonstandard} with respect to $A_2$ becomes (up to a boundary term)
\begin{align}
 \delta S_5 = \frac{\abs{\mu_5}}{4}\int_{\Sigma_5} \Bigl[\,\partial_{\Wa}\bigl(\sqrt{-h}\,\mathbb{C}^{[\Wa}{}_{\Wc}\,H^{\Wb_1\Wb_2]\Wc}\bigr)
    \EPSminus \frac{\sigma_5}{3!}\,\epsilon^{\Wa_1\cdots\Wa_4\Wb_1\Wb_2}\,\partial_{\Wa_1}\BA_{\Wa_2\Wa_3\Wa_4}\Bigr]\,\delta A_{\Wb_1\Wb_2} \,,
\end{align}
where
\begin{align}
 \mathbb{C}^{\Wa}{}_{\Wb} \equiv \frac{\tr(\theta^{-\frac{1}{2}})}{3}\,\delta^{\Wa}_{\Wb} - (\theta^{-\frac{1}{2}})^\Wa{}_\Wb \,,\qquad 
 \sigma_5\equiv \frac{\mu_5}{\abs{\mu_5}}\,.
\end{align}
By using the covariant derivative $D_{\Wa}$ associated with $h_{\Wa\Wb}$, the equation of motion becomes
\begin{align}
 D_\Wa \bigl[\,\mathbb{C}^{[\Wa}{}_{\Wd}\,H^{\Wb\Wc]\Wd} \EPSminus \sigma_5\,(*_hH_3)^{\Wa\Wb\Wc}\,\bigr] = 0 \,. 
\end{align}
This is consistent with the non-linear self-duality relation \cite{Sakatani:2016sko}
\begin{align}
 \mathbb{C}_{[\Wa}{}^{\Wd}\,H_{\Wb\Wc]\Wd} = \EPSpos \sigma_5\,(*_hH_3)_{\Wa\Wb\Wc} \,,
\label{eq:our-self-dual}
\end{align}
although the self-duality relation cannot be derived from our action.
On the other hand, under a simultaneous variation, $\delta X^i=v^i$ and $\delta A_2=\frac{1}{2!}\,v^i\,\BA_{ij_1j_2}\,\rmd X^{j_1j_2}$ (see \cite{Sezgin:1998tm}), the action \eqref{eq:M5-action-nonstandard} changes (up to a boundary term) as
\begin{align}
 \delta_v S_5 &=\int_{\Sigma_6}\rmd^6\sigma\,\Bigl\{\, \abs{\mu_5}\sqrt{-h}\,\Bigl[\,G_{ij}\,\bigl(\mathbb{G}^{\Wa\Wb}\,\nabla_{\Wa} \partial_{\Wb} X^j + D_{\Wa} \mathbb{G}^{\Wa\Wb}\,\partial_{\Wb} X^j\bigr) 
 +\frac{1}{2\cdot 3!}\,F_{i\Wa\Wb\Wc}\, \mathbb{C}^{[\Wa}{}_{\Wd}\,H^{\Wb\Wc]\Wd}\,\Bigr]
\nn\\
 &\qquad\qquad\quad\ \EPSminus \mu_5\, \epsilon^{\Wa_1\cdots\Wa_6}\Bigl(\frac{1}{6!}\,F_{i\Wa_1\cdots \Wa_6} +\frac{1}{2\cdot 3!\,3!}\,F_{i\Wa_1\Wa_2\Wa_3}\,H_{\Wa_4\Wa_5\Wa_6} \Bigr) \Bigr\}\,v^i\,,
\end{align}
where we have defined
\begin{align}
\begin{split}
 &\mathbb{G}^{\Wa\Wb} \equiv \frac{\tr(\theta^{\frac{1}{2}})}{6}\,h^{\Wa\Wb}
 +\frac{(\theta^{-\frac{1}{2}})^\Wc{}_\Wd}{6}\, 
 \Bigl[-\frac{1}{2}\,(H^2)^{\Wa\Wb}\,\delta^\Wd_\Wc + \frac{1}{2}\, (H^2)^{(\Wa}{}_\Wc\,h^{\Wb)\Wd}+ H^{\Wa\Wd\We}\,H^\Wb{}_{\Wc\We}\Bigr]\,,
\\
 &\nabla_{\Wa} \partial_\Wb X^i \equiv D_{\Wa} \partial_{\Wb} X^i + \Gamma^i_{kl}\,\partial_{\Wa}X^k\,\partial_{\Wb}X^l\,, \quad 
 \Gamma^i_{jk}\equiv \frac{1}{2}\,G^{il}\,\bigl(\partial_j G_{kl}+\partial_k G_{jl}-\partial_l G_{jk}\bigr) \,,
\\
 &F_4 \equiv \rmd\BA_3\,,\quad F_7\equiv \rmd\BA_6 + \frac{1}{2}\,\BA_3\wedge F_4 \,,\quad F_{i\Wa_1\cdots \Wa_p} \equiv F_{ij_1\cdots j_p}\,\partial_{\Wa_1}X^{j_1}\cdots \partial_{\Wa_p}X^{j_p}\,.
\end{split}
\end{align}
Namely, we obtain the equations of motion
\begin{align}
 \mathbb{G}^{\Wa\Wb}\,\nabla_{\Wa} \partial_\Wb X^i &= \EPSpos \sigma_5\, \frac{\epsilon^{\Wa_1\cdots\Wa_6}}{\sqrt{-h}}\Bigl(\frac{1}{6!}\,F^i{}_{\Wa_1\cdots \Wa_6} +\frac{1}{2\cdot 3!\,3!}\,F^i{}_{\Wa_1\Wa_2\Wa_3}\,H_{\Wa_4\Wa_5\Wa_6} \Bigr)
\nn\\
 &\quad -\frac{1}{2\cdot 3!}\,F^i{}_{\Wa\Wb\Wc} \, \mathbb{C}^{[\Wa}{}_{\Wd}\,H^{\Wb\Wc]\Wd}
        -D_{\Wa} \mathbb{G}^{\Wa\Wb}\,\partial_{\Wb} X^i\,. 
\label{eq:our-M5-eom_o}
\end{align}
In order to evaluate the last term, we recall the invariance of the action under a worldvolume diffeomorphism, $\delta_\xi X^i=\xi^{\Wa}\,\partial_{\Wa}X^i$ and $\delta_\xi A_2=\Lie_\xi A_2$,
\begin{align}
 \delta_\xi S_5 = \abs{\mu_5}\int_{\Sigma_6}\sqrt{-h}\,\Bigl(D_{\Wa} \mathbb{G}^{\Wa\Wb} \EPSminus \frac{\sigma_5}{3!\,3!}\,\frac{\epsilon^{\Wa_1\cdots\Wa_6}}{\sqrt{-h}}\,h^{\Wb\Wc}\,F_{\Wc\Wa_1\Wa_2\Wa_3}\,H_{\Wa_4\Wa_5\Wa_6} \Bigr)\,\xi_\Wb = 0 \,,
\end{align}
where a boundary term is neglected because it is irrelevant. 
Since the diffeomorphism parameter $\xi^\Wa(\sigma)$ is arbitrary, we obtain \cite{Sezgin:1998tm}
\begin{align}
 D_{\Wa} \mathbb{G}^{\Wa\Wb} = \EPSpos \frac{\sigma_5}{3!\,3!}\,\frac{\epsilon^{\Wa_1\cdots\Wa_6}}{\sqrt{-h}}\,h^{\Wb\Wc}\,F_{\Wc\Wa_1\Wa_2\Wa_3}\,H_{\Wa_4\Wa_5\Wa_6}\,. 
\end{align}
Using this identity and the non-linear self-duality relation \eqref{eq:our-self-dual}, we can express the equations of motion \eqref{eq:our-M5-eom_o} as
\begin{align}
 \mathbb{G}^{\Wa\Wb}\,\nabla_{\Wa} \partial_\Wb X^i = \EPSpos \sigma_5\, \frac{\epsilon^{\Wa_1\cdots\Wa_6}}{\sqrt{-h}}\Bigl(\frac{1}{6!}\,F^j{}_{\Wa_1\cdots \Wa_6} +\frac{1}{3!\,3!}\,F^j{}_{\Wa_1\Wa_2\Wa_3}\,H_{\Wa_4\Wa_5\Wa_6} \Bigr)\,P_j{}^i \,, 
\label{eq:our-M5-eom}
\end{align}
where we defined a projection,
\begin{align}
 P_j{}^i \equiv \delta_j^i - G_{jk}\,h^{\Wa\Wb}\,\partial_{\Wa} X^k\,\partial_{\Wb} X^i \,,
\end{align}
satisfying $P_i{}^k\, P_k{}^j= P_i{}^j$ (where we used $h^{\Wa\Wb}\equiv (h^{-1})^{\Wa\Wb}$ and $h_{\Wa\Wb}=G_{ij}\,\partial_{\Wa} X^i\,\partial_{\Wb} X^j$).

In order to compare the above equations of motion with the known ones, let us review the familiar results obtained in the superembedding approach \cite{Howe:1997vn,Howe:1997fb}. 
In the superembedding approach, we introduce a self-dual 3-form field satisfying
\begin{align}
 (*_hh)_{\Wa\Wb\Wc} = \EPSpos h_{\Wa\Wb\Wc}\,. 
\label{eq:rel1}
\end{align}
We then define
\begin{align}
 m^{\Wa\Wb} \equiv h^{\Wa\Wb} -2\,k^{\Wa\Wb}\,,\qquad k^{\Wa\Wb} = h^\Wa{}_{\Wc\Wd}\,h^{\Wb\Wc\Wd} \,. 
\end{align}
An important relation that relates $h_{\Wa\Wb\Wc}$ and the 3-form field $H_3\equiv F_3-\BA_3$ is
\begin{align}
 h_{\Wa\Wb\Wc} = \frac{1}{4}\,m_{[\Wa}{}^\Wd\,H_{\Wb\Wc]\Wd} \,. 
\label{eq:rel2}
\end{align}
The above quantities satisfy \cite{Howe:1997vn}
\begin{align}
\begin{split}
 &h_{\Wa_1\Wa_2\Wc}\,h^{\Wb_1\Wb_2\Wc} = \delta_{\Wa_1\Wa_2}^{\Wb_1\Wb_2}\,,\qquad 
 (m^{-1})^{\Wa\Wb}=\frac{1}{1-\frac{2}{3}\,k^2}\,\bigl(h^{\Wa\Wb}+2\,k^{\Wa\Wb}\bigr) \,,
\\
 &k^\Wa{}_\Wa=0\,,\quad k_\Wa{}^\Wc\,k_\Wc{}^\Wb = \frac{k^2}{6}\,\delta_\Wa^\Wb \,. 
\end{split}
\end{align}
We also define
\begin{align}
 K\equiv \frac{1+\frac{2}{3}\,k^2}{1-\frac{2}{3}\,k^2} \,, \qquad 
 Q\equiv 1-\frac{2}{3}\,k^2 = \frac{2}{K+1} \,,
\end{align}
and then we can show the following nontrivial relations \cite{Gibbons:2000ck,VanderSchaar:2001ay}:
\begin{align}
\begin{split}
 &k^{\Wa\Wb} = \frac{1}{8\,K\,(K+1)}\,\Bigl[(H^2)^{\Wa\Wb}-\frac{\tr(H^2)}{6}\,h^{\Wa\Wb}\Bigr]\,,
\\
 &H_{\Wa_1\Wa_2\Wc}\,H^{\Wb_1\Wb_2\Wc} = 2\, (K^2-1) \,\delta_{\Wa_1\Wa_2}^{\Wb_1\Wb_2} 
  +8\,(K+1)^2\,k^{[\Wb_1}_{[\Wa_1}\,k^{\Wb_2]}_{\Wa_2]} 
  +8\,K\,(K+1)\,\delta^{[\Wb_1}_{[\Wa_1}\,k^{\Wb_2]}_{\Wa_2]} \,,
\\
 &(H^2)^{\Wa\Wb} = 8\, K\,(K+1) \,k^{\Wa\Wb} +4\,(K^2-1)\,h^{\Wa\Wb} \,,\qquad 
 \tr(H^2)=24\,(K^2-1) \,,
\\
 &(H^4)^{\Wa\Wb} \equiv (H^2)^{\Wa\Wc}\,(H^2)_{\Wc}{}^\Wb
 = \frac{2}{3}\,\tr(H^2)\,\Bigl[h^{\Wa\Wb}+\frac{1}{2}\,(H^2)^{\Wa\Wb}\Bigr]\,,
\\
 &\tr(H^4) = 4\,\tr(H^2)\,\Bigl[1+\frac{1}{12}\,\tr(H^2)\Bigr]\,, \qquad
 H_{\Wa_1\Wa_2\Wc}\, H_{\Wb_1\Wb_2}{}^\Wc\,(H^2)^{\Wa_2\Wb_2} = (H^4)_{\Wa_1\Wb_1}\,.
\end{split}
\label{eq:H-relations}
\end{align}
If we introduce the 5-brane co-metric as \cite{Howe:1997vn,Howe:1997fb,Sezgin:1998tm,Gibbons:2000ck}
\begin{align}
 C^{\Wa\Wb} \equiv Q^{-1}\, m^{\Wa}{}_{\Wc}\, m^{\Wc\Wb}
 = K\,h^{\Wa\Wb} -2\,(K+1)\,k^{\Wa\Wb} \,,
\end{align}
it satisfies the following relations \cite{Gibbons:2000ck,VanderSchaar:2001ay}:
\begin{align}
\begin{split}
 &C^{\Wa\Wb} = K^{-1}\,\Bigl[\Bigl(1+\frac{1}{12}\,\tr(H^2)\Bigr)\,h^{\Wa\Wb} - \frac{1}{4}\,(H^2)^{\Wa\Wb} \Bigr] \,, \qquad 
 \det \bigl(C_\Wa{}^\Wb\bigr)=1\,,
\\
 &(C^{-1})_{\Wa\Wb} = K^{-1}\,\Bigl[ h_{\Wa\Wb} + \frac{1}{4}\,(H^2)_{\Wa\Wb} \Bigr] \,,\qquad 
  \tr C = \tr C^{-1} = 6\, K \,,
\\
 &(C^{-2})_{\Wa\Wb} = h_{\Wa\Wb} + \frac{1}{2}\,(H^2)_{\Wa\Wb} \,, \qquad
  (C^{2})^{\Wa\Wb} = \Bigl(1+\frac{\tr(H^2)}{6}\Bigr)\,h^{\Wa\Wb} - \frac{1}{2}\,(H^2)^{\Wa\Wb} \,,
\\
 &(C^{-1})_{\Wa\Wb} = -C_{\Wa\Wb} +2\,K\,h_{\Wa\Wb}\,, \qquad
 H_{\Wa_1\Wa_2\Wc}\, H_{\Wb_1\Wb_2}{}^\Wc\,(C^{-1})^{\Wa_2\Wb_2}
 = K^{-1}\, \Bigl(H^2+\frac{1}{4}\,H^4\Bigr)_{\Wa_1\Wb_1}\,.
\label{eq:C-relations}
\end{split}
\end{align}
Note that the 5-brane co-metric is proportional to the open membrane co-metric studied in \cite{Bergshoeff:2000jn,Berman:2001rka}. 
Using the co-metric, we can express the non-linear self-duality relation for $H_3$ as \cite{Howe:1997vn,Howe:1997fb,Sezgin:1998tm,Gibbons:2000ck}
\begin{align}
 C^\Wd{}_{[\Wa}\,H_{\Wb\Wc]\Wd} = \EPSpos (*_hH)_{\Wa\Wb\Wc} \,. 
\label{eq:H3-self-dual}
\end{align}
The equations of motion for scalar fields are obtained as \cite{Howe:1997vn,Howe:1997fb}
\begin{align}
 C^{\Wa\Wb}\,\nabla_{\Wa} \partial_\Wb X^i = \EPSpos \frac{\epsilon^{\Wa_1\cdots\Wa_6}}{\sqrt{-h}}\Bigl(\frac{1}{6!}\,F^j{}_{\Wa_1\cdots \Wa_6} +\frac{1}{3!\,3!}\,F^j{}_{\Wa_1\Wa_2\Wa_3}\,H_{\Wa_4\Wa_5\Wa_6} \Bigr)\,P_j{}^i \,. 
\label{eq:usual-M5-eom}
\end{align}

From the relations \eqref{eq:H-relations}, we can easily see
\begin{align}
 (\theta^{\frac{1}{2}})^{\Wa}{}_{\Wb}= C^{\Wa}{}_{\Wb}\,,\qquad 
 \mathbb{C}^{\Wa}{}_{\Wb} = 2\,K\,\delta^{\Wa}_{\Wb} - (C^{-1})^\Wa{}_\Wb = C^{\Wa}{}_{\Wb} \,,
\end{align}
and the known non-linear self-duality relation \eqref{eq:H3-self-dual} is equivalent to our relation \eqref{eq:our-self-dual}. 
We can also show the nontrivial relation
\begin{align}
 \mathbb{G}^{\Wa\Wb} &= K \,h^{\Wa\Wb}
  -\frac{\tr(\theta^{-\frac{1}{2}})}{24}\,(H^2)^{\Wa\Wb} + \frac{1}{12}\, (\theta^{-\frac{1}{2}})^{\Wc(\Wb} \,(H^2)^{\Wa)}{}_\Wc + \frac{1}{6}\,(\theta^{-\frac{1}{2}})^\Wc{}_\Wd \,H^{\Wa\Wd\We}\,H^\Wb{}_{\Wc\We} 
\nn\\
 &= K \,h^{\Wa\Wb}
  -\frac{K}{2}\,(H^2)^{\Wa\Wb} 
 + \frac{1}{4\,K} \,\Bigl(H^2+\frac{1}{4}\,H^4\Bigr)^{\Wa\Wb} 
\nn\\
 &= \frac{2K^2-1}{K} \,h^{\Wa\Wb} -\frac{1}{4\,K}\,(H^2)^{\Wa\Wb} 
  = C^{\Wa\Wb} \,. 
\end{align}
This indicates that the known equations of motion \eqref{eq:usual-M5-eom} are equivalent to ours \eqref{eq:our-M5-eom}. 
Namely, as long as the relations \eqref{eq:rel1} and \eqref{eq:rel2} are satisfied at an initial configuration, the equations of motion of our theory describe the same time evolution as the conventional M5-brane theory. 
It is also interesting to note that the intrinsic metric naturally reproduced the 5-brane metric or the open membrane metric (up to a Weyl rescaling)
\begin{align}
 (\Exp{\frac{1}{2}\omega}\gamma)_{\Wa\Wb} = \abs{\mu_5}^{\frac{1}{2}}\, (C^{-1})_{\Wa\Wb} \,,
\end{align}
as a result of the equations of motion. 
Moreover, it is interesting to note that, by using \eqref{eq:C-relations}, our action \eqref{eq:M5-action-nonstandard} becomes
\begin{align}
 S_5 = -\abs{\mu_5} \int_{\Sigma_6} \Bigl[\,\EPSpos *_h K - \sigma_5 \Bigl(\BA_6 + \frac{1}{2}\,\BA_3\wedge F_3 -F_6\Bigr)\Bigr] \,,
\end{align}
which takes the same form as the action studied in \cite{Cederwall:1997gg,Sezgin:1998tm}.

\subsubsection{Self-duality relation for M5-brane}

In this subsection, we show the self-duality relation for the M5-brane
\begin{align}
 \fbox{\quad$\displaystyle \bbeta^{(\text{\tiny M5})}_{IJ}\wedge \cP^J = \EPSpos \cM_{IJ}\,*_\gamma \cP^J \,.$\quad}
\end{align}
Instead of directly showing the relation, in the following we show an equivalent relation,
\begin{align}
 (L^{-\rmT}\,\bbeta^{(\text{\tiny M5})}\,L^{-1})_{IJ}\wedge \hat{\cP}^J = \EPSpos \hat{Z}_I \,. 
\end{align}
Using the equations of motion, the left-hand side becomes
\begin{align}
 &(L^{-\rmT}\,\bbeta^{(\text{\tiny M5})}\,L^{-1})_{IJ} \wedge \hat{\cP}^J
 = \frac{\mu_5}{5!}\,\bigl[L^{-\rmT}\,(\eta_{k_1\cdots k_4}+4\,F_{[k_1k_2k_3}\,\eta_{k_4]})\,L^{-1}\bigr]_{IJ}\,\rmd X^{k_1\cdots k_4} \wedge \hat{\cP}^J
\nn\\
 &= \frac{\mu_5}{5!}\,\bigl(\eta_{k_1\cdots k_4}+4\,H_{[k_1k_2k_3}\,\eta_{k_4]}\bigr)_{IJ}\,\rmd X^{k_1\cdots k_4} \wedge \hat{\cP}^J
\nn\\
 &= \begin{pmatrix}
 \EPSpos\frac{\mu_5^2\Exp{-\omega}}{5}\,\Bigl[G_{j_1j_2,\,k_1k_2}\,H_3\wedge\rmd X^{[j_1}\,\delta_i^{j_2]}\wedge *_\gamma (\rmd X^{k_1k_2}\wedge H_3)
 +\frac{G_{j_1\cdots j_5,\,k_1\cdots k_5}\,\delta_i^{[j_1} \rmd X^{j_2\cdots j_5]} \wedge *_\gamma\rmd X^{k_1\cdots k_5}}{4!}\Bigr] 
\\
 \frac{2\,\mu_5\,\rmd X^{i_1i_2}\wedge H_3 \EPSplus \frac{1}{2}\,\mu_5^2\Exp{-\omega} G_{j_1j_2,\,k_1k_2}\,\rmd X^{i_1i_2j_1j_2}\wedge *_\gamma (\rmd X^{k_1k_2}\wedge H_3)}{5 \sqrt{2!}} 
\\
 \frac{\mu_5\,\rmd X^{i_1\cdots i_5}}{\sqrt{5!}} \\
 0
 \end{pmatrix} . 
\end{align}
Then, our task is to show that this generalized vector is equal to
\begin{align}
 \EPSpos *_\gamma\hat{\cZ}_I = \begin{pmatrix}
 \EPSpos \Exp{\omega}G_{ij}\,*_\gamma\rmd X^j 
\\
 \frac{\mu_5\, \rmd X^{i_1i_2}\wedge H_3}{\sqrt{2!}} 
\\
 \frac{\mu_5\, \rmd X^{i_1\cdots i_5}}{\sqrt{5!}}
\\
 0
 \end{pmatrix} . 
\label{eq:M5-Zhat}
\end{align}
The nontrivial relations are the first and the second rows,
\begin{align}
\begin{split}
 &G_{ij}\,*_\gamma\rmd X^j
 =\frac{\mu_5^2\Exp{-2\omega}}{5}\, \underbrace{G_{j_1j_2,\,k_1k_2}\,H_3\wedge\rmd X^{[j_1}\,\delta_i^{j_2]}\wedge *_\gamma (\rmd X^{k_1k_2}\wedge H_3)}_{N_1}
\\
 &\qquad\qquad\qquad +\frac{\mu_5^2\Exp{-2\omega}}{5}\,\underbrace{\frac{G_{j_1\cdots j_5,\,k_1\cdots k_5}\,\delta_i^{[j_1} \rmd X^{j_2\cdots j_5]} \wedge *_\gamma\rmd X^{k_1\cdots k_5}}{4!}}_{N_2} \,,
\\
 &\rmd X^{i_1i_2}\wedge H_3 
 =\frac{2}{5}\,\rmd X^{i_1i_2}\wedge H_3 \underbrace{\EPSplus \frac{\mu_5 \Exp{-\omega}}{10}\,G_{j_1j_2,\,k_1k_2}\,\rmd X^{i_1i_2j_1j_2}\wedge *_\gamma (\rmd X^{k_1k_2}\wedge H_3)}_{N_3} \,.
\end{split}
\label{eq:non-trivial-rel}
\end{align}
We show that
\begin{align}
\begin{split}
 N_1 &= - \mu_5^{-2}\Exp{2\omega} \, G_{ik}\,\partial_{\Wb}X^k\, (H^2)^\Wb_{\Wd}\, *_\gamma \rmd\sigma^{\Wd}\,,
\\
 N_2 &= \mu_5^{-2}\Exp{2\omega}\, \bigl[5\,G_{ik}\,*_\gamma\rmd X^k + G_{ik}\,\partial_\Wb X^k\,(H^2)^\Wb_\Wd \,*_\gamma \rmd \sigma^\Wd \bigr] \,,
\\
 N_3 &= \frac{3}{5}\, \rmd X^{i_1i_2}\wedge H_3 \,,
\end{split}
\label{eq:N1-N2-N3}
\end{align}
and then the relations \eqref{eq:non-trivial-rel} are proven.

In order to show \eqref{eq:N1-N2-N3}, we need to use various relations displayed in Sect.~\ref{eq:superembedding}, such as \eqref{eq:C-relations} and \eqref{eq:H3-self-dual}. 
By using
\begin{align}
 \rmd\sigma^{\Wa_1\cdots\Wa_5} = \varepsilon^{\Wa_1\cdots\Wa_5\Wc}\,(*_\gamma\rmd\sigma_\Wc) \,,\qquad 
 *_h\rmd\sigma_\Wa = \frac{\sqrt{-h}}{\sqrt{-\gamma}}\,*_\gamma \rmd\sigma_\Wa \,,
\end{align}
we can simplify $N_1$ as
\begin{align}
 N_1&\equiv G_{j_1j_2,\,k_1k_2}\,H_3\wedge \rmd X^{[j_1}\,\delta_i^{j_2]}\wedge *_\gamma (\rmd X^{k_1k_2}\wedge H_3)
\nn\\
 &=G_{j_1k_1}\,G_{ik_2}\,\frac{1}{3!}\,H_{\Wa_1\Wa_2\Wa_3}\,\partial_{\Wa_4}X^{j_1}\,\partial_{\Wb_1}X^{k_1}\,\partial_{\Wb_2}X^{k_2}\,\frac{1}{3!}\,H_{\Wb_3\Wb_4\Wb_5}\,\rmd\sigma^{\Wa_1\cdots \Wa_4}\wedge *_\gamma \rmd\sigma^{\Wb_1\cdots\Wb_5}
\nn\\
 &=\frac{5!}{3!\,3!}\, G_{ik}\,\partial_{\Wb_2}X^k\,H_{\Wa_1\Wa_2\Wa_3}\,H_{\Wb_3\Wb_4\Wb_5}\,h^{\Wa_1\Wa_2\Wa_3\Wc\We,\,\Wb_2\cdots\Wb_5\Wd}\,\gamma_{\Wc\Wd}\,\frac{\sqrt{-h}}{\sqrt{-\gamma}}\,*_h \rmd\sigma_\We
\nn\\
 &= -\frac{20}{3!}\, G_{ik}\,\partial_{\Wb_2}X^k\,H^{\Wa_1\Wa_2\Wa_3}\,H_{\Wb_3\Wb_4\Wb_5}\,\delta_{\Wa_1\Wa_2\Wa_3\Wc\We_1}^{\Wb_3\Wb_4\Wb_5\Wb_2\Wd}\,\gamma^\Wc_\Wd\,\gamma^{\We_1}_{\We_2}\,\frac{\det h}{\det\gamma}\,*_\gamma \rmd\sigma^{\We_2}
\nn\\
 &= -\frac{\mu_5^{-2}\Exp{2\omega}}{3}\, G_{ik}\,\partial_{\Wb}X^k\,\Bigl[\tr(H^2)\,\delta^{\Wb\Wd}_{\Wc\We_1}-6\,(H^2)^{[\Wb}_{[\Wc}\,\delta^{\Wd]}_{\We_1]} +3\,H^{\Wa\Wb\Wd}\,H_{\Wa\Wc\We_1}\Bigr]\,\gamma^\Wc_\Wd\,\gamma^{\We_1}_{\We_2}\,*_\gamma\rmd\sigma^{\We_2}
\nn\\
 &= -\frac{\mu_5^{-2}\Exp{2\omega}}{3}\, G_{ik}\,\partial_{\Wb}X^k\,\biggl[\tr(H^2)\,\frac{(\gamma^2)^\Wb_{\We_2}-(\tr\gamma)\,\gamma^\Wb_{\We_2}}{2}
\nn\\
 &\qquad\qquad\qquad\qquad\qquad -\frac{3}{2}\,\Bigl[(H^2\,\gamma^2)^\Wb_{\We_2} -\tr(\gamma\,H^2)\,\gamma^\Wb_{\We_2}-(\tr\gamma)\,(H^2\,\gamma)^\Wb_{\We_2} + (\gamma\,H^2\,\gamma)^\Wb_{\We_2}\Bigr]
\nn\\
 &\qquad\qquad\qquad\qquad\qquad -3\,H^{\Wa\Wd\Wb}\,H_{\Wa\Wc\We_1}\,\gamma^\Wc_\Wd\,\gamma^{\We_1}_{\We_2}\biggr]\,*_\gamma\rmd\sigma^{\We_2}
\nn\\
 &= - \mu_5^{-2}\Exp{2\omega} \, G_{ik}\,\partial_{\Wb}X^k\, (H^2)^\Wb_{\Wd}\, *_\gamma\rmd\sigma^{\Wd} \,. 
\end{align}
Similarly, $N_2$ becomes
\begin{align}
 N_2&\equiv \frac{1}{4!}\,G_{j_1\cdots j_5,\,k_1\cdots k_5}\,\delta_i^{[j_1}\,\rmd X^{j_2\cdots j_5]}\wedge *_\gamma \rmd X^{k_1\cdots k_5}
\nn\\
 &= \frac{1}{4!}\,G_{ik_1}\,G_{j_2\cdots j_5,\,k_2\cdots k_5}\,\partial_{\Wa_1}X^{j_2}\cdots\partial_{\Wa_4}X^{j_5}\,\partial_{\Wb_1}X^{k_1}\cdots \partial_{\Wb_5}X^{k_5}\,\rmd \sigma^{\Wa_1\cdots\Wa_4}\wedge *_\gamma \rmd \sigma^{\Wb_1\cdots\Wb_5}
\nn\\
 &= \frac{1}{4!}\,G_{ik}\,\partial_{\Wb_1}X^k\,h_{\Wa_1\cdots\Wa_4,\,\Wb_2\cdots \Wb_5}\,\varepsilon^{\Wa_1\cdots\Wa_4\Wc}{}_\Wd *_h\rmd\sigma^\Wd\,\frac{\sqrt{-h}}{\sqrt{-\gamma}}\,\varepsilon^{\Wb_1\cdots\Wb_5\We}\,\gamma_{\We\Wc} 
\nn\\
 &= 2\,G_{ik}\,\partial_{\Wb_1}X^k\,h^{\Wc\Wd,\,\Wb_1\We}\,\frac{\sqrt{-h}}{\sqrt{-\gamma}}\,\gamma_{\We\Wc} *_h\rmd\sigma_\Wd
\nn\\
 &=\frac{\det h}{\det\gamma}\,G_{ik}\,\partial_\Wb X^k\,\bigl[-h^{-1}\,\gamma\,h^{-1}\,\gamma +(\tr\gamma)\,h^{-1}\,\gamma\bigr]^\Wb_\Wd \, *_\gamma \rmd\sigma^\Wd
\nn\\
 &= \mu_5^{-2}\Exp{2\omega} G_{ik}\,\partial_\Wb X^k\,\bigl[-\delta^\Wb_\Wd-\tfrac{1}{2}\,(H^2)^\Wb_\Wd +6\,\,(\delta^\Wb_\Wd + \tfrac{1}{4}\,(H^2)^\Wb_\Wd)\bigr]\,*_\gamma\rmd\sigma^\Wd
\nn\\
 &=\mu_5^{-2}\Exp{2\omega} \bigl[5\,G_{ik}\,*_\gamma\rmd X^k + G_{ik}\,\partial_\Wb X^k\,(H^2)^\Wb_\Wd\,*_\gamma \rmd \sigma^\Wd\bigr]\,. 
\end{align}
Finally, $N_3$ becomes
\begin{align}
 N_3&\equiv \EPSpos \frac{\mu_5 \Exp{-\omega}}{10}\, G_{j_1j_2,\,k_1k_2}\,\rmd X^{i_1i_2j_1j_2}\wedge *_\gamma\bigl(\rmd X^{k_1k_2}\wedge H_3\bigr) 
\nn\\
 &=\EPSpos \frac{\mu_5 \Exp{-\omega}}{10}\, G_{j_1j_2,\,k_1k_2}\,\partial_{\Wa_1}X^{i_1}\,\partial_{\Wa_2}X^{i_2}\,\partial_{\Wa_3}X^{j_1}\,\partial_{\Wa_4}X^{j_2}\,\partial_{\Wb_1}X^{k_1}\,\partial_{\Wb_2}X^{k_2}\,\frac{1}{3!}\,H_{\Wb_3\Wb_4\Wb_5}\,\rmd\sigma^{\Wa_1\cdots\Wa_4}\wedge *_\gamma \rmd\sigma^{\Wb_1\cdots\Wb_5}
\nn\\
 &= \EPSpos \frac{\mu_5 \Exp{-\omega}}{10}\, \frac{\sqrt{-h}}{\sqrt{-\gamma}}\, \partial_{\Wa_1}X^{i_1}\,\partial_{\Wa_2}X^{i_2}\,h_{\Wa_3\Wa_4,\,\Wb_1\Wb_2}\,\frac{1}{3!}\,H_{\Wb_3\Wb_4\Wb_5}\,\varepsilon^{\Wa_1\cdots\Wa_4\Wc_2\Wd}\,\varepsilon^{\Wb_1\cdots \Wb_5\Wc_1}\,\gamma_{\Wc_1\Wc_2}\,*_h\rmd\sigma_\Wd 
\nn\\
 &= \EPSpos \frac{\mu_5 \Exp{-\omega}}{5}\, \frac{\sqrt{-h}}{\sqrt{-\gamma}}\, \partial_{\Wa_1}X^{i_1}\,\partial_{\Wa_2}X^{i_2}\,\bigl(H^{\Wa_1\Wa_2\Wd}\,h^{\Wc_1\Wc_2}-3\,H^{\Wc_1[\Wa_1\Wa_2}\,h^{\Wd]\Wc_2}\bigr)\,\gamma_{\Wc_1\Wc_2}\,*_h\rmd\sigma_\Wd 
\nn\\
 &= \frac{3}{5}\, \rmd X^{i_1i_2}\wedge H_3 \,. 
\end{align}
In this way, we have shown the nontrivial self-duality relation for the M5-brane. 

\subsection{Action for a Kaluza--Klein Monopole}
\label{sec:mKKM}

As the last example, let us consider a KKM in M-theory. 
In fact, a KKM couples to the mixed-symmetry potential $\BA_{i_1\cdots i_8,\,j}$, but this potential appears in the generalized metric $\cM_{IJ}$ of the $E_{d(d)}$ exceptional spacetime only when $d\geq 8$. 
Therefore, we cannot reproduce the whole brane action for a KKM \cite{Bergshoeff:1997gy,Bergshoeff:1998ef} due to our limitation, $d\leq 7$\,. 
In this section, by neglecting the gauge fields, we demonstrate that our action can reproduce the dominant part of the action for a KKM,
\begin{align}
 S \sim - \int_{\Sigma_7} \rmd^7\sigma\,k^2\,\sqrt{-\det\bigl(G_{ij}\,D_\Wa X^i\,D_\Wb X^j\bigr)} \,. 
\end{align}

The main difference from the previously considered M-branes is that a KKM requires the existence of an isometry direction generated by a generalized Killing vector $k^I$. 
In this case, employing the standard procedure in the gauged sigma model, we introduce an additional 1-form gauge field $a_1(\sigma)$ and include it in $\cA^I$,
\begin{align}
 \cA^I \ \to \ \cA^I + a_1\, k^I \,. 
\end{align}
In other words, the generalized vector $\cP^I$ is modified as
\begin{align}
 \cP^I \ \to\ \cP^I - a_1\,k^I\,. 
\end{align}
Supposing that the generalized Killing vector takes the form $k^I=(k^i, 0,\dotsc, 0)$, we have
\begin{align}
 (\cP^I) = {\footnotesize\begin{pmatrix}
 D X^i \\ \frac{\cP_{i_1i_2}}{\sqrt{2!}}\\ \frac{\cP_{i_1\cdots i_5}}{\sqrt{5!}}\\ \frac{\cP_{i_1\cdots i_7,i}}{\sqrt{7!}} 
 \end{pmatrix}} ,\qquad D X^i \equiv \rmd X^i - a_1\, k^i \,.
\end{align}
We then consider the action
\begin{align}
 S_{\text{KKM}} = \EPSneg \frac{1}{7}\int_{\Sigma_7} \Bigl[\,\frac{1}{2}\,\cM_{IJ}(X)\, \cP^I\wedge *_\gamma \cP^J \EPSminus \cP^I \wedge \bbeta^{(\text{\tiny KKM})}_{IJ}\, \wedge \EE^J \,\Bigr] \,,
\end{align}
where $\bbeta^{(\text{\tiny KKM})}_{IJ}$ takes the following form by neglecting the gauge fields:
\begin{align}
 \bbeta^{(\text{\tiny KKM})}_{IJ} \equiv \frac{\mu_{\text{\tiny K}}}{6!}\, \eta_{IJ;\,k_1\cdots k_5i,\,j} \,DX^{k_1}\wedge\cdots \wedge DX^{k_5}\,k^i\,k^j \,. 
\end{align}
More explicitly, we consider the following action:
\begin{align}
 S_{\text{KKM}} = \EPSneg\frac{1}{7}\int_{\Sigma_7} \Bigl[\,\frac{1}{2}\,\cM_{IJ}\,\cP^I\wedge *_\gamma\cP^J \EPSminus \frac{\mu_{\text{\tiny K}}}{6!}\,\cP_{i_1\cdots i_7,\,i}\wedge D X^{i_1\cdots i_6}\,k^{i_7}\,k^i\,\Bigr] \,.
\end{align}
Since we are neglecting the background gauge fields, the first term simply becomes
\begin{align}
 \cM_{IJ}\,\cP^I\wedge *_\gamma\cP^J
 &= \Exp{\omega}\Bigl[G_{ij}\,\rmd X^i\wedge *_\gamma \rmd X^j
 +\frac{1}{2!}\,G^{i_1i_2,\,j_1j_2}\, \cP_{i_1i_2}\wedge *_\gamma \cP_{j_1j_2}
\nn\\
 &\qquad 
 +\frac{1}{5!}\,G^{i_1\cdots i_5,\,j_1\cdots j_5}\, \cP_{i_1\cdots i_5}\wedge *_\gamma \cP_{j_1\cdots j_5}
\nn\\
 &\qquad 
 +\frac{1}{7!}\,G^{i_1\cdots i_7,\,j_1\cdots j_7}\,G^{ij}\, \cP_{i_1\cdots i_7,\,i}\wedge *_\gamma \cP_{j_1\cdots j_7,\,j} \Bigr]\,.
\end{align}

The equation of motion for $\cP_{i_1\cdots i_7,\,i}$ gives
\begin{align}
 \EPSneg \Exp{\omega}G^{i_1\cdots i_7,\,j_1\cdots j_7}\,G^{ij}\, * \cP_{j_1\cdots j_7,\,j} + 7\,\mu_{\text{\tiny K}}\,D X^{[i_1\cdots i_6}\,k^{i_7]}\,k^i = 0 \,,
\end{align}
and the equations of motion for $\cP_{i_1\cdots i_5}$ and $\cP_{i_1i_2}$ give
\begin{align}
 \cP_{j_1\cdots j_5}=0 \,,\qquad \cP_{i_1i_2} =0 \,. 
\end{align}
Using these, the equation of motion for $\gamma_{\Wa\Wb}$ becomes
\begin{align}
 &G_{ij}\,D_\Wa X^i\,D_\Wb X^j 
 = - \frac{1}{7!}\,G^{i_1\cdots i_7,\,j_1\cdots j_7}\,G^{ij}\, \cP_{\Wa;\, i_1\cdots i_7,\,i}\, \cP_{\Wb;\, j_1\cdots j_7,\,j} 
\nn\\
 &= - \frac{7}{6!}\,\abs{\mu_{\text{\tiny K}}}^2\Exp{-2\,\omega}\,k^2\,G_{i_1\cdots i_7,\,j_1\cdots j_7}\, \varepsilon^{\Wc_1\cdots\Wc_6}{}_\Wa \,\varepsilon^{\Wd_1\cdots\Wd_6}{}_\Wb\,
\nn\\
 &\quad \times D_{\Wc_1}X^{i_1}\cdots \,D_{\Wc_6}X^{i_6}\,k^{i_7}\,D_{\Wd_1}X^{j_1}\cdots \,D_{\Wd_6}X^{j_6} \,k^{j_7} \,,
\end{align}
where $k^2\equiv G_{ij}\,k^i\,k^j$\,. 
If we define
\begin{align}
 \Pi_{ij} \equiv G_{ij} - \frac{k_i\,k_j}{k^2} \,,\qquad 
 \pi_{\Wa\Wb}\equiv G_{ij}\,D_\Wa X^i\,D_\Wb X^j =\Pi_{ij}\,D_\Wa X^i\,D_\Wb X^j \,,
\end{align}
the above equation can be expressed as
\begin{align}
 \pi_{\Wa\Wb} &= - \frac{\abs{\mu_{\text{\tiny K}}}^2\Exp{-2\,\omega}\,(k^2)^2}{6!}\,\Pi_{i_1\cdots i_6,\,j_1\cdots j_6}\, \varepsilon^{\Wc_1\cdots\Wc_6}{}_\Wa \,\varepsilon^{\Wd_1\cdots\Wd_6}{}_\Wb \,D_{\Wc_1}X^{i_1}\cdots \,D_{\Wc_6}X^{i_6}\, D_{\Wd_1}X^{j_1}\cdots \,D_{\Wd_6}X^{j_6} 
\nn\\
 &= \abs{\mu_{\text{\tiny K}}}^2\Exp{-2\,\omega}\,(k^2)^2\,\frac{\det \pi}{\det\gamma}\, (\gamma\,\pi^{-1}\,\gamma)_{\Wa\Wb} \,,
\end{align}
and we obtain
\begin{align}
 (\gamma^{-1}\,\pi\,\gamma^{-1}\,\pi)^\Wa{}_\Wb = \abs{\mu_{\text{\tiny K}}}^2\Exp{-2\,\omega}\,(k^2)^2\,\frac{\det\pi}{\det\gamma}\,\delta^\Wa_\Wb \,.
\end{align}
This leads to
\begin{align}
\begin{split}
 &\sqrt{-\gamma}\,(\gamma^{-1}\,\pi)^\Wa{}_\Wb = \abs{\mu_{\text{\tiny K}}} \Exp{-\omega} k^2 \sqrt{-\pi}\,\delta^\Wa_\Wb \,, 
\\
 &\sqrt{-\gamma}\,\gamma^{\Wa\Wb}\,G_{ij}\,D_\Wa X^i\,D_\Wb X^j
 =\sqrt{-\gamma}\,(\gamma^{-1}\,\pi)^\Wa{}_\Wa = 7\,\abs{\mu_{\text{\tiny K}}} \Exp{-\omega} k^2 \sqrt{-\pi} \,, 
\end{split}
\end{align}
and we finally obtain
\begin{align}
 S_{\text{KKM}} &= \frac{1}{7} \int_{\Sigma_7} \frac{\mu_{\text{\tiny K}}}{6!}\,\cP_{i_1\cdots i_7,\,i}\wedge D X^{i_1\cdots i_6}\,k^{i_7}\,k^i 
\nn\\
 &= \EPSneg \frac{1}{7} \int_{\Sigma_7} \Exp{\omega} G_{ij}\, D X^i\wedge *_\gamma D X^j 
 = - \abs{\mu_{\text{\tiny K}}} \int_{\Sigma_7} \rmd^7\sigma\, k^2 \sqrt{-\pi} \,. 
\end{align}
In this way, we can reproduce the well-known action for a KKM. 

In order to introduce the worldvolume gauge fields, we need to modify the $\eta$-form,
\begin{align}
 \bbeta^{(\text{\tiny KKM})}_{IJ} = \eta_{IJ;\,\sMA}\, \bar{\cQ}_{(\text{\tiny KKM})}^\sMA \,, \qquad 
 \bar{\cQ}_{(\text{\tiny KKM})}^\sMA \equiv \frac{\mu_{\text{\tiny K}}}{6}{\footnotesize\begin{pmatrix} 0 \\ 0 \\ \frac{6\,\rmd X^{[i_1\cdots i_5}\,k^{i_6]}\,k^k}{\sqrt{6!}}\\ 0\\ 0 \end{pmatrix}} ,
\end{align}
by performing the active diffeomorphism $\bar{\cQ}_{(\text{\tiny KKM})}^\sMA \to \cQ_{(\text{\tiny KKM})}^\sMA \equiv \cL^\sMA{}_\sMB\,\bar{\cQ}_{(\text{\tiny KKM})}^\sMB$, where $\cL^\sMA{}_\sMB$ is defined in \eqref{eq:cL-R2}.
The resulting $\eta$-form, $\bbeta^{(\text{\tiny KKM})}_{IJ} \equiv \eta_{IJ;\,\sMA}\, \cQ_{(\text{\tiny KKM})}^\sMA$, transforms covariantly under generalized diffeomorphisms. 
In our approach, the action is invariant under gauge transformations, and we expect that by introducing all of the gauge fields in the $E_{8(8)}$ case, we will straightforwardly reproduce the whole action for a KKM. 

In the $E_{7(7)}$ case, we cannot consider exotic branes since there are no winding coordinates (or auxiliary fields $\cP_{i_1\cdots i_8,\,j_1j_2j_3}$ and $\cP_{i_1\cdots i_8,\,j_1\cdots j_6}$) for these branes. 
However, in the $E_{8(8)}$ case, we can consider similar actions like
\begin{align}
\begin{split}
 S_{5^3} &= \EPSneg \frac{1}{6}\int_{\Sigma_6} \Bigl[\,\frac{1}{2}\,\cM_{IJ}(X)\, \cP^I\wedge *_\gamma \cP^J \EPSminus \cP^I \wedge \bbeta^{({\tiny 5^3})}_{IJ}\, \wedge \EE^J\, \Bigr] \,,
\\
 S_{2^6} &= \EPSneg \frac{1}{3}\int_{\Sigma_3} \Bigl[\,\frac{1}{2}\,\cM_{IJ}(X)\, \cP^I\wedge *_\gamma \cP^J \EPSminus \cP^I \wedge \bbeta^{({\tiny 2^6})}_{IJ}\, \wedge \EE^J\, \Bigr] \,,
\end{split}
\end{align}
although the explicit forms of the $\eta$-symbols, $\bbeta^{({\tiny 5^3})}_{IJ}$ and $\bbeta^{({\tiny 2^6})}_{IJ}$, are not yet determined. 
In the $E_{8(8)}$ case, the generalized metric does not contain the potentials $\BA_{i_1\cdots i_9,\,i_1i_2i_3}$ and $\BA_{i_1\cdots i_9,\,i_1\cdots i_6}$ that couple to the exotic $5^3$-brane and the $2^6$-brane, but we can consider the truncated action like the KKM action presented in this subsection. 
In order to reproduce the whole action for a $5^3$-brane and a $2^6$-brane, we are led to consider the $E_{9(9)}$ exceptional spacetime. 
Another possibility to describe a KKM or exotic branes in $d\leq 7$ is discussed in Sect.~\ref{sec:exotic}.

\subsection{Comments on duality symmetry}
\label{sec:duality}

In the previous sections, we have discussed our sigma model actions only in the usual section, where the set of null vectors $\lambda^a$ take the simple form, $(\lambda^a_I)=(\delta^a_i,0,\dotsc,0)$. 
In such cases, $\EE^I$, $\cA^I$, and $\cP^I$ transform covariantly under generalized diffeomorphisms (which do not change the section $\lambda^a$), and our action was manifestly invariant. 
Since a subgroup of the $T$- or $U$-duality group, known as the geometric subgroup, can be realized as a rigid part of generalized diffeomorphisms, invariance of our action under the geometric subgroup is also manifest. 
In this subsection, we consider global duality transformations that change the section $\lambda^a$, and show that $\EE^I$, $\cA^I$, and $\cP^I$ transform covariantly. 
In the conventional formulation of string theory/M-theory, such duality symmetry exists only in constant background, and we assume here that the supergravity fields are constant (unless otherwise stated). 

\subsubsection{Obstacle to manifest $U$-duality covariance}

Let us begin with a brief review of the obstacle to describing the equations of motion in a manifestly duality-covariant form \cite{Percacci:1994aa,Duff:2015jka}. 

In the DSM defined in a constant background, the equation of motion for $\cP_i$ gives
\begin{align}
 \cP_i = \EPSneg G_{ij}\,*_\gamma \rmd X^j + B_{ij}\, \rmd X^j \,,
\end{align}
and taking the exterior derivative, we obtain
\begin{align}
 \rmd\cP_i = \EPSneg G_{ij}\,\rmd *_\gamma \rmd X^j = 0 \,,
\end{align}
where we used the equation of motion for $X^i$ in the last equality. 
Namely, for a given solution, we can (at least locally) find $\tilde{X}_i(\sigma)$ that satisfies $\cP_i=\rmd \tilde{X}_i$\,. 
Then, we can express $\cP^I$ as $\cP^I=\rmd X^I$, where $(X^I)\equiv (X^i,\,\tilde{X}_i)$, and the equations of motion become
\begin{align}
 \eta_{IJ}\, \rmd X^J = \EPSneg \cH_{IJ}\, *_\gamma \rmd X^J \,. 
\end{align}
This is manifestly covariant under a global $\OO(d,d)$ rotation \cite{Duff:1989tf}
\begin{align}
 \rmd X^I\to (\Lambda^{-1})^I{}_J\,\rmd X^J\,,\quad 
 \cH_{IJ}\to \Lambda^K{}_I\,\Lambda^L{}_J\,\cH_{KL}\qquad 
 \bigl(\Lambda^K{}_I\,\Lambda^L{}_J\,\eta_{KL}=\eta_{IJ}\bigr)\,.
\end{align}

On the other hand, in the case of an M2/M5-brane in a constant background, as we can easily see from \eqref{eq:M2-Zhat} or \eqref{eq:M5-Zhat}, the equations of motion give $\rmd *_\gamma \hat{\cZ}_I=0$ and thus $\rmd *_\gamma \cP^I =0$. 
Then, the self-duality relation, $\bbeta^{(\text{\tiny Mp})}_{IJ}\wedge \cP^J = \EPSpos \cM_{IJ}\,*_\gamma \cP^J$, and $\rmd\bbeta^{(\text{\tiny Mp})}_{IJ}=0$ lead to
\begin{align}
 \bbeta^{(\text{\tiny Mp})}_{IJ}\wedge \rmd \cP^J=0 \,. 
\label{eq:eta-dP}
\end{align}
Remarkably, unlike the case of the DSM, this does not mean $\rmd \cP^I=0$\,. 
Indeed, as was pointed out in \cite{Duff:2015jka}, if we consider a solution of an M2-brane (for $d\geq 4$)
\begin{align}
 \{X^i\} = \{ \sigma^0,\, \alpha\,\sigma^1 \cos(\omega\,\sigma^0),\,\alpha\,\sigma^1 \sin(\omega\,\sigma^0),\, \beta\,\sigma^2,0,\dotsc,0\}\qquad 
 (\alpha,\beta,\omega:\text{ constant})\,,
\end{align}
we find that $\rmd\cP_{i_1i_2}\neq 0$ although \eqref{eq:eta-dP} is satisfied. 
The only exception is the M2-brane in $d=3$, called the topological membrane \cite{Duff:2015jka}. 
In that case, the equations of motion give
\begin{align}
 (\cP^I) = \begin{pmatrix} \rmd X^i \\ \frac{\BA_{i_1i_2 j}\,\rmd X^j \EPSminus G_{i_1i_2,\,j_1j_2}*_h \rmd X^{j_1j_2}}{\sqrt{2!}} 
 \end{pmatrix} = \begin{pmatrix} \rmd X^i \\ \frac{\BA_{i_1i_2 j}\,\rmd X^j\EPSminus \varepsilon_{i_1i_2j}\,\rmd X^j}{\sqrt{2!}}
 \end{pmatrix} ,
\end{align}
where $\varepsilon_{ijk}\equiv \sqrt{-G}\,\epsilon_{ijk}$, and $\rmd \cP^I=0$ is automatically satisfied. 
Then, at least locally, we can find the dual coordinates $Y_{i_1i_2}$ satisfying $\cP_{i_1i_2}=\rmd Y_{i_1i_2}$ and the self-duality equation becomes
\begin{align}
 \bbeta^{(\text{\tiny M2})}_{IJ}\wedge \rmd X^J = \EPSpos \cM_{IJ}\,*_\gamma \rmd X^J\,,\qquad 
 (X^I) \equiv \bigl(X^i,\,\tfrac{Y_{i_1i_2}}{\sqrt{2!}}\bigr) \,.
\end{align}
This is covariant under the whole $U$-duality group $E_{3(3)}=\SL(3)\times \SL(2)$ \cite{Duff:2015jka}. 
In general cases with $d\geq 4$, although we cannot express $\cP_{i_1i_2}$ as $\cP_{i_1i_2}=\rmd Y_{i_1i_2}$, the self-duality relation
\begin{align}
 \bbeta^{(\text{\tiny M2})}_{IJ}\wedge \cP^J = \EPSpos \cM_{IJ}\,*_\gamma \cP^J 
\label{eq:M2-self-duality-again}
\end{align}
is itself still satisfied, and it is formally covariant under $E_{d(d)}$ transformations
\begin{align}
 \cP^I\to (\Lambda^{-1})^I{}_J\,\cP^J\,,\quad 
 \cM_{IJ}\to \Lambda^K{}_I\,\Lambda^L{}_J\,\cM_{KL}\,,\quad 
 \bbeta^{(\text{\tiny M2})}_{IJ}\to \Lambda^K{}_I\,\Lambda^L{}_J\,\bbeta^{(\text{\tiny M2})}_{KL}\,. 
\end{align}
In particular, under global $U$-duality transformations generated by $R_{i_1i_2i_3}$ and $R_{i_1\cdots i_6}$, which we call the $\omega$-transformations, $\cP^I$ is transformed as
\begin{align}
 \cP^I \ \to \ \cP'^I \equiv \bigl(\Exp{\frac{1}{3!}\,\omega^{i_1i_2i_3}\,R_{i_1i_2i_3}}\Exp{\frac{1}{6!}\,\omega^{i_1\cdots i_6}\, R_{i_1\cdots i_6}}\bigr)^I{}_J\, \cP^J\,,
\end{align}
and, for example in $d=4$, we have
\begin{align}
 \cP^I= \begin{pmatrix} \rmd X^i \\ \frac{\cP_{i_1i_2}}{\sqrt{2!}} \end{pmatrix} \ \to \ 
 (\cP'^I) = \begin{pmatrix} \rmd X^i - \frac{1}{2}\,\omega^{ij_1j_2}\,\cP_{i_1i_2} \\ \frac{\cP_{i_1i_2}}{\sqrt{2!}} 
 \end{pmatrix} .
\label{eq:SL5-P}
\end{align}
The problem discussed in \cite{Duff:2015jka} is basically that if we continue to use the parameterization
\begin{align}
 (\cP'^I) = \begin{pmatrix} \rmd X'^i \\ \frac{\cP'_{i_1i_2}}{\sqrt{2!}} \end{pmatrix} ,
\end{align}
the non-closedness $\rmd\cP_{i_1i_2}\neq 0$ leads to the non-integrability of $\rmd X'^i$
\begin{align}
 \rmd\cP'^i = \rmd^2 X'^i=- \frac{1}{2}\,\omega^{ij_1j_2}\,\rmd\cP_{j_1j_2}\neq 0\,. 
\end{align}
Then, the conclusion of \cite{Duff:2015jka} was that $\omega$-transformations are not allowed and \eqref{eq:M2-self-duality-again} is covariant only under the ``geometric subgroup'' generated by $\{K_i{}^j\,,\ R^{i_1i_2i_3}\,,\ R^{i_1\cdots i_6}\}$ (i.e., coordinate transformations $\GL(d)$ and constant shift of $\BA_3$ and $\BA_6$). 
In the following, we stress that the parameterization $(\cP^I)= (\rmd X^i,\ \frac{\cP_{i_1i_2}}{\sqrt{2!}})$ should be changed under $\omega$-transformations, and the integrability condition $\rmd\cP'^i=0$ should be modified as
\begin{align}
 \rmd \bigl(\lambda^a_I\,\cP^I\bigr) = 0\,,
\label{eq:integrability-condition}
\end{align}
which is important to allow for the whole duality symmetry. 

\subsubsection{Duality covariance}

Let us consider the DSM, where the duality group is $\OO(d,d)$. 
The $\OO(d,d)$ group is generated by $\frac{2d(2d-1)}{2}$ generators, $\{T_{\bm{\alpha}}\}\equiv \bigl\{K_i{}^j\,,\ R^{ij}\,, \ R_{ij} \bigr\}$, whose matrix representations are
\begin{align}
 {\footnotesize
 (K_k{}^l)^I{}_J \equiv \begin{pmatrix} \delta_k^i\,\delta^l_j & 0 \\ 0 & -\delta_k^j\,\delta^l_i \end{pmatrix} ,\quad 
 (R^{k_1k_2})^I{}_J \equiv \begin{pmatrix} 0 & 0 \\ -2\,\delta^{k_1k_2}_{ij} & 0 \end{pmatrix} ,\quad 
 (R_{k_1k_2})^I{}_J \equiv \begin{pmatrix} 0 & 2\,\delta_{k_1k_2}^{ij} \\ 0 & 0 \end{pmatrix} .}
\end{align}
Here, the $K_i{}^j$ correspond to general coordinate transformations $\GL(d)$ and the $R^{ij}=R^{[ij]}$ correspond to the $B$-field gauge transformations, and these generate the geometric subgroup. 
The correspondents of the $\omega$-transformations, which change the section $\lambda^a$\,, are called $\beta$-transformations that are generated by the remaining generators $R_{ij}=R_{[ij]}$\,. 

In the following, we show that $\cE^I(\sigma)$ transforms covariantly
\begin{align}
 \cE^I(\sigma) = \begin{pmatrix} \rmd X^i \\ F_{ij}\,\rmd X^j \end{pmatrix}\ \to \ 
 \cE'^I(\sigma) = \begin{pmatrix} \delta^i_j & \beta^{ij} \\ 0 & \delta_i^j \end{pmatrix} \begin{pmatrix} \rmd X^j \\ F_{jk}\,\rmd X^k \end{pmatrix} ,
\end{align}
under a global $\beta$-transformation. 
In order to determine the transformation rule, let us rewrite the definition of $\cE^I(\sigma)$ in terms of the $\beta$-rotated frame. 
Since the original section has been specified by $(\lambda^a_I) = (\delta^a_i,\, 0)$, in the $\beta$-rotated frame $\lambda^a$ takes the form
\begin{align}
 (\lambda'^a_I) = \begin{pmatrix} \delta_i^j & 0 \\ \beta^{ij} & \delta^i_j \end{pmatrix} \begin{pmatrix} \delta^a_j \\ 0 \end{pmatrix} = \begin{pmatrix} \delta^a_i \\ \beta^{ia} \end{pmatrix} ,
\end{align}
and the linear section equations \eqref{eq:double-linear-section} give
\begin{align}
 \lambda^a_I\,\eta^{IJ}\,\partial_J T(x) = \bigl(\tilde{\partial}^a - \beta^{ai}\,\partial_i\bigr)\,T(x) = 0 \,,
\label{eq:LSE}
\end{align}
where $T(x)$ represents a supergravity field or a diffeomorphism parameter in the doubled spacetime. 
Originally, $\EE^I$ was defined as $\EE^I=\Exp{\gLie_\xi}\bar{\EE}^I$ by using the static $\bar{\EE}^I$ defined in \eqref{eq:DSM-Ebar}. 
In the $\beta$-rotated frame, $\bar{\EE}^I$ and the diffeomorphism parameter $\xi^I$ take the form
\begin{align}
 \bar{\EE}'^I = \begin{pmatrix} \delta^i_j & \beta^{ij} \\ 0 & \delta_i^j \end{pmatrix} \bar{\EE}^J 
 = {\small \begin{pmatrix} \rmd \sigma^a \\ 0 \\ \vdots \\ 0 \end{pmatrix}} ,\qquad 
 \xi'^I = \begin{pmatrix} \delta^i_j & \beta^{ij} \\ 0 & \delta_i^j \end{pmatrix} \xi^J = \begin{pmatrix} \xi^i + \beta^{ij}\,\tilde{\xi}_j \\ \tilde{\xi}_i \end{pmatrix} .
\end{align}
In addition, the structure of the generalized diffeomorphism is also different according to the change of the section. 
By employing a convention, where $\tilde{\partial}^i$ is replaced by $\beta^{ij}\,\partial_j$ due to the linear section equations \eqref{eq:LSE}, a derivative in the $\beta$-rotated frame becomes
\begin{align}
 \partial_I T(x) = \begin{pmatrix} \partial_i T \\ \tilde{\partial}^i T \end{pmatrix}
 = \begin{pmatrix} \partial_i T \\ \beta^{ij}\,\partial_j T \end{pmatrix}
 = \begin{pmatrix} \delta_i^j & 0 \\ \beta^{ij} & \delta^i_j \end{pmatrix} \begin{pmatrix} \partial_j T \\ 0 \end{pmatrix}.
\end{align}
Then, the generalized Lie derivative of an arbitrary generalized vector $W^I$ becomes
\begin{align}
\begin{split}
 \gLie_V W^I &\equiv V^J\,\partial_J W^I - \bigl(\partial_J V^I -\partial^I V_J\bigr)\,W^J
\\
 &= \begin{pmatrix} \delta^i_j & \beta^{ij} \\ 0 & \delta_i^j \end{pmatrix}
    \begin{pmatrix} \Lie_{v'} w'^j \\ \Lie_{v'} \tilde{w}_j - 2\,w'^k\,\partial_{[k} \tilde{v}_{j]} \end{pmatrix},
\end{split}
\end{align}
where $(V'^I)\equiv (v^i-\beta^{ij}\,\tilde{v}_j,\,\tilde{v}_i)$ and $(W'^I)\equiv (w^i-\beta^{ij}\,\tilde{w}_j,\,\tilde{w}_i)$\,. 
Therefore, we obtain
\begin{align}
\begin{split}
 \gLie_{\xi'} W^I = \begin{pmatrix} \delta^i_j & \beta^{ij} \\ 0 & \delta_i^j \end{pmatrix}
    \begin{pmatrix} \Lie_{\xi} w'^j \\ \Lie_{\xi} \tilde{w}_j - 2\,w'^k\,\partial_{[k} \tilde{\xi}_{j]} \end{pmatrix},
\end{split}
\end{align}
and we can show that a finite generalized diffeomorphism takes the form,
\begin{align}
 \Exp{\gLie_\xi} = \begin{pmatrix} \delta^i_p & \beta^{ip} \\ 0 & \delta_i^p \end{pmatrix} \begin{pmatrix} \delta^p_q & 0 \\ F_{pq}(x') & \delta_p^q \end{pmatrix} \begin{pmatrix} \frac{\partial x'^q}{\partial x^k} & 0 \\ 0 & \frac{\partial x^k}{\partial x'^q} \end{pmatrix} \begin{pmatrix} \delta^k_j & -\beta^{kj} \\ 0 & \delta_k^j \end{pmatrix} ,
\end{align}
which is precisely the $\beta$-rotated version of \eqref{eq:finite-transf} [where the usual diffeomorphism and the $B$-field gauge transformation have precisely the same form as \eqref{eq:finite-transf}]. 
Then, the components of $\EE^I$ described in the $\beta$-rotated frame become
\begin{align}
 \EE'^I(\sigma) = \begin{pmatrix} \delta^i_j & \beta^{ij} \\ 0 & \delta_i^j \end{pmatrix} \begin{pmatrix} \rmd X^j \\ F_{jk}\,\rmd X^k \end{pmatrix} . 
\end{align}
This shows that components of $\EE^I$ are covariantly transformed under $\beta$-transformations. 
Since the $\OO(d,d)$ symmetry is generated by the geometric subgroup and $\beta$-transformations, we have shown the covariance of $\EE^I$ under the whole $\OO(d,d)$ transformations. 
Moreover, since $\cA^I(\sigma)$ also transforms covariantly by its definition, $\cP^I(\sigma)$ also should, and our action is invariant under $\OO(d,d)$ transformations. 
Note that, in the $\beta$-rotated frame, $\cP^I$ takes the form
\begin{align}
 \cP'^I(\sigma)=\begin{pmatrix} \delta^i_j & \beta^{ij} \\ 0 & \delta_i^j \end{pmatrix} \begin{pmatrix} \rmd X^j \\ \cP_j \end{pmatrix} ,
\end{align}
and $\rmd X^i$ can be extracted from $\cP^I$ as $\rmd X^a=\lambda^a_I\,\cP^I$ by using $\lambda^a_I$\,. 
Therefore, the correct integrability condition (or the Bianchi identity) to require is $\rmd (\lambda^a_I\,\cP^I)=0$ as advocated in \eqref{eq:integrability-condition}. 
Note also that if the original background is not constant, the generalized metric after the $\beta$-transformation includes the dual-coordinate dependence from \eqref{eq:LSE}. 
Since scalar fields $\tilde{X}_i(\sigma)$ are not introduced in our DSM, we cannot define our DSM in such background. 
This is the reason why we have supposed the background to be constant. 
Of course, since the supergravity fields are functions only of $x'^i\equiv x^i-\beta^{ij}\,\tilde{x}_j$ in the $\beta$-rotated background, instead of $X^i(\sigma)$, we can introduce $X'^i(\sigma)$ as the fundamental variables in our DSM, but it is equivalent to going back to the usual section $(\lambda^a_I) = (\delta^a_i,\, 0)$.

We can straightforwardly also apply the above discussion to M-brane sigma models. 
For example, in the case of $d=4$ discussed around Eq.~\eqref{eq:SL5-P}, the $\omega$-transformation rotates the usual section $\lambda^a_I=(\delta^a_i,\,0)$ as
\begin{align}
 \lambda^a_I \ \to \ \lambda'^a_I = \begin{pmatrix} \delta^a_i \\ \frac{\omega^{ai_1i_2}}{\sqrt{2!}} \end{pmatrix} . 
\end{align}
There, the linear section equations \eqref{eq:EFT-linear-section} show $\tilde{\partial}^{ij} = \omega^{ijk}\, \partial_k$\,, and a derivative becomes
\begin{align}
 \partial_I T(x) = \begin{pmatrix} \partial_i T \\ \frac{\tilde{\partial}^{i_1i_2} T}{\sqrt{2!}} \end{pmatrix}
 = \begin{pmatrix} \partial_i T \\ \frac{\omega^{i_1i_2j}\,\partial_j T}{\sqrt{2!}} \end{pmatrix}
 = \begin{pmatrix} \delta_i^j & 0 \\ \frac{\omega^{i_1i_2j}}{\sqrt{2!}} & \delta^{i_1i_2}_{j_1j_2} \end{pmatrix} \begin{pmatrix} \partial_j T \\ 0 \end{pmatrix}.
\end{align}
Accordingly, the generalized Lie derivative becomes
\begin{align}
\begin{split}
 \gLie_{V} W^I = \begin{pmatrix} \delta^i_j & -\frac{\omega^{i_1i_2j}}{\sqrt{2!}} \\ 0 & \delta_{i_1i_2}^{j_1j_2} \end{pmatrix} 
  \begin{pmatrix} \Lie_{v'} w'^j \\ \frac{\Lie_{v'} \tilde{w}_{j_1j_2} - 3\,w'^k\,\partial_{[k} \tilde{v}_{j_1j_2]}}{\sqrt{2!}} \end{pmatrix},
\end{split}
\end{align}
where $(V'^I)\equiv (v^i+\frac{1}{2}\,\omega^{ij_1j_2}\,\tilde{v}_{j_1j_2},\,\tilde{v}_i)$ and $(W'^I)\equiv (w^i+\frac{1}{2}\,\omega^{ij_1j_2}\,\tilde{w}_{j_1j_2},\,\tilde{w}_i)$\,. 
Then, components of $\EE^I$ described in the $\omega$-rotated frame become
\begin{align}
 \EE'^I(\sigma) = \begin{pmatrix} \delta^i_j & -\frac{\omega^{i_1i_2j}}{\sqrt{2!}} \\ 0 & \delta_{i_1i_2}^{j_1j_2} \end{pmatrix} \begin{pmatrix} \rmd X^j \\ \frac{F_{j_1j_2k}\,\rmd X^k}{\sqrt{2!}} \end{pmatrix} ,
\end{align}
and we see that $\EE^I$ transforms covariantly under the $\omega$-transformation. 
Moreover, the correct parameterization of $\cP^I$ in the $\omega$-rotated frame is
\begin{align}
 \cP'^I(\sigma) = \begin{pmatrix} \delta^i_j & -\frac{\omega^{i_1i_2j}}{\sqrt{2!}} \\ 0 & \delta_{i_1i_2}^{j_1j_2} \end{pmatrix} \begin{pmatrix} \rmd X^j \\ \frac{\cP_{j_1j_2}}{\sqrt{2!}} \end{pmatrix} ,
\end{align}
and the integrability condition in the $\omega$-rotated frame is
\begin{align}
 \rmd \bigl(\lambda'^a_I\,\cP^I\bigr) = \rmd^2 X^a = 0\,. 
\end{align}
Therefore, the self-duality relation \eqref{eq:M2-self-duality-again} is covariant under the whole $\SL(5)$ $U$-duality symmetry. 

Even for the higher-dimensional case $d\geq 5$, from a similar argument, it will be possible to show that $\EE^I$ transforms covariantly,
\begin{align}
 \EE^I(\sigma) \ \to \ \bigl(\Exp{\frac{1}{3!}\,\omega^{i_1i_2i_3}\,R_{i_1i_2i_3}}\Exp{\frac{1}{6!}\,\omega^{i_1\cdots i_6}\, R_{i_1\cdots i_6}}\bigr)^I{}_J\, \EE^J\,,
\end{align}
as is clear from the construction. 

\subsubsection{On dual coordinates}

For completeness, we also comment on a section $\lambda_a\equiv (\lambda_{aI})=(0,\,\delta_a^i)$, where supergravity fields depend only on the dual coordinates $\tilde{x}_i$\,. 
On this section, generalized diffeomorphisms are combinations of the usual Lie derivative (with opposite indices) and $\beta$-transformations. 
Then, starting from a static configuration, $\tilde{X}_0(\sigma)=\sigma_0$ and $\tilde{X}_1(\sigma)=\sigma_1$, we obtain the parameterization of $\EE^I$,
\begin{align}
 \EE^I(\sigma) = \begin{pmatrix} 2\,\tilde{\partial}^{[i} V^{j]}(\tilde{X})\,\rmd \tilde{X}_j \\ \rmd \tilde{X}_i(\sigma) \end{pmatrix} . 
\end{align}
In this case, $\cA^I$ and $\cP^I$ take the form
\begin{align}
 \cA^I = \begin{pmatrix} \cA_i \\ 0 \end{pmatrix} ,\qquad 
 \cP^I = \EE^I-\cA^I = \begin{pmatrix} \cP_i \\ \rmd \tilde{X}_i \end{pmatrix} . 
\end{align}
The scalar fields $\tilde{X}_i$ describe fluctuations along the dual directions, while the $V^i$ describe fluctuations along the $x^i$-directions. 
Our action then becomes
\begin{align}
\begin{split}
 S &= \EPSneg \frac{1}{2} \int_{\Sigma_2} \Bigl[\,\frac{1}{2}\,\cH_{IJ}(X)\, \cP^I\wedge *_\gamma \cP^J \EPSplus \eta_{IJ}\,\cP^I \wedge \EE^J \,\Bigr] 
\\
 &= \EPSneg \frac{1}{2} \int_{\Sigma_2} \Bigl[\,\frac{1}{2}\,\cH_{IJ}(X)\, \cP^I\wedge *_\gamma \cP^J \EPSplus \cP^i \wedge \rmd \tilde{X}_i \,\Bigr] - \int_{\Sigma_2} \rmd V_1 \,,
\end{split}
\end{align}
where we defined $\rmd\equiv \rmd\sigma_{\Wa}\wedge\partial^{\Wa}$ ($\Wa=1,2$) and $V_1(\sigma)\equiv V^i\bigl(\tilde{X}(\sigma)\bigr)\,\rmd \tilde{X}_i(\sigma)$ is regarded as a fundamental variable. 
By parameterizing the generalized metric as
\begin{align}
 \cH_{IJ} = \begin{pmatrix} \tilde{g}_{mn} & (\tilde{g}\, \beta)_m{}^{n} \\
 -(\beta\,\tilde{g})^m{}_{n} & (\tilde{g}^{-1}-\beta\,\tilde{g}\,\beta)^{mn} 
 \end{pmatrix},
\end{align}
and eliminating the auxiliary fields $\cP_i$, we obtain the action
\begin{align}
 S = \EPSneg \frac{1}{2} \int_{\Sigma_2} \Bigl[\,\tilde{g}^{ij}(\tilde{X})\, \rmd \tilde{X}_i\wedge *_\gamma \rmd \tilde{X}_j + \beta^{ij}(\tilde{X})\, \rmd \tilde{X}_i\wedge \rmd \tilde{X}_j \,\Bigr] - \int_{\Sigma_2} \rmd V_1 \,,
\end{align}
which is the well-known dual action \cite{Duff:1989tf} if the background is constant. 
Again, note that the integrability condition becomes $\rmd(\lambda_{aI}\,\cP^I)=\rmd^2 \tilde{X}_a=0$\,. 

We can also consider similar parameterizations of $\EE^I$ in the M-brane actions by choosing non-standard sections. 
Unlike the conventional DSM, our sigma model does not include all of the generalized coordinates $X^I(\sigma)$ as the fundamental variables, but we can choose a part of generalized coordinates depending on the choice of the section. 

\section{Type IIB branes in exceptional spacetime}
\label{sec:IIB-branes}

In this section, we explain how to reproduce worldvolume actions for type IIB branes. 
The detailed analysis will be reported elsewhere, but here we explain the basic procedure and demonstrate that we can reproduce the action for a $(p,q)$-string. 

Before considering brane actions, let us review the parameterization of the generalized coordinates that are suitable for describing type IIB branes. 
We begin with the M-theory parameterization of the generalized coordinates in the $E_{d(d)}$ EFT for $d\leq 8$,
\begin{align}
 (x^I) = (\underbrace{x^i_{\vphantom{o}}}_{\mathrm{P}},\,\underbrace{y_{i_1i_2}}_{\mathrm{M}2},\,\underbrace{y_{i_1\cdots i_5}}_{\mathrm{M}5},\,\underbrace{y_{i_1\cdots i_7,\,j}}_{\mathrm{KKM/M}8},\, \underbrace{y_{i_1\cdots i_8,\,j_1j_2j_3}}_{5^3},\,\underbrace{y_{i_1\cdots i_8,\,j_1\cdots j_6}}_{2^6},\,\underbrace{y_{i_1\cdots i_8,\,j_1\cdots j_8,\,k}}_{0^{(1,7)}}) \,. 
\end{align}
Each coordinate is the winding coordinate associated with the brane specified below. 
For $d=8$, $y_{i_1\cdots i_7,\,j}$ includes 64 coordinates, and among these, 56 coordinates with $j\in \{i_1,\dotsc ,i_7\}$ correspond to the KKM while the remaining 8 coordinates with $j\not\in \{i_1,\dotsc, i_7\}$ may correspond to 8-branes (known as M8-branes). 
If we decompose the physical coordinates $x^i$ as $(x^i)=(x^r,\,x^\rmM)$ $(r=1,\dotsc,d-1)$ where $x^\rmM$ represents the M-theory direction, we can decompose the above generalized coordinates as those suitable for type IIA branes,
\begin{align}
 {\small (x^I)} &{\small = (\underbrace{x^r_{\vphantom{o}}}_{\mathrm{P}},\underbrace{y_{\vphantom{o}}}_{\mathrm{D0}},\underbrace{y_{r}}_{\mathrm{F}1},\underbrace{y_{r_1r_2}}_{\mathrm{D}2},\underbrace{y_{r_1\cdots r_4}}_{\mathrm{D}4},\underbrace{y_{r_1\cdots r_5}}_{\mathrm{NS}5},\underbrace{y_{r_1\cdots r_6,s}}_{\mathrm{KKM}/7_2},\underbrace{y_{r_1\cdots r_6}}_{\mathrm{D}6},\underbrace{y_{r_1\cdots r_7,s}}_{6^1_3}, \underbrace{y_{r_1\cdots r_7}}_{7_2},}
\nn\\
 &\quad {\small \underbrace{y_{r_1\cdots r_7,s_1s_2}}_{5^2_2},\underbrace{y_{r_1\cdots r_7,s_1s_2s_3}}_{4^3_3},\underbrace{y_{r_1\cdots r_7,s_1\cdots s_5}}_{2^5_3},\underbrace{y_{r_1\cdots r_7,r_1\cdots r_6}}_{1^6_4},\underbrace{y_{r_1\cdots r_7,r_1\cdots r_7}}_{0^7_3} ,\underbrace{y_{r_1\cdots r_7,r_1\cdots r_7,s}}_{0_4^{(1,6)}})\,, }
\end{align}
where we defined the type IIA coordinates (where the index $\rmM$ is removed)
\begin{align}
\begin{split}
 &y\equiv x^\rmM \,,\quad 
 y_r\equiv y_{r\rmM} \,,\quad 
 y_{r_1\cdots r_4}\equiv y_{r_1\cdots r_4\rmM}\,,\quad 
 y_{r_1\cdots r_6,s}\equiv y_{r_1\cdots r_6\rmM,s}\,,\quad 
 y_{r_1\cdots r_6}\equiv y_{r_1\cdots r_6\rmM,\rmM}\,,
\\
 &y_{r_1\cdots r_7}\equiv y_{r_1\cdots r_7\rmM}\,,\quad 
 y_{r_1\cdots r_7,s_1s_2s_3}\equiv y_{r_1\cdots r_7\rmM,s_1s_2s_3}\,,\quad 
 y_{r_1\cdots r_7,s_1s_2}\equiv y_{r_1\cdots r_7\rmM,s_1s_2\rmM}\,,
\\
 &y_{r_1\cdots r_7,r_1\cdots r_6}\equiv y_{r_1\cdots r_7\rmM,r_1\cdots r_6}\,,\quad 
 y_{r_1\cdots r_7,s_1\cdots s_5}\equiv y_{r_1\cdots r_7\rmM,s_1\cdots s_5\rmM}\,,
\\
 &y_{r_1\cdots r_7,r_1\cdots r_7,s}\equiv y_{r_1\cdots r_7\rmM,r_1\cdots r_7\rmM,s}\,,\quad 
 y_{r_1\cdots r_7,r_1\cdots r_7}\equiv y_{r_1\cdots r_7\rmM,r_1\cdots r_7\rmM,\rmM} \,.
\end{split}
\end{align}

In order to obtain the generalized coordinates for type IIB branes, we further decompose the physical coordinates in the type IIA side as $(x^r)=(x^a,\,x^y)$ ($a=1,\dotsc,d-2$) and perform a $T$-duality along the $x^y$-direction. 
Under $T$-dualities, dependence of brane tensions on the string coupling constant $g_s$ does not change, and we summarize the mapping between the winding coordinates \cite{Sakatani:2017nfr} in the following way. 
The type II branes with tension proportional to $g_s^0$ are the fundamental string (F1) and the Kaluza--Klein momentum (P) while those with tension proportional to $g_s^{-1}$ are D-branes. 
By employing the convention of \cite{Sakatani:2017nfr}, their winding coordinates are mapped under the $T$-duality as follows:
\begin{align}
\vcenter{
\xy
\xymatrix "M"@C=3pt@R=12pt{
 x^a \ar[d] & x^y \vphantom{y_a} \ar[rd] & \ar[ld] y_y & y_a \vphantom{x^y} \ar[d] &\qquad & 
 y \vphantom{x^y} \ar[d] & \ar[d] y_{ay}& y_{a_1a_2} \vphantom{x^y} \ar[d] & y_{a_1a_2a_3y} \ar[d] & y_{a_1\cdots a_4} \vphantom{x^y} \ar[d] & y_{a_1\cdots a_5y} \ar[d] & y_{a_1\cdots a_6} \vphantom{x^y} \ar[d] \\
 \sfx^a \vphantom{\sfy_\By^1} & \sfx^\By \vphantom{\sfy_\By^1} & \sfy_\By^1 & \sfy_a^1 & & 
 \sfy_\By^2 & -\sfy_a^2 \vphantom{\sfy_\By^1} & \sfy_{a_1a_2\By} \vphantom{\sfy_\By^1} & \sfy_{a_1a_2a_3} \vphantom{\sfy_\By^1} & \sfy^1_{a_1\cdots a_4\By} \vphantom{\sfy_\By^1} & \sfy_{a_1\cdots a_5}^1 \vphantom{\sfy_\By^1} & \sfy^{(11)}_{a_1\cdots a_6\By} \vphantom{\sfy_\By^1}
}%
 \POS"M1,1"."M1,2"!C*\frm{^\}},+U*++!D\txt{P} 
 ,"M1,3"."M1,4"!C*\frm{^\}},+U*++!D\txt{F1} 
 ,"M1,6"."M1,6"!C*\frm{^\}},+U*++!D\txt{D0} 
 ,"M1,7"."M1,8"!C*\frm{^\}},+U*++!D\txt{D2} 
 ,"M1,9"."M1,10"!C*\frm{^\}},+U*++!D\txt{D4} 
 ,"M1,11"."M1,12"!C*\frm{^\}},+U*++!D\txt{D6} 
 ,"M2,1"."M2,2"!C*\frm{_\}},+D*++!U\txt{P} 
 ,"M2,3"."M2,4"!C*\frm{_\}},+D*++!U\txt{F1} 
 ,"M2,6"."M2,7"!C*\frm{_\}},+D*++!U\txt{D1} 
 ,"M2,8"."M2,9"!C*\frm{_\}},+D*++!U\txt{D3} 
 ,"M2,10"."M2,11"!C*\frm{_\}},+D*++!U\txt{D5} 
 ,"M2,12"."M2,12"!C*\frm{_\}},+D*++!U\txt{D7} 
\endxy
}\,. 
\end{align}

The type II branes with tension proportional to $g_s^{-2}$ include the NS5-brane, KKM, and the exotic $5^2_2$-brane. 
Their winding coordinates are mapped as follows from type IIA theory to type IIB theory:
\begin{align}
\vcenter{
\xy
\xymatrix "M"@C=3pt@R=12pt{
 y_{a_1\cdots a_4y} \ar[d] & y_{a_1\cdots a_5} \ar[rd] & y_{a_1\cdots a_5y,y} \ar[ld] & y_{a_1\cdots a_5y,\bar{b}} \ar[d] & y_{a_1\cdots a_6,b} \ar[rd] & y_{a_1\cdots a_6y,by} \ar[ld] & y_{a_1\cdots a_6y,b_1b_2} \ar[d] \\
 -\sfy^2_{a_1\cdots a_4\By} & -\sfy_{a_1\cdots a_5}^2 & \sfy_{a_1\cdots a_5\By,\By}\vphantom{\mathsf{y}^1} & \sfy_{a_1\cdots a_5\By,\bar{b}} & \pm \sfy_{a_1\cdots a_6,b} \vphantom{\sfy^1} & \pm \sfy^1_{a_1\cdots a_6\By,b\By} & \sfy^1_{a_1\cdots a_6\By,b_1b_2} 
}%
 \POS"M1,1"."M1,2"!C*\frm{^\}},+U*++!D\txt{NS5} 
 ,"M1,3"."M1,5"!C*\frm{^\}},+U*++!D\txt{KKM} 
 ,"M1,6"."M1,7"!C*\frm{^\}},+U*++!D\txt{$5^2_2$} 
 ,"M2,1"."M2,2"!C*\frm{_\}},+D*++!U\txt{NS5} 
 ,"M2,3"."M2,5"!C*\frm{_\}},+D*++!U\txt{KKM} 
 ,"M2,6"."M2,7"!C*\frm{_\}},+D*++!U\txt{$5^2_2$} 
\endxy
} \,. 
\end{align}
Here, the bar, as in $y_{a_1\cdots a_5y,\bar{b}}$ represents that $\bar{b}\in\{a_1,\cdots, a_5\}$ and $\pm$ represents that the sign is not determined yet in \cite{Sakatani:2017nfr}. 

There are another set of 7-branes that also have tension proportional to $g_s^{-2}$ but are not connected to other branes under $T$-dualities. 
The winding coordinates for the eight 7-branes in the type IIA side are $y_{a_1\cdots a_5y,b}$ ($b\not\in\{a_1,\cdots, a_5\}$), $y_{a_1\cdots a_6,y}$, and $y_{a_1\cdots a_6y}$\,. 
Although we have not identified their transformation rule yet, a natural expectation is as follows:
\begin{align}
\begin{pmatrix}
 y_{a_1\cdots a_5y,b}\\
 y_{a_1\cdots a_6,y}\\
 y_{a_1\cdots a_6y}
\end{pmatrix}
\quad \longrightarrow \quad 
\begin{pmatrix}
 \sfy_{a_1\cdots a_5\By,b} \\
 \sfy_{a_1\cdots a_6,\By} \\
 \sfy^{(12)}_{a_1\cdots a_6\By} 
\end{pmatrix} .
\end{align}
Here, the type IIB coordinates, $\sfy_{a_1\cdots a_5\By,b}$ ($b\not\in\{a_1,\cdots, a_5\}$), $\sfy_{a_1\cdots a_6,\By}$, and $\sfy^{(12)}_{a_1\cdots a_6\By}$ correspond to seven $7_2$-branes and a 7-brane that (together with the D7-brane and the $7_3$-brane) behaves as a triplet under $\SL(2)$ $S$-duality transformations. 
The detailed properties of these 7-branes are not well known, but they are necessary to construct a $U$-duality multiplet. 

The type II branes with tension proportional to $g_s^{-3}$ are the exotic $p$-branes $p^{7-p}_3$\,. 
Under the $T$-duality, their winding coordinates are mapped as
\begin{align}
{\tiny\vcenter{
\xy
\xymatrix "M"@C=1pt@R=12pt{
 y_{a_1\cdots a_6y,y} \ar[d] & y_{a_1\cdots a_6y,b} \ar[d] & y_{a_1\cdots a_6y,b_1b_2y} \ar[d] & y_{a_1\cdots a_6y,b_1b_2b_3} \ar[d] & y_{a_1\cdots a_6y,b_1\cdots b_4y} \ar[d] & y_{a_1\cdots a_6y,b_1\cdots b_5} \ar[d] & y_{a_1\cdots a_6y,b_1\cdots b_6y} \ar[d] \\
 \sfy^{(22)}_{a_1\cdots a_6\By} & \pm \sfy^2_{a_1\cdots a_6\By,b\By} & - \sfy^2_{a_1\cdots a_6\By,b_1b_2} & \pm \sfy_{a_1\cdots a_6\By,b_1b_2b_3\By} & \sfy_{a_1\cdots a_6\By,b_1\cdots b_4} & \pm \sfy^1_{a_1\cdots a_6\By,b_1\cdots b_5\By} & \sfy^1_{a_1\cdots a_6\By,b_1\cdots b_6}
}%
 \POS"M1,1"."M1,2"!C*\frm{^\}},+U*++!D\txt{{\normalsize$6^1_3$}} 
 ,"M1,3"."M1,4"!C*\frm{^\}},+U*++!D\txt{{\normalsize$4^3_3$}} 
 ,"M1,5"."M1,6"!C*\frm{^\}},+U*++!D\txt{{\normalsize$2^5_3$}} 
 ,"M1,7"."M1,7"!C*\frm{^\}},+U*++!D\txt{{\normalsize$0^7_3$}} 
 ,"M2,1"."M2,1"!C*\frm{_\}},+D*++!U\txt{{\normalsize$7_3$}} 
 ,"M2,2"."M2,3"!C*\frm{_\}},+D*++!U\txt{{\normalsize$5^2_3$}} 
 ,"M2,4"."M2,5"!C*\frm{_\}},+D*++!U\txt{{\normalsize$3^4_3$}} 
 ,"M2,6"."M2,7"!C*\frm{_\}},+D*++!U\txt{{\normalsize$1^6_3$}} 
\endxy
}} \,. 
\end{align}

Finally, the type II branes with tension proportional to $g_s^{-4}$ are called the $1^6_4$-brane and the $0_4^{(1,6)}$-brane. 
The transformation rules for the corresponding winding coordinates are
\begin{align}
\vcenter{
\xy
\xymatrix "M"@C=3pt@R=12pt{
 y_{a_1\cdots a_6y,b_1\cdots b_5y} \ar[d] & y_{a_1\cdots a_6y,b_1\cdots b_6} \ar[rrd] &\quad & y_{a_1\cdots a_6y,b_1\cdots b_6y,y} \ar[lld] & y_{a_1\cdots a_6y,b_1\cdots b_6y,c} \ar[d] \\
 \pm \sfy^2_{a_1\cdots a_6\By,b_1\cdots b_5\By} & - \sfy^2_{a_1\cdots a_6\By,b_1\cdots b_6} &&\pm \sfy_{a_1\cdots a_6\By,b_1\cdots b_6\By,\By} \vphantom{\sfy^2_{a_1\cdots a_6\By,b_1\cdots b_5\By}} & \sfy_{a_1\cdots a_6\By,b_1\cdots b_6\By,c}
}%
 \POS"M1,1"."M1,2"!C*\frm{^\}},+U*++!D\txt{$1^6_4$} 
 ,"M1,4"."M1,5"!C*\frm{^\}},+U*++!D\txt{$0_4^{(1,6)}$} 
 ,"M2,1"."M2,2"!C*\frm{_\}},+D*++!U\txt{$1^6_4$} 
 ,"M2,4"."M2,5"!C*\frm{_\}},+D*++!U\txt{$0_4^{(1,6)}$} 
\endxy
}\,. 
\end{align}

It is interesting to note that, as has been uncovered in \cite{Lombardo:2016swq} (see also \cite{Bergshoeff:2017gpw}), the $T$-duality transformation rules for the winding coordinates are very simple (up to the convention-dependent sign factor). 
For a type II brane with tension proportional to $g_s^{-n}$, if we consider the winding coordinate with $m$-number of $y$-indices, after a $T$-duality along the $y$-direction, we obtain a winding coordinate with $(n-m)$-number of $y$-indices with other indices unchanged. 
For example, a $0^7_3$-brane $(T_{0^7_3}\propto g_s^{-3})$ associated with the winding coordinate $y_{a_1\cdots a_6y,b_1\cdots b_6y}$ that includes two $y$ is mapped to a $1^6_3$-brane with the winding coordinate $\sfy^1_{a_1\cdots a_6\By,b_1\cdots b_6}$\,.

According to the above dictionary, the whole M-theory coordinates $(x^I)$ are mapped to the type IIB coordinates,
\begin{align}
\begin{split}
 (x^\sfM) = &(\underbrace{x^\sfm_{\vphantom{o}}}_{\mathrm{P}},\,\underbrace{\sfy^\alpha_\sfm}_{\mathrm{F}1/\mathrm{D}1},\,\underbrace{\sfy_{\sfm_1\sfm_2\sfm_3}}_{\mathrm{D}3},\,\underbrace{\sfy^\alpha_{\sfm_1\cdots \sfm_5}}_{\mathrm{NS}5/\mathrm{D}5},\,\underbrace{\sfy_{\sfm_1\cdots \sfm_6,\,\sfn}}_{\mathrm{KKM}/7_2},\, \underbrace{\sfy^{(\alpha\beta)}_{\sfm_1\cdots \sfm_7}}_{\mathrm{Q}7}, 
\\
 &\, \underbrace{\sfy^\alpha_{\sfm_1\cdots \sfm_7,\,\sfn_1\sfn_2}}_{5^2_2/5^2_3},\, \underbrace{\sfy_{\sfm_1\cdots \sfm_7,\,\sfn_1\cdots \sfn_4}}_{3^4_3},\, \underbrace{\sfy^\alpha_{\sfm_1\cdots \sfm_7,\,\sfn_1\cdots \sfn_6}}_{1^6_4/1^6_3},\, \underbrace{\sfy_{\sfm_1\cdots \sfm_7,\,\sfn_1\cdots \sfn_7,\,\mathsf{p}}}_{0_4^{(1,6)}}) \,,
\end{split}
\end{align}
where $\sfm,\sfn,\mathsf{p}=1,\dotsc,d-1$ and $\alpha,\,\beta=1,2$\,. 
In \cite{Sakatani:2017nfr}, the map between the M-theory coordinates and the type IIB coordinates was expressed as
\begin{align}
 x^I= S^I{}_\sfM\,x^\sfM\,,\qquad x^\sfM = (S^{-1})^\sfM{}_I\,x^I\,,
\end{align}
and by using the same matrix $S^I{}_\sfM$, the generalized metric was also transformed as
\begin{align}
 \sfM_{\sfM\sfN} = S^I{}_{\sfM}\,S^J{}_{\sfN}\,\cM_{IJ}\,. 
\end{align}
Then, with the help of Buscher-like transformation rules for supergravity fields, the generalized metric $\sfM_{\sfM\sfN}$ is nicely parameterized with the type IIB supergravity fields (see \cite{Sakatani:2017nfr} for the details). 
In the following, we use the parameterization of $\sfM_{\sfM\sfN}$ and obtain the brane action for a $(p,q)$-string. 
More detailed discussions and actions for other type IIB branes will be reported elsewhere. 

\subsection{Action for a $(p,q)$-string}
\label{sec:pq-string}

When we considered M-branes we chose the M-theory section \eqref{eq:M-theory-section}, but here we choose the type IIB section,
\begin{align}
 (\lambda^a_\sfM) = \begin{pmatrix} \lambda^a_{\sfm} \\ (\lambda^a)_\alpha^{\sfm} \\ \frac{(\lambda^a)^{\sfm_1\sfm_2\sfm_3}}{\sqrt{3!}} \\ \frac{(\lambda^a)_\alpha^{\sfm_1\cdots \sfm_5}}{\sqrt{5!}} \\ \frac{(\lambda^a)^{\sfm_1\cdots \sfm_6,\,\sfm}}{\sqrt{6!}} \end{pmatrix} = \begin{pmatrix} \delta^a_{\sfm} \\ 0 \\ 0 \\ 0 \\ 0 \end{pmatrix} ,
\end{align}
where the supergravity fields depend only on the physical coordinates $\sfx^\sfm$. 
Similar to the M-brane case, $\EE^\sfM$ and $\cP^\sfM \equiv \EE^\sfM -\cA^\sfM$ take the form
\begin{align}
 \EE^\sfM = {\footnotesize
 \begin{pmatrix}
 \rmd X^\sfm \\
 -F_{\sfm\sfn}^\alpha\, \rmd X^\sfn \\
 -\frac{F_{\sfm_1\sfm_2\sfm_3\sfn}+\frac{3}{2}\,\epsilon_{\gamma\delta}\, F^\gamma_{\sfn[\sfm_1}\,F^\delta_{\sfm_2\sfm_3]}}{\sqrt{3!}}\,\rmd X^\sfn \\
 -\frac{F_{\sfm_1\cdots \sfm_5\sfn}^\alpha +10\, F_{\sfn[\sfm_1\sfm_2\sfm_3}\,F_{\sfm_4\sfm_5]}^\alpha - 5\,\epsilon_{\gamma\delta}\, F^\gamma_{\sfn[\sfm_1}\,F^\delta_{\sfm_2\sfm_3}\,F^\alpha_{\sfm_4\sfm_5]}}{\sqrt{5!}} \,\rmd X^\sfn \\
 \cdots \end{pmatrix}} , \qquad 
 \cP^\sfM = \begin{pmatrix} \rmd X^{\sfm} \\ \cP^\alpha_{\sfm} \\ \frac{\cP_{\sfm_1\sfm_2\sfm_3}}{\sqrt{3!}} \\ \frac{\cP^\alpha_{\sfm_1\cdots \sfm_5}}{\sqrt{5!}} \\ \frac{\cP_{\sfm_1\cdots \sfm_6,\,\sfm}}{\sqrt{6!}} \end{pmatrix} ,
\end{align}
where the last row in $\EE^\sfM$ has been abbreviated for simplicity and we have defined
\begin{align}
 F_{\sfm_1\sfm_2}^\alpha \equiv 2\,\partial_{[\sfm_1} A^\alpha_{\sfm_2]} \,,\quad
 F_{\sfm_1\sfm_2\sfm_3} \equiv 3\,\partial_{[\sfm_1} A_{\sfm_2\sfm_3]} \,,\quad
 F_{\sfm_1\cdots \sfm_6}^\alpha \equiv 6\,\partial_{[\sfm_1} A^\alpha_{\sfm_2\cdots\sfm_6]} \,.
\end{align}
Furthermore, similar to \eqref{eq:GM-scalar-M}, we introduce the scalar field $\Exp{\omega(\sigma)}$ into the generalized metric $\sfM_{\sfM\sfN}$ instead of the overall factor $\abs{\BEG}^{\frac{1}{9-d}}$ (see Eqs.~(2.12)--(2.17) in \cite{Sakatani:2017nfr}). 
In the case of a string, which corresponds to the $\eta$-symbol $\eta_{\sfM\sfN}^\alpha$ (see \cite{Sakatani:2017nfr}), the $\eta$-form becomes
\begin{align}
 \bbeta^{(1)}_{\sfM\sfN} = \bbeta_{\sfM\sfN;\,\sBA}\,\cQ_{(1)}^{\sBA} = \mu_1\,q_\alpha \eta_{\sfM\sfN}^\alpha \,,\qquad 
 \cQ_{(1)}^{\sBA} \equiv \mu_1 {\footnotesize \begin{pmatrix} q_\alpha \\[-1mm] 0 \\[-1mm] 0 \\[-1mm] 0 \\[-1mm] 0 \\[-1mm] 0 \\[-1mm] 0 \\[-1mm] 0 \end{pmatrix}},
\end{align}
where $(q_\alpha)\equiv (p,q)$ are constants. 
Then, we consider a string action
\begin{align}
\begin{split}
 S_1 &= \EPSneg \frac{1}{2} \int_{\Sigma_2} \Bigl[\,\frac{1}{2}\,\sfM_{\sfM\sfN}(X)\,\cP^{\sfM}\wedge *_\gamma \cP^{\sfN} \EPSminus \cP^\sfM \wedge \bbeta^{(1)}_{\sfM\sfN}\wedge \EE^\sfN\,\Bigr] 
\\
 &=\EPSneg \frac{1}{2} \int_{\Sigma_2} \Bigl[\, \frac{1}{2}\,\sfM_{\sfM\sfN}(X)\,\cP^{\sfM}\wedge *_\gamma\cP^{\sfN} \EPSminus \mu_1 \,q_\alpha\, \cP_\sfm^\alpha\wedge \rmd X^\sfm \,\Bigr] - \mu_1 \int_{\Sigma_2} q_\alpha \,F_2^\alpha \,,
\end{split}
\label{eq:pq-covariant-action}
\end{align}
where the fundamental fields are
\begin{align}
 \{X^\sfm(\sigma),\,\cP_{\sfm}^\alpha(\sigma),\,\cP_{\sfm_1\sfm_2\sfm_3}(\sigma),\,\cP_{\sfm_1\cdots \sfm_5}^\alpha(\sigma),\,\cP_{\sfm_1\cdots\sfm_6,\, \sfm}(\sigma),\,\gamma_{\Wa\Wb}(\sigma),\, \omega(\sigma) ,\,\GA_1^\alpha(\sigma)\} \,. 
\end{align}
We can eliminate the auxiliary fields, $\cP_{\sfm}^\alpha$, $\cP_{\sfm_1\sfm_2\sfm_3}$, $\cP_{\sfm_1\cdots \sfm_5}^\alpha$, and $\cP_{\sfm_1\cdots\sfm_6,\, \sfm}$, by using their equations of motion, and the action becomes
\begin{align}
 S_1 = \EPSneg \frac{1}{2}\,\int_{\Sigma_2} \Bigl(\frac{\Exp{\omega} + \Exp{-\omega}\mu_1^2\,\abs{q}^2}{2} \Bigr) \, \BEG_{\sfm\sfn}\, \rmd X^\sfm \wedge *_\gamma \rmd X^\sfn
 + \mu_1 \int_{\Sigma_2} q_\alpha\, \bigl(\sfB_2^\alpha- F_2^\alpha\bigr) \,,
\end{align}
where we defined
\begin{align}
 \abs{q} \equiv \sqrt{\Exp{-\Bphi}q^2 + \Exp{\Bphi}(p-q\,\sfC_0)^2} \,. 
\end{align}
As in the case of the usual string action, the equation of motion for $\gamma_{\Wa\Wb}$ gives
\begin{align}
 \gamma_{\Wa\Wb}\propto h_{\Wa\Wb}\equiv \BEG_{\sfm\sfn}\,\partial_\Wa X^\sfm\,\partial_\Wb X^\sfn \,,
\end{align}
and by using this, the action for $X^\sfm$ and $A_1^\alpha$ becomes
\begin{align}
 S_1 &= - \int_{\Sigma_2} \rmd^2\sigma\,\Bigl(\frac{\Exp{\omega}+ \Exp{-\omega}\mu_1^2\,\abs{q}^2}{2} \Bigr) \, \sqrt{-\det h} 
 + \mu_1 \int_{\Sigma_2} q_\alpha\, \bigl(\sfB_2^\alpha- F_2^\alpha\bigr) 
\nn\\
 &= - \abs{\mu_1}\int_{\Sigma_2} \rmd^2\sigma\,\abs{q}\cosh\omega' \, \sqrt{-\det h} 
 + \mu_1 \int_{\Sigma_2} q_\alpha\, \bigl(\sfB_2^\alpha- F_2^\alpha\bigr) \,,
\end{align}
where $\Exp{\omega'}\equiv \frac{\Exp{\omega}}{\abs{\mu_1}\,\abs{q}}$\,. 
The equations of motion for $\omega$ show that $\Exp{\omega}=\abs{\mu_1}\,\abs{q}$, and we finally obtain
\begin{align}
 S_1 = - \abs{\mu_1}\int_{\Sigma_2} \rmd^2\sigma\, \abs{q} \, \sqrt{-\det h} 
 + \mu_1 \int_{\Sigma_2} q_\alpha\, \bigl(\sfB_2^\alpha- F_2^\alpha\bigr) \,. 
\end{align}
This is the well-known $(p,q)$-string action \cite{Schwarz:1995dk,Bergshoeff:2006gs} for $(q_\alpha)=(p,q)$. 
We can also show that the self-duality relation is satisfied
\begin{align}
 \fbox{\quad$\displaystyle \bbeta^{(1)}_{\sfM\sfN}\wedge \cP^\sfN = \EPSpos \sfM_{\sfM\sfN}\,*_\gamma \cP^\sfN \,.$\quad} 
\end{align}

\section{Exotic branes and gauge fields in the external space}
\label{sec:exotic}

In the previous sections, we have considered only the internal components of the supergravity fields, such as $\BA_{i_1i_2i_3}$ and $\BA_{i_1\cdots i_6}$. 
Here, we also consider the external components, such as $\BA_{\mu i_1i_2}$ and $\BA_{\mu_1\mu_2i_1\cdots i_4}$, where $\mu$ runs over the external $(11-d)$ dimensions. 
In fact, the external $n$-form gauge fields make up the so-called $R_n$-representation of the $E_{d(d)}$ group (see \cite{Riccioni:2007au}). 
We denote these external fields as
\begin{align}
 \mathscr{A}_\mu^{I^1}\,,\ \, 
 \mathscr{B}_{\mu_1\mu_2;\,I^2}\,,\ \, 
 \mathscr{C}_{\mu_1\mu_2\mu_3;\,I^3}\,,\ \, 
 \mathscr{D}_{\mu_1\cdots\mu_4;\,I^4}\,,\ \, 
 \mathscr{E}_{\mu_1\cdots\mu_5;\,I^5}\,,\ \, 
 \mathscr{F}_{\mu_1\cdots\mu_6;\,I^6}\,,\ \, 
 \mathscr{G}_{\mu_1\cdots\mu_7;\,I^7}\,,\dotsc,
\end{align}
where the index $I^n$ ($n=1,2,3,\ldots$) transforms in the $R_n$-representation of the $E_{d(d)}$ (note that $I^1= I$ and $I^2=\sMA$ in our M-theory parameterization).

Recently, while this manuscript was being prepared, \cite{Arvanitakis:2017hwb} appeared on arXiv that constructed a $U$-duality-covariant action for strings, including the external fields as well. 
In our convention, their action takes the form
\begin{align}
 S &= \frac{1}{2} \int_{\Sigma_2} T\, \Bigl[\,\frac{1}{2}\,\cM_{IJ}\,D Y^I\wedge *_\gamma D Y^J +g_{\mu\nu}\,\rmd X^\mu\wedge *_\gamma \rmd X^\nu\,\Bigr] 
\nn\\
 &\quad - \frac{1}{2} \int_{\Sigma_2} q^{I^2}\, \Bigl[\eta_{IJ;\,I^2}\, \cA^I\wedge \rmd Y^J + \eta_{IJ;\,I^2}\, \mathscr{A}^I\wedge D Y^J + \mathscr{B}_{\mu\nu;\,I^2}\,\rmd X^\mu\wedge \rmd X^\nu\Bigr] \,,
\end{align}
where $D Y^I \equiv \rmd Y^I -\cA^I + \mathscr{A}^I$ and $\mathscr{A}^I\equiv \mathscr{A}_\mu^I\,\rmd X^\mu$\,. 
If the external fields (i.e.~$g_{\mu\nu}$, $\mathscr{A}^{I}$, and $\mathscr{B}_{\mu\nu;\,I^2}$) are ignored, their action reproduces our 1-brane action \eqref{eq:pq-covariant-action} by identifying $\rmd Y^I$ with our $\EE^I$.
As a natural extension, it is important to introduce external fields $\{\mathscr{A}_\mu^{I^1}\,, \, \mathscr{B}_{\mu_1\mu_2;\,I^2}\,, \ldots\}$ up to the $(p+1)$-form into our $p$-brane actions. 
If all of the external fields are introduced in a gauge-invariant manner, it will be possible to reproduce the actions for a KKM and exotic branes as we discuss below. 

In order to argue that the extension of our $p$-brane action can completely reproduce the Wess--Zumino couplings for exotic branes, let us review which potentials are included in the external fields. 
For simplicity, let us first consider branes with co-dimension higher than two. 
In M-theory, this means 7-branes or lower-dimensional branes. 
They are standard objects in M-theory (i.e.~M2, M5, and KKM) that couple to standard fields (i.e.~$\BA_3$, $\BA_6$, and $\BA_{7,1}$). 
As one can see from Table \ref{tab:external}, an external $p$-form contains only the standard fields when we consider the $E_{d(d)}$ group with $d\leq 8-p$. 
\begin{table}[bt]
\begin{align*}
\begin{split}
 &\vcenter{
\xymatrix "M"@C=-3pt@R=4pt{
 &\llap{{$E_{3(3)}\,[6] \atop E_{4(4)}\,[10]$}}\ar[d]& &\llap{{$E_{5(5)} \,[16]\atop E_{6(6)}\,[27]$}}\ar[d] & &\llap{{$\scriptstyle E_{7(7)}\,[56]$}}\ar[d] & & \llap{{$\scriptstyle E_{8(8)}\,[248]$}}\ar[d]
\\
 \mathscr{A}_\mu:\,\{ \BA_\mu^i,\, \BA_{\mu i_1i_2},\, && \BA_{\mu i_1\cdots i_5}, && \BA_{\mu i_1\cdots i_7,\,j}, && \BA_{\mu i_1\cdots i_8,\,j_1j_2j_3},\, \BA_{\mu i_1\cdots i_8,\,j_1\cdots j_6},\, \BA_{\mu i_1\cdots i_8,\,j_1\cdots j_8,\,k}, && \cdots \}\,,
}} 
\\
 &\vcenter{
\xymatrix "M"@C=-3pt@R=4pt{
 &\llap{{$\scriptstyle E_{3(3)}\,[3]$}}\ar[d]&
 &\llap{{$E_{4(4)}\,[5]\atop E_{5(5)}\,[10]$}}\ar[d]& &\llap{{$\scriptstyle E_{6(6)}\,[27]$}}\ar[d] & &\llap{{$\scriptstyle E_{7(7)}\,[133]$}}\ar[d] 
\\
 \mathscr{B}_{\mu_1\mu_2}:\,\{ \BA_{\mu_1\mu_2 i},&&\BA_{\mu_1\mu_2 i_1\cdots i_4}, && \BA_{\mu_1\mu_2 i_1\cdots i_6,\,j}, && \BA_{\mu_1\mu_2 i_1\cdots i_7,\,j_1j_2j_3},\,\BA_{\mu_1\mu_2 i_1\cdots i_7,\,j_1\cdots j_6}, && \cdots \}\,,
}} 
\\
 &\vcenter{
\xymatrix "M"@C=-3pt@R=4pt{
 &\llap{{$E_{3(3)}\,[2]\atop E_{4(4)}\,[5]$}}\ar[d]& &\llap{{$\scriptstyle E_{5(5)}\,[16]$}}\ar[d] & &\llap{{$\scriptstyle E_{6(6)}\,[78]$}}\ar[d] 
\\
 \mathscr{C}_{\mu_1\mu_2\mu_3}:\,\{ \BA_{\mu_1\mu_2\mu_3},\, \BA_{\mu_1\mu_2\mu_3 i_1i_2i_3}, && \BA_{\mu_1\mu_2\mu_3 i_1\cdots i_5,\,j}, && \BA_{\mu_1\mu_2\mu_3 i_1\cdots i_6,\,j_1j_2j_3},\,\BA_{\mu_1\mu_2\mu_3 i_1\cdots i_6,\,j_1\cdots j_6}, && \cdots \}\,,
}} 
\\
 &\vcenter{
\xymatrix "M"@C=-3pt@R=4pt{
 &\llap{{$\scriptstyle E_{3(3)}\,[3]$}}\ar[d]& 
 &\llap{{$\scriptstyle E_{4(4)}\,[10]$}}\ar[d]& &\llap{{$\scriptstyle E_{5(5)}\,[45]$}}\ar[d]& 
\\
 \mathscr{D}_{\mu_1\cdots\mu_4}:\,\{ \BA_{\mu_1\cdots\mu_4 i_1i_2},&& \BA_{\mu_1\cdots\mu_4 i_1\cdots i_4,\,j}, && \BA_{\mu_1\cdots\mu_4 i_1\cdots i_5,\,j_1j_2j_3}, && \cdots \}\,,
}} 
\\
 &\vcenter{
\xymatrix "M"@C=-3pt@R=4pt{
 &\llap{{$\scriptstyle E_{3(3)}\,[6]$}}\ar[d]& 
 &\llap{{$\scriptstyle E_{4(4)}\,[24]$}}\ar[d]& 
\\
 \mathscr{E}_{\mu_1\cdots\mu_5}:\,\{ \BA_{\mu_1\cdots\mu_5 i},\, \BA_{\mu_1\cdots\mu_5 i_1i_2i_3,\,j},&& \BA_{\mu_1\cdots\mu_5 i_1\cdots i_4,\,j_1j_2j_3}, && \cdots \}\,,
}} 
\\
 &\vcenter{
\xymatrix "M"@C=-3pt@R=4pt{
 &\llap{{$\scriptstyle E_{3(3)}\,[11]$}}\ar[d]& 
\\
 \mathscr{F}_{\mu_1\cdots\mu_6}:\,\{ \BA_{\mu_1\cdots\mu_6},\, \BA_{\mu_1\cdots\mu_6 i_1i_2,\,j},\, \BA_{\mu_1\cdots\mu_6 i_1i_2i_3,\,j_1j_2j_3}, && \cdots \}\,,
}} 
\\
 &\vcenter{
\xymatrix "M"@C=-3pt@R=4pt{
 \mathscr{G}_{\mu_1\cdots\mu_7}:\,\{ \BA_{\mu_1\cdots\mu_7 i,\,j}, \cdots \}\,, }}
\\
 &\vcenter{
\xymatrix "M"@C=-3pt@R=4pt{
 \mathscr{H}_{\mu_1\cdots\mu_8}:\,\{ \BA_{\mu_1\cdots\mu_8,\,i},\cdots \}\,,}}
\\
 &\vcenter{
\xymatrix "M"@C=-3pt@R=4pt{
 \mathscr{I}_{\mu_1\cdots\mu_9}:\,\{ \cdots \}\,.}}
\end{split}
\end{align*}
\caption{Contents of the external fields in M-theory for each $E_{d(d)}$. Only the supergravity fields that couple to branes with co-dimension higher than one are explicitly shown.}
\label{tab:external}
\end{table}
The external $p$-form field comes to contain non-standard supergravity fields when we consider the $E_{d(d)}$ group with $d=9-p$. 
The non-standard potentials, $\BA_{9,3}$ and $\BA_{9,6}$, are known to couple to defect branes (i.e.~co-dimension 2-branes) known as the exotic $5^3$-brane and $2^6$-brane, respectively. 

In order to reproduce whole actions for a KKM in the $E_{d(d)}$ exceptional spacetime with $1\leq d\leq 8$, we need to include external fields up to the 8-form,
\begin{align}
 \{\mathscr{A}_1^{I^1}\,,\ \, 
 \mathscr{B}_{2;\,I^2}\,,\ \, 
 \mathscr{C}_{3;\,I^3}\,,\ \, 
 \mathscr{D}_{4;\,I^4}\,,\ \, 
 \mathscr{E}_{5;\,I^5}\,,\ \, 
 \mathscr{F}_{6;\,I^6}\,,\ \, 
 \mathscr{G}_{7;\,I^7}\,,\ \,
 \mathscr{H}_{8;\,I^8}\}\,,
\end{align}
and write down a gauge-invariant action. 
Since all components of the field $\BA_{\hat{\mu}_1\cdots\hat{\mu}_8,\,i}$, where $\{\hat{\mu}\}=\{\mu,\,i\}$, are contained in these external fields, the Wess--Zumino term for a KKM will be completely reproduced. 
On the other hand, in order to consider the exotic $5^3$-brane, we need to consider $3\leq d\leq 8$. 
In this case, naively, we may write down an action using only
\begin{align}
 \{\mathscr{A}_1^{I^1}\,,\ \, 
 \mathscr{B}_{2;\,I^2}\,,\ \, 
 \mathscr{C}_{3;\,I^3}\,,\ \, 
 \mathscr{D}_{4;\,I^4}\,,\ \, 
 \mathscr{E}_{5;\,I^5}\,,\ \, 
 \mathscr{F}_{6;\,I^6}\}\,. 
\end{align}
These external fields include all components of $\BA_{\hat{\mu}_1\cdots\hat{\mu}_9,\,i_1i_2i_3}$\,. 
Similarly, the exotic $2^6$-brane appears only for $6\leq d\leq 8$, and in order to write down the action, we may only need
\begin{align}
 \{\mathscr{A}_1^{I^1}\,,\ \, \mathscr{B}_{2;\,I^2}\,,\ \, \mathscr{C}_{3;\,I^3}\}\,. 
\end{align}
If our expectation is correct, the exotic $2^6$-brane will be the most tractable example. 
We may also consider co-dimension-1 branes and co-dimension-0 branes that couple to non-standard supergravity fields hidden in the ellipses in Table \ref{tab:external} (see \cite{Bakhmatov:2017les} for a recent study on mixed-symmetry potentials and the associated co-dimension-1 branes). 
Further investigation along this direction will be interesting.

\section{Conclusion}
\label{sec:Conclusion}

In this paper, we showed that the action of the form
\begin{align}
 S_p = \EPSneg \frac{1}{p+1}\int_{\Sigma_{p+1}} \Bigl[\,\frac{1}{2}\,\cM_{IJ}(X)\, \cP^I\wedge *_\gamma \cP^J \EPSminus \cP^I \wedge \bbeta_{IJ}\, \wedge \EE^J\, \Bigr] \,,
\end{align}
can reproduce the conventional M-brane actions in a uniform manner. 
In the case of the M5-brane, the intrinsic metric $\gamma_{\Wa\Wb}$ naturally reproduced the 5-brane metric as a result of the equations of motion, and by using this metric, the self-duality relation,
\begin{align}
 \bbeta^{(\text{\tiny M5})}_{IJ}\wedge \cP^J = \EPSpos \cM_{IJ}\,*_\gamma \cP^J\,,
\end{align}
was realized. 
In contrast to the conventional formulations of extended sigma models (i.e.~double/exceptional sigma model), the worldvolume gauge fields, such as $A_2$ and $A_5$, are naturally introduced inside $\EE^I$, which essentially plays the role of $\rmd X^I$ in the conventional formulations. 
In order to show the applicability of our formalism to type IIB branes, we demonstrated that the well-known $(p,q)$-string action can be correctly reproduced. 
An extension of our $p$-brane action which includes external fields and actions for exotic branes was discussed in Sect.~\ref{sec:exotic}. 

It will be interesting future work to reproduce all of the known brane actions in M-theory and type IIB theory. 
So far, actions of exotic branes are constructed only for the exotic $5^2_2$-branes and $5^2_3$-branes in type II theory \cite{Chatzistavrakidis:2013jqa,Kimura:2014upa,Blair:2017hhy} and the $5^3$-brane in M-theory \cite{Kimura:2016anf}. 
By considering the $E_{9(9)}$ exceptional spacetime or including external fields, it will be possible to reproduce the actions for these branes as well as the other exotic branes discussed in Sect.~\ref{sec:exotic}. 
Extended sigma models play an important role in describing string/brane dynamics in ``stringy'' backgrounds, such as non-Riemannian backgrounds (see \cite{Ko:2015rha} for the detailed analysis) and backgrounds with non-geometric fluxes called $U$-folds. 
It will be interesting to study concrete applications. 

Finally, let us discuss the global $U$-duality rotations of our M-brane actions, assuming the existence of $n$ isometries in the physical $d$-torus. 
For concreteness, we suppose $n=3$; $E_{d(d)}$ exceptional spacetime with three isometries. 
We decompose the coordinates as $(x^i)=(x^m, \,y^p)$ ($m=1,\dotsc,d-3$, $p=1,2,3$) and the $y^p$ directions are isometric. 
In this case, the physical duality group is $\SL(3)\times \SL(2)$ and an M2-brane and an M5-brane wrapped on the isometric 3-torus should transform with each other as an $\SL(2)$ doublet. 
One may realize this symmetry in our formulation in the following manner. 

Similar to the case of the KKM discussed in Sect.~\ref{sec:mKKM}, we introduce three 1-form gauge fields $a_1^{(p)}$ associated with the Killing vectors $k_{(p)}^I\,\partial_I \equiv \partial_p$. 
We then replace $\cP^I$ in the M2/M5 action with $\cP^I - a_1^{(p)}\,k_{(p)}^I$\,, and after eliminating the gauge fields we obtain the actions for M2/M5-branes that fluctuate in the $(d-3)$-dimensional spacetime. 
Thanks to the isometries, we can trivially integrate the wrapped M5-brane action over the 3-torus, and the M5-brane action will become an effective 2-brane action. 
Then, a natural expectation (at least if we ignore the gauge field $A_2$ for simplicity) is that the wrapped M5-brane action will take the form of the 2-brane action \eqref{eq:Mp-action} with the following $\eta$-form:
\begin{align}
 \bbeta_{IJ} \equiv \eta_{IJ;\,\sMA}\, \cQ^{\sMA}\,,\qquad 
 \cQ_{(\text{\tiny wM5})}^{\sMA} \sim {\footnotesize\begin{pmatrix} 0 \\ \frac{4\,DX^{[i_1}\,k_{(1)}^{i_2}\,k_{(2)}^{i_3}\,k_{(3)}^{i_4]}}{\sqrt{4!}} \\ 0 \\ 0 \\ 0 \end{pmatrix}} . 
\end{align}
In fact, this kind of charge appears if we consider a duality ration of the M2 charge $\cQ_{(\text{\tiny M2})}^{\sMA}$,
\begin{align}
 &\cQ_{(\text{\tiny M2})}^{\sMA} 
 = \frac{\mu_2}{2} {\footnotesize\begin{pmatrix} D X^i \\ 0 \\ 0 \\ 0 \\ 0 \end{pmatrix}} 
\ \to\ \cQ_{(\text{\tiny M2'})}^{\sMA} \equiv (\Exp{\frac{1}{3!}\,q^{i_1i_2i_3}\,R_{i_1i_2i_3}})^{\sMA}{}_{\sMB}\,\cQ_{(\text{\tiny M2})}^{\sMB} 
 = \frac{\mu_2}{2} {\footnotesize\begin{pmatrix} D X^i \\ \frac{4\,DX^{[i_1}\,q^{i_2i_3i_4]}}{\sqrt{4!}} \\ 0 \\ 0 \\ 0 \end{pmatrix}} ,
\end{align}
where $q^{i_1i_2i_3}$ is proportional to $k_{(1)}^{[i_1}\,k_{(2)}^{i_2}\,k_{(3)}^{i_3]}$ and $(R_{i_1i_2i_3})^{\sMA}{}_{\sMB}$ is an $E_{d(d)}$ generator in the $R_2$-representation (see Appendix \ref{app:Edd}). 
This 2-brane with the charge $\cQ_{(\text{\tiny M2'})}^{\sMA}$ may be interpreted as a bound state of an M2-brane and wrapped M5-branes like the $(p,q)$-string. 
It will be interesting to perform a more detailed analysis and clarify its relation to the $(p,q)$-membrane discussed in \cite{Bengtsson:2004nj}. 

\appendix

\section{Conventions}
\label{app:conventions}

\subsection{Differential forms}

We employ the following conventions for differential forms on a worldvolume:
\begin{align}
\begin{split}
 &\varepsilon^{0\cdots p}=\EPSneg \frac{1}{\sqrt{-\gamma}}\,,\quad 
 \varepsilon_{0\cdots p}=\EPSpos \sqrt{-\gamma} \,, \quad 
 \epsilon_{0\cdots p}= \EPSpos 1 = -\epsilon^{0\cdots p} \,, \quad 
 \rmd^{p+1}\sigma = \rmd\sigma^0\wedge\cdots\wedge\rmd\sigma^p \,,
\\
 &(*_\gamma w_q)_{\Wa_1\cdots\Wa_{p+1-q}} =\frac{1}{q!}\,\varepsilon^{\Wb_1\cdots\Wb_q}{}_{\Wa_1\cdots\Wa_{p+1-q}}\,w_{\Wb_1\cdots\Wb_q} \,,
\\
 & *_\gamma (\rmd \sigma^{\Wa_1}\wedge \cdots \wedge \rmd \sigma^{\Wa_q}) = \frac{1}{(p+1-q)!}\,\varepsilon^{\Wa_1\cdots\Wa_q}{}_{\Wb_1\cdots\Wb_{p+1-q}}\,\rmd \sigma^{\Wb_1}\wedge \cdots \wedge \rmd \sigma^{\Wb_{p+1-q}} \,. 
\end{split}
\end{align}

\subsection{$E_{d(d)}$ algebra and the $R_n$-representation}
\label{app:Edd}

In the M-theory parameterization, we decompose the $E_{d(d)}$ $(d\leq 7)$ generators as follows:
\begin{align}
 \{T_{\bm{\alpha}}\} = \{K_i{}^j\,,\ R^{i_1i_2i_3}\,,\ R^{i_1\cdots i_6}\,,\ R_{i_1i_2i_3}\,,\ R_{i_1\cdots i_6} \}\qquad (\bm{\alpha}=1,\dotsc,\dim E_{d(d)})\,.
\end{align}
Their commutation relations are given as follows \cite{Berman:2011jh}:
\begin{align}
\begin{split}
 &\bigl[K_i{}^j,\, K_k{}^l\bigr] 
 = \delta_k^j\, K_i{}^l - \delta_i^l\,K_k{}^j\,,\quad \ \ \ 
 \bigl[K_i{}^j,\, R^{k_1k_2k_3} \bigr] 
 = - 3\,\delta_i^{[k_1\vert} \,R^{j\vert k_2k_3]}\,, 
\\
 &\bigl[K_i{}^j,\, R_{k_1k_2k_3} \bigr] = 3\,\delta_{[k_1\vert}^j \, R_{i\vert k_2k_3]}\,,\quad 
 \ \bigl[K_i{}^j,\, R^{k_1\cdots k_6} \bigr] 
 = - 6\,\delta_i^{[k_1\vert} \, R^{j\vert k_2\cdots k_6]}\,,
\\
 &\bigl[K_i{}^j,\, R_{k_1\cdots k_6} \bigr] = 6\,\delta^j_{[k_1\vert}\, R_{i\vert k_2\cdots k_6]}\,,\quad 
 \bigl[ R^{i_1i_2i_3},\,R^{i_4i_5i_6} \bigr] = - R^{i_1\cdots i_6}\,, 
\\
 &\bigl[ R^{i_1i_2i_3},\,R_{j_1j_2j_3} \bigr] = - \frac{3!\cdot 3!}{2!}\,\delta_{[j_1j_2}^{[i_1i_2}\, K_{j_3]}{}^{i_3]} + \frac{1}{3}\,3!\, \delta_{j_1j_2j_3}^{i_1i_2i_3}\, D \,, 
\\
 &\bigl[ R^{i_1i_2i_3},\, R_{j_1\cdots j_6} \bigr] = \frac{6!}{3!}\, \,\delta_{[j_1j_2j_3}^{i_1i_2i_3}\,R_{j_4j_5j_6]} \,, 
\\
 & \bigl[ R_{i_1i_2i_3},\,R_{i_4i_5i_6} \bigr] = R_{i_1\cdots i_6}\,, \qquad \qquad 
 \bigl[ R_{i_1i_2i_3},\,R^{j_1\cdots j_6} \bigr] = -\frac{6!}{3!}\,\delta^{[j_1j_2j_3}_{i_1i_2i_3}\,R^{j_4j_5j_6]} \,, 
\\
 &\bigl[R^{i_1\cdots i_6},\,R_{j_1\cdots j_6} \bigr] = - \frac{6!\cdot 6!}{5!}\, \delta_{[j_1\cdots j_5}^{[i_1\cdots i_5}\, K^{i_6]}{}_{j_6]} + \frac{2}{3}\,6!\,\delta_{j_1\cdots j_6}^{i_1\cdots i_6}\,D \,,
\end{split}
\end{align}
where $D\equiv \sum_i K^i{}_i$\,. 

We denote the representation of the $E_{d(d)}$ group that is composed of the external $n$-form fields as the $R_n$-representation, whose dimensions are determined as follows \cite{Riccioni:2007au}:
\begin{align}
 \begin{array}{|c||c|c|c|c|c|c|c|} \hline
  E_{d(d)} & R_1 & R_2 & R_3 & R_4 & R_5 & R_6 & \cdots \\ \hline\hline
  \SL(5) & 10 & 5 & 5 & 10 & 24 & 40+15 & \cdots \\ \hline
  \text{SO}(5,5) & 16 & 10 & 16 & 45 & 144 & 320+126+10 & \cdots \\ \hline
  E_{6(6)} & 27 & 27 & 78 & 351 & 1728+27 & & \\ \hline
  E_{7(7)} & 56 & 133 & 912 & 8645+133 & & & \\ \hline
 \end{array} \,. 
\end{align}
The $R_{9-d}$-representation is always the adjoint representation and there is a symmetry in the dimensions, $\dim R_{n}=\dim R_{9-d-n}$\,. 
In the M-theory parameterization, we decompose the index $I^n$ of the $R_{n}$-representation as
\begin{align}
\begin{split}
 (V^{I^1}) &= \bigl(v^i,\, \tfrac{v_{i_1i_2}}{\sqrt{2!}},\, \tfrac{v_{i_1\cdots i_5}}{\sqrt{5!}},\, \tfrac{v_{i_1\cdots i_7,\,k}}{\sqrt{7!}},\, \tfrac{v_{i_1\cdots i_8,\,k_1k_2k_3}}{\sqrt{8!\,3!}},\, \tfrac{v_{i_1\cdots i_8,\,k_1\cdots k_6}}{\sqrt{8!\,6!}},\, \tfrac{v_{i_1\cdots i_8,\,k_1\cdots k_8,\,k}}{\sqrt{8!\,8!}}, \cdots \bigr)\,,
\\
 (V_{I^2}) &= \bigl(v_{i},\, \tfrac{v_{i_1\cdots i_4}}{\sqrt{4!}},\, \tfrac{v_{i_1\cdots i_6,\,k}}{\sqrt{6!}},\, \tfrac{v_{i_1\cdots i_7,\,k_1k_2k_3}}{\sqrt{7!\,3!}},\, \tfrac{v_{i_1\cdots i_7,\,k_1\cdots k_6}}{\sqrt{7!\,6!}}, \cdots\bigr) \,,
\\
 (V_{I^3}) &= \bigl(v,\, \tfrac{v_{i_1i_2i_3}}{\sqrt{3!}},\, \tfrac{v_{i_1\cdots i_5,\,k}}{\sqrt{5!}},\, \tfrac{v_{i_1\cdots i_6,\,k_1k_2k_3}}{\sqrt{6!\,3!}},\, \tfrac{v_{i_1\cdots i_6,\,k_1\cdots k_6}}{\sqrt{6!\,6!}}, \cdots \bigr)\,,
\\
 (V_{I^4}) &= \bigl(\tfrac{v_{i_1i_2}}{\sqrt{2!}},\, \tfrac{v_{i_1\cdots i_4,\,k}}{\sqrt{4!}},\, \tfrac{v_{i_1\cdots i_5,\,k_1k_2k_3}}{\sqrt{5!\,3!}}, \cdots \bigr) \,,
\\
 (V_{I^5}) &= \bigl(v_{i},\, \tfrac{v_{i_1i_2i_3,\,k}}{\sqrt{3!}},\, \tfrac{v_{i_1\cdots i_4,\,k_1k_2k_3}}{\sqrt{4!\,3!}}, \cdots \bigr)\,,
\\
 (V_{I^6}) &= \bigl(v,\, \tfrac{v_{i_1i_2,\,k}}{\sqrt{2!}},\, \tfrac{v_{i_1i_2i_3,\,k_1k_2k_3}}{\sqrt{3!\,3!}}, \cdots \bigr)\,,
\end{split}
\end{align}
where the ellipses are not necessary when we consider the $E_{d(d)}$ group with $d\geq 9-n$. 
We may simply denote $I^1$ and $I^2$ as $I$ and $\sMA$, respectively. 

The matrix representations of the $E_{d(d)}$ generators in the $R_1$-representation are given as follows \cite{Berman:2011jh}:
\begin{align}
 &(K_{k_1}{}^{k_2})^I{}_J \equiv {\footnotesize
 \begin{pmatrix}
 \delta_{k_1}^i \delta_j^{k_2} & 0 & 0 & 0 \\
 0 & -\frac{\bdelta_{i_1i_2}^{k_2l} \bdelta_{k_1l}^{j_1j_2}}{\sqrt{2!\,2!}} & 0 & 0 \\
 0 & 0 & -\frac{\bdelta_{i_1\cdots i_5}^{k_2l_1\cdots l_4} \bdelta_{k_1l_1\cdots l_4}^{j_1\cdots j_5}}{4!\sqrt{5!\,5!}} & 0 \\
 0 & 0 & 0 & -\frac{\frac{1}{6!}\bdelta_{i_1\cdots i_7}^{k_2l_1\cdots l_6} \bdelta_{k_1l_1\cdots l_6}^{j_1\cdots j_7}\delta_i^j +\bdelta_{i_1\cdots i_7}^{j_1\cdots j_7} \delta_{i}^{k_2}\delta_{k_1}^j}{\sqrt{7!\,7!}}
 \end{pmatrix} + \frac{\delta_{k_1}^{k_2}}{9-d}}\,\delta^I_J \,, 
\\
 &(R_{k_1k_2k_3})^I{}_J \equiv {\footnotesize
 \begin{pmatrix}
 0 & -\frac{\bdelta^{i j_1j_2}_{k_1k_2k_3}}{\sqrt{2!}} & 0 & 0 \\
 0 & 0 & \frac{\bdelta_{i_1i_2 k_1k_2k_3}^{j_1\cdots j_5}}{\sqrt{2!\,5!}} & 0 \\
 0 & 0 & 0 & \frac{\bdelta_{i_1\cdots i_5 l_1l_2}^{j_1\cdots j_7}\,\bdelta^{l_1l_2j}_{k_1k_2k_3}}{2!\sqrt{5!\,7!}} \\
 0 & 0 & 0 & 0
 \end{pmatrix} } , 
\\
 &(R^{k_1k_2k_3})^I{}_J \equiv {\footnotesize
 \begin{pmatrix}
 0 & 0 & 0 & 0 \\
 -\frac{\bdelta_{i_1i_2 j}^{k_1k_2k_3}}{\sqrt{2!}} & 0 & 0 & 0 \\
 0 & \frac{\bdelta^{j_1j_2 k_1k_2k_3}_{i_1\cdots i_5}}{\sqrt{2!\,5!}} & 0 & 0 \\
 0 & 0 & \frac{\bdelta^{j_1\cdots j_5 l_1l_2}_{i_1\cdots i_7}\,\bdelta_{l_1l_2i}^{k_1k_2k_3}}{2!\sqrt{5!\,7!}} & 0
 \end{pmatrix} } , 
\\
 &(R_{k_1\cdots k_6})^I{}_J \equiv {\footnotesize
 \begin{pmatrix}
 0 & 0 & \frac{\bdelta^{j_1\cdots j_5 i}_{k_1\cdots k_6}}{\sqrt{5!}} & 0 \\
 0 & 0 & 0 & \frac{\bdelta_{i_1i_2 l_1\cdots l_5}^{j_1\cdots j_7}\,\bdelta^{l_1\cdots l_5j}_{k_1\cdots k_6}}{5!\sqrt{2!\,7!}} \\
 0 & 0 & 0 & 0 \\
 0 & 0 & 0 & 0
 \end{pmatrix} } ,
\\
 &(R^{k_1\cdots k_6})^I{}_J \equiv {\footnotesize
 \begin{pmatrix}
 0 & 0 & 0 & 0 \\
 0 & 0 & 0 & 0 \\
 \frac{\bdelta_{i_1\cdots i_5 j}^{k_1\cdots k_6}}{\sqrt{5!}} & 0 & 0 & 0\\
 0 & \frac{\bdelta^{j_1j_2 l_1\cdots l_5}_{i_1\cdots i_7}\,\bdelta_{l_1\cdots l_5i}^{k_1\cdots k_6}}{5!\sqrt{2!\,7!}} & 0 & 0
 \end{pmatrix} } ,
\end{align}
where we defined $\bdelta_{i_1\cdots i_p}^{j_1\cdots j_p}\equiv p!\,\delta_{i_1\cdots i_p}^{j_1\cdots j_p}$\,. 

Using the $\eta$-symbols, we can also find the matrix representations of the $E_{d(d)}$ generators $(T_{\bm{\alpha}})^{\sMA}{}_{\sMB}$ in the $R_2$-representation through
\begin{align}
 (\Exp{h^{\bm{\alpha}}\,T_{\bm{\alpha}}})^K{}_I\,(\Exp{h^{\bm{\alpha}}\,T_{\bm{\alpha}}})^L{}_J\,\eta_{KL;\,\sMB} \,(\Exp{h^{\bm{\alpha}}\,T_{\bm{\alpha}}})^{\sMB}{}_{\sMA}
 = \eta_{IJ;\,\sMA} \,.
\label{eq:eta-invariant-tensor}
\end{align} 
The explicit matrix forms are obtained as follows:
\begin{align}
 &\!\!\!\!{\tiny 
 (K_{s_1}^{~~s_2})^{\sMA}{}_{\sMB} \equiv 
 {\arraycolsep=0mm 
 \left(\begin{array}{ccccc}
 \delta_{s_1}^i \delta^{s_2}_j & 0 & 0 & 0 & 0 \\
 0 & \!\!\!\!\!\! \frac{\bdelta^{i_1\cdots i_4}_{s_1t_1t_2t_3}\bdelta^{s_2t_1t_2t_3}_{j_1\cdots j_4}}{3!\,\sqrt{4!\,4!}}\!\!\!\!\!\!\!\!\!\!\!\!\!\!\! & 0 & 0 & 0 \\
 0 & 0 & \!\!\!\!\!\!\!\!\!\!\!\!\! \frac{\frac{1}{5!}\bdelta^{i_1\cdots i_6}_{s_1t_1\cdots t_5}\bdelta^{s_2t_1\cdots t_5}_{j_1\cdots j_6} \delta_l^k + \bdelta_{j_1\cdots j_6}^{i_1\cdots i_6} \delta_{s_1}^k \delta^{s_2}_l}{\sqrt{6!\,6!}} \!\!\!\!\!\!\!\!\!\!\!\!\!\!\!\!\!\!\!\!\!\!\!\!\!\!\!\!\!\!\!\!\!\!\!\! & 0 & 0 \\
 0 & 0 & 0 & \!\!\!\!\!\!\!\!\!\!\!\!\!\!\!\!\!\!\!\!\!\!\!\!\!\! \frac{\frac{1}{6!} \bdelta^{i_1\cdots i_7}_{s_1t_1\cdots t_6}\bdelta_{j_1\cdots j_7}^{s_2t_1\cdots t_6} \bdelta^{k_1k_2k_3}_{l_1l_2l_3} +\frac{1}{2!}\bdelta_{j_1\cdots j_7}^{i_1\cdots i_7}\bdelta^{k_1k_2k_3}_{s_1t_1t_2}\bdelta_{l_1l_2l_3}^{s_2t_1t_2}}{\sqrt{7!\,3!\,7!\,3!}} \!\!\!\!\!\!\!\!\!\!\!\!\!\!\!\!\!\!\!\!\!\!\!\!\!\!\!\!\!\!\!\!\!\!\!\!\!\!\!\!\! & 0 \\
 0 & 0 & 0 & 0 & \!\!\!\!\!\!\!\!\!\!\!\!\!\!\!\!\!\!\!\!\!\!\!\!\!\!\!\!\!\!\!\!\!\!\!\!\!\!\! \frac{\frac{1}{6!}\bdelta^{i_1\cdots i_7}_{s_1t_1\cdots t_6}\bdelta_{j_1\cdots j_7}^{s_2t_1\cdots t_6} \bdelta^{k_1\cdots k_6}_{l_1\cdots l_6} +\frac{1}{5!} \bdelta^{i_1\cdots i_7}_{j_1\cdots j_7}\bdelta^{k_1\cdots k_6}_{s_1t_1\cdots t_5}\bdelta_{l_1\cdots l_6}^{s_2t_1\cdots t_5}}{\sqrt{7!\,6!\,7!\,6!}} 
\end{array}\right)}}
\nn\\
 &\!\!\!\!\qquad\qquad\quad - \frac{2\,\delta_{s_1}^{s_2}}{9-d}\, \delta^{\sMA}_{\sMB} \,, 
\\
 &\!\!\!\!{\footnotesize 
 (R^{s_1s_2s_3})^{\sMA}{}_{\sMB} \equiv 
 {\arraycolsep=0mm 
 \left(\begin{array}{ccccc}
 ~0~ & \frac{\bdelta^{is_1s_2s_3}_{j_1\cdots j_4}}{\sqrt{4!}} & 0 & 0 & 0 \\
 0 & 0 & \!\!\!\frac{\frac{3-\sqrt{2}}{7} \bdelta^{i_1\cdots i_4s_1s_2s_3}_{j_1\cdots j_6l}-\frac{1}{2!} \bdelta^{i_1\cdots i_4t_1t_2}_{j_1\cdots j_6} \bdelta^{s_1s_2s_3}_{t_1t_2l}}{\sqrt{4!\,6!}}\!\!\!\!\!\!\!\!\!\! & 0 & 0 \\
 0 & 0 & 0 & \!\!\!\!\!\!\!\!\!\!\!\!\!\!\!\!\!\!\!\!\!\!\!\!\!\!\!\!\frac{\frac{3-\sqrt{2}}{7} \bdelta^{i_1\cdots i_6k}_{j_1\cdots j_7} \bdelta^{s_1s_2s_3}_{l_1l_2l_3}-\frac{1}{2!} \bdelta^{i_1\cdots i_6r}_{j_1\cdots j_7} \bdelta^{kt_1t_2}_{l_1l_2l_3} \bdelta^{s_1s_2s_3}_{t_1t_2r}}{\sqrt{6!\,7!\,3!}}\!\!\!\!\!\!\!\!\!\!\!\!\!\!\!\! & 0 \\
 0 & 0 & 0 & 0 & \frac{\bdelta^{i_1\cdots i_7}_{j_1\cdots j_7} \bdelta^{k_1k_2k_3s_1s_2s_3}_{l_1\cdots l_6}}{\sqrt{7!\,3!\,7!\,6!}} \\
 0 & 0 & 0 & 0 & 0 
\end{array}\right)}} , 
\\
 &\!\!\!\!{\footnotesize 
 (R_{s_1s_2s_3})^{\sMA}{}_{\sMB} \equiv 
 {\arraycolsep=0mm 
 \left(\begin{array}{ccccc}
 0 & 0 & 0 & 0 & 0 \\
 \frac{\bdelta_{js_1s_2s_3}^{i_1\cdots i_4}}{\sqrt{4!}} & 0 & 0 & 0 & 0 \\
 0 & \!\!\!\frac{\frac{3-\sqrt{2}}{7} \bdelta_{j_1\cdots j_4s_1s_2s_3}^{i_1\cdots i_6k} -\frac{1}{2!} \bdelta_{j_1\cdots j_4t_1t_2}^{i_1\cdots i_6} \bdelta_{s_1s_2s_3}^{t_1t_2k}}{\sqrt{6!\,4!}}\!\!\!\!\!\!\!\!\!\!\!\!\!\! & 0 & 0 & 0 \\
 0 & 0 & \!\!\!\!\!\!\!\!\!\!\!\!\!\!\!\!\!\!\!\!\!\!\!\!\!\frac{\frac{3-\sqrt{2}}{7} \bdelta_{j_1\cdots j_6l}^{i_1\cdots i_7} \bdelta_{s_1s_2s_3}^{k_1k_2k_3}-\frac{1}{2!} \bdelta_{j_1\cdots j_6r}^{i_1\cdots i_7} \bdelta_{lt_1t_2}^{k_1k_2k_3} \bdelta_{s_1s_2s_3}^{t_1t_2r}}{\sqrt{7!\,3!\,6!}}\!\!\!\!\!\!\!\!\!\!\!\!\!\!\!\!\!\!\! & 0 & 0 
\\
 0 & 0 & 0 & ~~\frac{\bdelta_{j_1\cdots j_7}^{i_1\cdots i_7} \bdelta_{l_1l_2l_3s_1s_2s_3}^{k_1\cdots k_6}}{\sqrt{7!\,6!\,7!\,3!}}~~ & 0 
\end{array}\right)}} , 
\\
 &\!\!\!\!{\footnotesize (R^{s_1\cdots s_6})^{\sMA}{}_{\sMB} \equiv 
 \begin{pmatrix}
 0 & 0 & \frac{\frac{1+2\sqrt{2}}{7} \bdelta^{s_1\cdots s_6i}_{j_1\cdots j_6l}-\bdelta^{s_1\cdots s_6}_{j_1\cdots j_6} \delta^i_l}{\sqrt{6!}} & 0 & 0 \\
 0 & 0 & 0 & \frac{\bdelta^{s_1\cdots s_6r}_{j_1\cdots j_7} \bdelta^{i_1\cdots i_4}_{l_1l_2l_3r}}{\sqrt{7!\,3!\,4!}} & 0 \\
 0 & 0 & 0 & 0 & \frac{\frac{1+2\sqrt{2}}{7} \bdelta_{j_1\cdots j_7}^{i_1\cdots i_6k} \bdelta^{s_1\cdots s_6}_{l_1\cdots l_6}-\bdelta^{s_1\cdots s_6k}_{j_1\cdots j_7} \bdelta^{i_1\cdots i_6}_{l_1\cdots l_6}}{\sqrt{7!\,6!\,6!}} \\
 0 & 0 & 0 & 0 & 0 \\
 0 & 0 & 0 & 0 & 0 
 \end{pmatrix} } ,
\\
 &\!\!\!\!{\footnotesize (R_{s_1\cdots s_6})^{\sMA}{}_{\sMB} \equiv 
 \begin{pmatrix}
 0 & 0 & 0 & 0 & 0 \\
 0 & 0 & 0 & 0 & 0 \\
 \frac{\frac{1+2\sqrt{2}}{7} \bdelta_{s_1\cdots s_6j}^{i_1\cdots i_6k}-\bdelta_{s_1\cdots s_6}^{i_1\cdots i_6} \delta_j^k}{\sqrt{6!}} & 0 & 0 & 0 & 0 \\
 0 & \!\!\frac{\bdelta_{s_1\cdots s_6r}^{i_1\cdots i_7} \bdelta_{j_1\cdots j_4}^{k_1k_2k_3r}}{\sqrt{7!\,3!\,4!}}\!\! & 0 & 0 & 0 
\\
 0 & 0 & \frac{\frac{1+2\sqrt{2}}{7} \bdelta^{i_1\cdots i_7}_{j_1\cdots j_6l} \bdelta_{s_1\cdots s_6}^{k_1\cdots k_6}-\bdelta_{s_1\cdots s_6l}^{i_1\cdots i_7} \bdelta_{j_1\cdots j_6}^{k_1\cdots k_6}}{\sqrt{7!\,6!\,6!}} & 0 & 0 
 \end{pmatrix} } .
\end{align}

From the relation \eqref{eq:eta-invariant-tensor}, for generalized vectors $A^I$ and $B^J$ that transform in the $R_1$-representation and $C^{\sMA}$ that transforms in the $R_2$-representation, a combination, $A^I\,B^J\,\eta_{IJ;\,\sMA}\,C^{\sMA}$, is invariant under $U$-duality transformations:
\begin{align}
 A^I\,B^J\,\eta_{IJ;\,\sMA}\,C^{\sMA}
 \to (\Exp{h^{\bm{\alpha}}\,T_{\bm{\alpha}}})^I{}_K\,(\Exp{h^{\bm{\alpha}}\,T_{\bm{\alpha}}})^J{}_L\,A^K\,B^L\,\eta_{IJ;\,\sMA}\, (\Exp{h^{\bm{\alpha}}\,T_{\bm{\alpha}}})^{\sMA}{}_{\sMB}\, C^{\sMB}
 = A^K\,B^L\,\eta_{KL;\,\sMB}\,C^{\sMB}\,.
\end{align}

\subsection{$\eta$-symbols in the M-theory parameterization}
\label{app:eta-symbol}

The $\eta$-symbols $\eta^{\sMA} =(\eta^{IJ;\,\sMA})$ and $\eta_{\sMA} =(\eta_{IJ;\,\sMA})$ ($\eta^{IJ;\,\sMA}=\eta^{JI;\,\sMA}$ and $\eta_{IJ;\,\sMA}=\eta_{JI;\,\sMA}$) are constant matrices that connect the symmetric product of two $R_1$-representations and the $R_2$-representation. 
When we consider M-theory, we decompose the $R_2$-representation as
\begin{align}
 (\eta_{\sMA}) = \biggl(\eta_i,\, \frac{\eta_{i_1\cdots i_4}}{\sqrt{4!}},\, \frac{\eta_{i_1\cdots i_6,\,k}}{\sqrt{6!}},\, \frac{\eta_{i_1\cdots i_7,\,k_1k_2k_3}}{\sqrt{7!\,3!}},\, \frac{\eta_{i_1\cdots i_7,\,k_1\cdots k_6}}{\sqrt{7!\,6!}} \biggr)\,. 
\end{align}
The two types of $\eta$-symbols, $\eta^{\sMA}$ and $\eta_{\sMA}$, are simply related as
\begin{align}
 \eta^{IJ;\,\sMA} = \eta_{IJ;\,\sMA} \,, 
\end{align}
as matrices. 
Their explicit matrix forms are determined in \cite{Sakatani:2017xcn} and are given as follows (see \cite{Sakatani:2017xcn} for the explicit form of $\eta^{\sMA}$):
\begin{align}
 \eta_k 
 &\equiv \begin{pmatrix}
 0 & \frac{\bdelta_{k i}^{j_1j_2}}{\sqrt{2!}} & 0 & 0 \\
 \frac{\bdelta_{k j}^{i_1i_2}}{\sqrt{2!}} & 0 & 0 & 0 \\
 0 & 0 & 0 & 0 \\
 0 & 0 & 0 & 0 
 \end{pmatrix} , 
\\
 \eta_{k_1\cdots k_4} 
 &\equiv \begin{pmatrix}
 0 & 0 & \frac{\bdelta_{i k_1\cdots k_4}^{j_1\cdots j_5}}{\sqrt{5!}} & 0 \\
 0 & \frac{\bdelta_{k_1\cdots k_4}^{i_1i_2j_1j_2}}{\sqrt{2!\,2!}} & 0 & 0 \\
 \frac{\bdelta_{j k_1\cdots k_4}^{i_1\cdots i_5}}{\sqrt{5!}} & 0 & 0 & 0 \\
 0 & 0 & 0 & 0
 \end{pmatrix} , 
\\
 \eta_{k_1\cdots k_6,\,l} &\equiv \eta^{\text{KKM}}_{k_1\cdots k_6,\,l} + \eta_{k_1\cdots k_6l} \,, 
\\
 &{}{\tiny\hspace*{-20mm}{\eta^{\text{KKM}}_{k_1\cdots k_6,\,l}} 
 {\arraycolsep=-2mm \equiv \left(\begin{array}{cccc}
 0 & 0 & 0 & \frac{\bdelta_{k_1\cdots k_6 i}^{j_1\cdots j_7} \delta_l^j- \frac{\bdelta_{k_1\cdots k_6 l}^{j_1\cdots j_7} \delta_i^j}{7}}{\sqrt{7!}}~~ \\
 0 & 0 & \frac{-\Bigl(\bdelta_{k_1\cdots k_6}^{j_1\cdots j_5k}\bdelta_{kl}^{i_1i_2}- \frac{2}{7}\bdelta_{k_1\cdots k_6 l}^{j_1\cdots j_5i_1i_2}\Bigr)}{\sqrt{2! 5!}} & 0 \\
 0 & \frac{-\Bigl(\bdelta_{k_1\cdots k_6}^{i_1\cdots i_5k}\bdelta_{kl}^{j_1j_2}-\frac{2}{7} \bdelta_{k_1\cdots k_6 l}^{i_1\cdots i_5j_1j_2}\Bigr)}{\sqrt{2! 5!}} & 0 & 0 \\
 ~~\frac{\bdelta_{k_1\cdots k_6 j}^{i_1\cdots i_7} \delta_l^i- \frac{\bdelta_{k_1\cdots k_6 l}^{i_1\cdots i_7} \delta_j^i}{7}}{\sqrt{7!}} & 0 & 0 & 0
 \end{array}\right) ,}}
\\
 \eta_{k_1\cdots k_7} 
 &\equiv \frac{1}{7\sqrt{2}} \begin{pmatrix}
 0 & 0 & 0 & 3\,\frac{\bdelta_{k_1\cdots k_7}^{j_1\cdots j_7}\,\delta_i^j}{\sqrt{7!}} \\
 0 & 0 & \frac{\bdelta_{k_1\cdots k_7}^{j_1\cdots j_5i_1i_2}}{\sqrt{2!\,5!}} & 0 \\
 0 & \frac{\bdelta_{k_1\cdots k_7}^{i_1\cdots i_5j_1j_2}}{\sqrt{2!\,5!}} & 0 & 0 \\
 3\,\frac{\bdelta_{k_1\cdots k_7}^{i_1\cdots i_7}\,\delta_j^i}{\sqrt{7!}} & 0 & 0 & 0
\end{pmatrix} , 
\\
 \eta_{k_1\cdots k_7,\,l_1l_2l_3} 
 &\equiv 
 {\arraycolsep=-1.0mm \left(\begin{array}{cccc}
 0 & 0 & 0 & 0 \\
 0 & 0 & 0 & \frac{-\bdelta_{l_1l_2l_3 m_1\cdots m_4}^{j_1\cdots j_7}\,\bdelta^{j i_1i_2 m_1\cdots m_4}_{k_1\cdots k_7}}{4!\sqrt{2!\,7!}} \\
 0 & 0 & \frac{\bdelta^{i_1\cdots i_5m_1m_2}_{k_1\cdots k_5k_6k_7}\,\bdelta_{m_1m_2 l_1l_2l_3}^{j_1\cdots j_5}}{2!\sqrt{5!\,5!}} & 0 \\
 ~~~0~~~ & \frac{-\bdelta_{l_1l_2l_3 m_1\cdots m_4}^{i_1\cdots i_7}\,\bdelta^{i j_1j_2 m_1\cdots m_4}_{k_1\cdots k_7}}{4!\sqrt{2!\,7!}} & 0 & 0
 \end{array}\right)} , 
\\
 \eta_{k_1\cdots k_7,\,l_1\cdots l_6} 
 &\equiv \begin{pmatrix}
 0 & 0 & 0 & 0 \\
 0 & 0 & 0 & 0 \\
 0 & 0 & 0 & \frac{\bdelta_{k_1\cdots k_7}^{j_1\cdots j_7}\,\bdelta_{l_1\cdots l_6}^{j i_1\cdots i_5}}{\sqrt{5!\,7!}} \\
 0 & 0 & \frac{\bdelta_{k_1\cdots k_7}^{i_1\cdots i_7}\,\bdelta_{l_1\cdots l_6}^{i j_1\cdots i_5}}{\sqrt{5!\,7!}} & 0
 \end{pmatrix} .
\end{align}
We also define the $\Omega$-tensor:
\begin{align}
 (\Omega_{IJ}) &\equiv \begin{pmatrix}
 0 & 0 & 0 & \frac{\epsilon^{j_1\cdots j_7}\,\delta_i^j}{\sqrt{7!}} \\
 0 & 0 & \frac{\epsilon^{i_1i_2j_1\cdots j_5}}{\sqrt{2!\,5!}} & 0 \\
 0 & -\frac{\epsilon^{i_1\cdots i_5j_1j_2}}{\sqrt{2!\,5!}} & 0 & 0 \\
 -\frac{\epsilon^{i_1\cdots i_7}\,\delta_j^i}{\sqrt{7!}} & 0 & 0 & 0
 \end{pmatrix} ,
\\
 (\Omega^{IJ}) &\equiv \begin{pmatrix}
 0 & 0 & 0 & \frac{\epsilon_{j_1\cdots j_7}\,\delta^i_j}{\sqrt{7!}} \\
 0 & 0 & \frac{\epsilon_{i_1i_2j_1\cdots j_5}}{\sqrt{2!\,5!}} & 0 \\
 0 & -\frac{\epsilon_{i_1\cdots i_5j_1j_2}}{\sqrt{2!\,5!}} & 0 & 0 \\
 -\frac{\epsilon_{i_1\cdots i_7}\,\delta^j_i}{\sqrt{7!}} & 0 & 0 & 0
 \end{pmatrix} .
\end{align}
The relation between the $\eta$-symbols (and the $\Omega$-tensor) and the $Y$-tensor known in the literature has been shown in detail in \cite{Sakatani:2017xcn} (see, in particular, Appendix B therein). 
Similar expressions for the $\eta$-symbols and the $\Omega$-tensor that are suitable for type IIB theory are given in \cite{Sakatani:2017xcn}.

\end{document}